\documentclass[a4paper,oneside,12pt]{memoir}

\usepackage[utf8]{inputenc}
\usepackage{graphicx}
\usepackage{amsmath,amssymb}
\usepackage[unicode,linktocpage,bookmarksopen,hypertexnames=false]{hyperref}
\newsubfloat{figure}
\usepackage{multicol}
\usepackage{slashed}

\subcaptionstyle{\centering}

\def\Tr{\text{Tr}\,}
\def\bra#1{\langle#1|}
\def\ket#1{|#1\rangle}
\def\braket#1#2{\langle#1|#2\rangle}

\def\zz{\mathbb{Z}}

\DeclareMathOperator*{\sprod}{\textstyle \prod}
\DeclareMathOperator*{\ssum}{\textstyle \sum}

\setsecnumdepth{subsection}
\maxtocdepth{subsection}

\OnehalfSpacing*

\checkandfixthelayout
\fixpdflayout

\newcommand{\faculty}[1]{%
   \gdef\fac{#1}}
\newcommand{\fac}{}

\newcommand{\tutor}[1]{%
   \gdef\tut{#1}}
\newcommand{\tut}{}

\begin{document}

\frontmatter

\begin{titlingpage}
\faculty{Jagiellonian University \\
  Faculty of Physics, Astronomy and Applied Computer Science\\[20pt]
  \includegraphics[width=2.5cm]{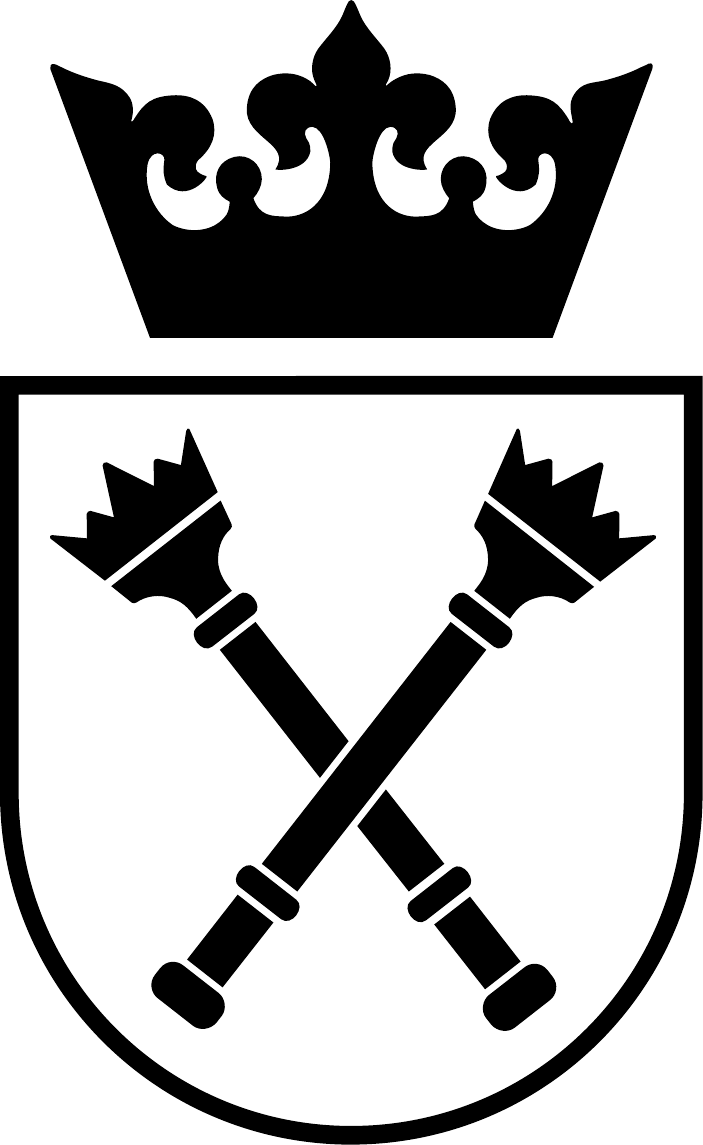}\\[40pt]}
\title{\HUGE\textbf{Volume reduction in large-$N$ lattice gauge theories}\\[15pt]}
\author{\LARGE Mateusz Koreń\\[12pt]}
\tutor{Ph.D. thesis under supervision of prof. Jacek Wosiek}
\date{\large Kraków, 2013}
\aliaspagestyle{titlingpage}{empty}
\setlength{\droptitle}{-100pt}
\maketitle
\end{titlingpage}
\tableofcontents

\chapter{Introduction}

\section*{Abstract}

This work covers volume reduction in quantum field theories on a lattice at large $N$ (number of colors), as first
described by Eguchi and Kawai in Ref.~\cite{ek82}. The volume reduction (or volume independence) means that the theory
defined on an arbitrarily small lattice is equivalent in the large-$N$ limit to the theory on an infinite lattice with
the same bare parameters.

We analyze the volume reduction by means of Monte Carlo simulations using the lattice model on a single site (or a small
fixed number of sites) with Wilson fermions in the adjoint representation, using $N$ up to 60. Most of the results focus
on two flavours of Dirac fermions and the single fermionic flavour is also discussed where there is a significant
difference of behaviour.

We find that the $(\zz_N)^4$ center symmetry, necessary for the realization of volume reduction, is unbroken in the
reduced model for a large range of parameters and, in particular, that the maximum admissible value of the adjoint
fermion mass is non-zero in the large-$N$ limit.

We calculate physical quantities, such as the plaquette, the static quark potential and the eigenvalues of the Dirac
operator. We analyze the finite-$N$ corrections and consider the practicality of volume-reduced models in supplementing
the large-volume calculations.

\section*{Thesis organization}

This work is organized as follows. In Chapter \ref{ch:intro} we introduce the basic notations and show the derivation of
Eguchi-Kawai reduction as well as the reason of its failure, together with some historical attempts to cure the problems
of the original construction. In Chapter \ref{ch:equiv} we give a pedagogical review of the modern way of understanding
volume reduction in the language of large-$N$ orbifold equivalences. Chapter \ref{ch:aek} contains a definition of the
lattice model we analyze, and a review of perturbative predictions for the volume-reduced models.

Chapters \ref{ch:mc}, \ref{ch:ph_diag} and \ref{ch:obs} contain the main part of this work which is the numerical
investigation of the large-$N$ volume-reduced lattice model with adjoint fermions. Chapter \ref{ch:mc} describes the
numerical setup for the Monte Carlo simulations used to generate our results, presented in the subsequent chapters.
Chapter \ref{ch:ph_diag} describes the phase diagram of the analyzed model and focuses on finding the range of
parameters where volume reduction holds. In Chapter \ref{ch:obs} we analyze several physical quantities in the
volume-reduced model.

Finally, we conclude with Chapter \ref{ch:sum} where we summarize the obtained results and give an outline of future
directions to extend the analysis performed in this work.

\section*{Publications}

A substantial part of the results presented in this thesis has already been published by the author and collaborators in
Refs.~\cite{bks11,bks12}.
\pagebreak
\section*{Acknowledgments}

I would like to thank to my supervisor prof.\ Jacek Wosiek for sharing his experience and providing guidance and
support. I am also greatly indebted to prof.\ Stephen Sharpe who was a mentor for me during my internship at the
University of Washington in Seattle where large part of this work was created. I have also greatly benefited from
discussions with Barak Bringoltz, as well as with Masanori Hanada, Piotr Korcyl and Mithat \"Unsal.

This work was supported by Foundation for Polish Science MPD Programme co-financed by the European Regional Development
Fund, agreement no. MPD/2009/6.

I am grateful to the University of Washington for hospitality. Large part of the numerical simulations was done using
\verb#Shiva# computing cluster and \verb#Deszno# supercomputer at the Faculty of Physics, Astronomy and Applied Computer
Science of Jagiellonian University.

Last but not least, I would like to thank to my parents and to my wife Donata for their continuous love and support
during my Ph.D. studies and the writing of this thesis.

\mainmatter

\chapter{Eguchi-Kawai reduction}
\label{ch:intro}

\section{Large-$N$ lattice gauge theory primer}
\label{sec:primer}

Lattice gauge theory \cite{wil74} is based on the euclidean path integral formulation of quantum field theories
\cite{feh65,ps95}. The spacetime is discretized as a 4-dimensional cubic lattice with spacing $a$. The lattice coordinates
are labeled by
\begin{equation}
x = an = a(n_1,n_2,n_3,n_4),\quad n_i\in\mathbb{Z}.
\end{equation}

To introduce gauge fields we assign a group element $U_\mu(x)\in\mathcal{G}$ (in this work $\mathcal{G}$ will typically
be $SU(N)$, unless otherwise stated) to lattice links connecting neighboring sites $(x, x+\mu)$ -- matrix notation is
imposed in gauge (``color'') indices\footnote{In this work we always understand the notation $x+\mu$ as lattice vector
$x$ plus unit lattice vector in direction $\mu$.}. Fermion fields are introduced by assigning a Dirac spinor
$\psi_\alpha(x)$ to each lattice site. The anticommuting nature of fermion fields is taken into account by using
Grassmann variables.

Gauge transformation of the fermion field under group $\mathcal{G}$ is given by
\begin{equation}
\psi_\alpha(x)\to\omega(x)\psi_\alpha(x),
\end{equation}
where $\omega(x)\in\mathcal{R}(\mathcal{G})$ is a matrix in some given representation of the group $\mathcal{G}$. The
gauge fields transform according to
\begin{equation}
U_\mu(x) \to \omega(x) U_\mu(x) \omega^{-1}(x+\mu).
\label{eq:trans}
\end{equation}

In continuum, parallel transporter along a curve $\Gamma$ is given by
\begin{equation}
U_\Gamma = \mathcal{P}\exp\{ig\int_\Gamma \hat{A}_\mu(x) dx_\mu\},
\end{equation}
where $\mathcal{P}$ is the Dyson's path ordering, $g$ is the bare coupling constant and $\hat{A}_\mu(x) =
A_\mu(x)^a\tau^a$ is the vector potential of the gauge field ($\tau^a$ are the generators of the proper gauge algebra).

The same quantity on the lattice is given by a product of link matrices along the contour $\Gamma$. Thus we identify the
lattice variables with their continuum counterparts via
\begin{equation}
U_\mu(x)=\mathrm{exp}\big\{iga\hat{A}_\mu(x+\tfrac12\mu) \big\},
\end{equation}
Moreover, it is easy to see that, in accordance with Eq.~\ref{eq:trans}, any parallel transporter along closed contour
$\Gamma$ transforms covariantly and its trace, called the Wilson loop\footnote{It is quite common in the literature to
use the term Wilson loop also for the untraced operator along a closed contour. We also use this convention in this
work, using the term ``Wilson loop matrix'' for the untraced operator whenever the meaning cannot be easily deduced from
the context.}
\begin{equation}
W_\Gamma\equiv \Tr U_\Gamma
\end{equation}
is gauge invariant.

Let us also mention about a special type of Wilson loop that will be of great importance in this work. Consider a finite
lattice of sizes $\{L_\mu\}$, with periodic boundary conditions. We can construct a closed contour by taking a straight
line of length $L_\mu$ in the proper direction -- a Wilson loop along such line is called the Polyakov loop (or Wilson
line) $P_\mu$ and is the simplest example of a non-contractible loop (i.e.\ loop whose winding number is different from
zero).

We can now introduce the action for the theory. It must be gauge invariant and it must give the continuum action
($\sim\Tr \hat{F}^2$) in the limit $a\to0$. The simplest choice for the gauge action, introduced by Wilson \cite{wil74},
is based on the simplest non-trivial Wilson loop, the so-called plaquette:
\begin{equation}
\Tr U^\square_{\mu\nu}(x) = \Tr U_\mu(x)U_\nu(x+\mu)U_\mu^{-1}(x+\nu)U_\nu^{-1}(x),
\end{equation}
which can be shown (see e.g.\ Ref.~\cite{cre91}) to correspond to $\,\Tr\!\exp\{iga^2\hat{F}_{\mu\nu}(x)\}$. Therefore
we define the action for the gauge group $SU(N)$ as
\begin{equation}
  S_{\text{gauge}}=\beta\sum_{x,\,\mu<\nu}\big\{1-\frac{1}{N}\mathrm{ReTr}U^\square_{\mu\nu}(x)\big\},
\label{eq:action}
\end{equation}
the sum goes over all plaquettes on the lattice and $\beta$ is the normalization factor which for conformity
with the continuum theory must be set to
\begin{equation}
  \beta=\frac{2N}{g^2}\,.
\end{equation}

The action for the fermion field
\begin{equation}
S_\text{ferm} = \bar{\psi}\,D_\text{lat\,}\psi
\end{equation}
requires more attention as the naive discretization of the Dirac operator leads to the famous doubling problem
\cite{cre91}. In this work we choose the simplest solution to this problem, namely the Wilson fermions which remove the
unwanted doublers at the cost of explicit breaking of the chiral symmetry which is only restored as $a\to0$
\cite{cre91,wil77}.
The Wilson Dirac operator (with the Wilson parameter $r=1$ and after customary rescaling) is equal to
\begin{equation}
D_\text{lat}^W(x,y) = \delta_{xy} - \kappa \sum_{\mu=1}^4 \Big[(1-\gamma_\mu) U^{\mathcal{R}}_\mu(x)\delta_{y,x+\mu}
+ (1+\gamma_\mu)U^{-1\,\mathcal{R}}_\mu(y)\delta_{y,x-\mu} \Big],
\end{equation}
where $U^\mathcal{R}$ is the gauge link in the chosen group representation\footnote{In this work we will be particularly
interested in the adjoint representation. The gauge theory with adjoint fermions will be labeled as QCD(Adj) to
distinguish it from the physical QCD with fundamental fermions. Another theory used in this work is QCD(AS), which
contains fermions in the two-index antisymmetric representation.}, $\gamma_\mu$ are the euclidean Dirac matrices and
$\kappa$ is the (dimensionless) hopping parameter related to the bare fermion mass by
\begin{equation}
\kappa=\frac{1}{2a m_0+8}.
\label{eq:kappa}
\end{equation}

Action defined in this way may now be quantized within the path integral formalism, using the generating functional
\begin{equation}
  \mathcal{Z}=\int[dU][d\psi][d\bar\psi^{\,}]e^{-S_\text{gauge}[U]-S_\text{ferm}[\psi,\bar\psi,U]}.
\end{equation}
To avoid Grassmann variables we use the bilinearity of the fermionic action and integrate out $\psi$, $\bar\psi$
to get
\begin{equation}
  \mathcal{Z}=\int[dU]e^{-S_\text{gauge}[U]-\ln\det D^W_\text{lat}[U]}.
\label{eq:z_gen}
\end{equation}

Note that, thanks to the lattice regulator, there is no need to introduce gauge fixing via Fadeev-Popov procedure.
That is particularly clear in the case of finite lattices (which are used in computer simulations) where the path
integral is nothing but a finite-dimensional Haar integral over the group space.

The theory given in Eq.~\ref{eq:z_gen} contains no dimensionful parameters. However, the renormalization group connects
the dimensionless bare coupling constant $g$ with the dimensionful lattice spacing $a$ (which plays the role of the
UV cutoff), in a process called dimensional transmutation -- see Ref.~\cite{cre91} for a comprehensive discussion. For
our purposes, it is sufficient to note that for asymptotically free theories, which are the topic of this work, the
continuum limit $a\to0$ is approached by taking $g\to0$ (or equivalently $\beta\to\infty$).

We now analyze the large-$N$ limit of the theory i.e.\ use gauge group $SU(N)$ with infinite (or, in computer practice,
finite but large) number of colors. That requires \cite{tho74} keeping the product
\begin{equation}
b=\frac1{g^2 N}
\label{eq:b}
\end{equation}
fixed\footnote{$b$ is the inverse of the 't Hooft coupling,  $b^{-1}=\lambda=g^2 N$, and is the quantity typically used
in the large-$N$ lattice literature. Note that $b=\frac{\beta}{2N^2}$.}. As first shown in Ref.~\cite{tho74}, the
large-$N$ limit results in a vast simplification of the perturbative expansion of the theory -- it allows only graphs
with topology of a sphere (the so-called planar graphs).

There are several possibilities of what can happen with the fermion fields. If we start from the QCD with $N=3$ colors
and $N_f$ fundamental fermions there are at least 3 interesting possibilities:
\begin{enumerate}
  \item{'t Hooft limit \cite{tho74} -- keep $N_f$ fixed,}
  \item{Veneziano limit \cite{ven76} -- keep the ratio $N_f/N$ fixed,}
  \item{Corrigan-Ramond limit \cite{cor79} -- keep $N_f$ fixed but use fermions in the antisymmetric representation
  (which coincides with the fundamental representation when $N=3$).}
\end{enumerate}
The first possibility quenches fermions in the large-$N$ limit making the dynamics dependent only on the gluon sector.
The two others allow dynamical fermions making the large-$N$ dynamics more reminiscent of the original QCD.

Another major simplification of the theory in the large-$N$ limit is the factorization of products
\begin{equation}
  \langle \hat{A} \hat{B}\rangle =
  \langle \hat{A}\rangle\langle \hat{B}\rangle + \mathcal{O}(1/N),
  \label{eq:fact}
\end{equation}
where $\hat{A}$ and $\hat{B}$ are quantum operators properly normalized to possess a finite limit as $N\to\infty$
\cite{mim79,yaf82}. Therefore, in the large-$N$ theories variances of operators vanish.

The large-$N$ lattice gauge theory is a very active field of work -- see Ref.~\cite{lp12} for a recent review of
research in this field.

\section{Derivation of the Eguchi-Kawai reduction}

The first notion of volume reduction in large-$N$ lattice gauge theory was introduced by Eguchi and Kawai \cite{ek82}.
Let us consider two theories:
\begin{enumerate}
\item{$U(N)$ (or equivalently $SU(N)$) pure gauge theory on infinite lattice (called the ``full model'' in the following),
with generating functional
\begin{equation}
  \mathcal{Z} = \int\left[dU\right]e^{-S[U]} =
  \int(\sprod_{x,\mu}dU_{x,x+\mu})e^{-\beta\sum_{x,\mu<\nu}\left(1-\frac{1}{N}
  \mathrm{ReTr}U^{\square}_{\mu\nu}(x)\right)},
\end{equation}}
\item{The same theory reduced to a single lattice site with periodic boundary conditions -- the so-called Eguchi-Kawai
model -- with generating functional
\begin{equation}
  \mathcal{Z}_{EK} =
  \int(\sprod_{\mu}dU_{\mu})e^{-\beta\sum_{\mu<\nu}\left(1-\frac{1}{N} \mathrm{ReTr}U^{\square}_{\mu\nu}\right)},
  \label{eq:z_ek}
\end{equation}
where the reduced plaquette is $U^{\square}_{\mu\nu}= U_\mu U_\nu U^\dagger_\mu U^\dagger_\nu$.}
\end{enumerate}
Eguchi and Kawai showed that in the large-$N$ limit these two theories satisfy the same Dyson-Schwinger equations and
are thus equivalent in the large-$N$ limit (for observables that are invariant under translational symmetry), provided
that some conditions are satisfied.

To see what these conditions are let us quickly sketch the proof here. We start with the derivation of the
Dyson-Schwinger equations for the expectation values of Wilson loops (also called loop equations) in the full model
\cite{ek82,mim79,foe79,wad81,kuy03}.

First we choose a closed contour $\Gamma$ such that the link $U_\mu(y)$ is only encountered once in the contour. For
ease of notation we label $\Gamma'$ as the contour $\Gamma$ without the link $(y,y+\mu)$ i.e.
\begin{equation}
W_\Gamma \equiv \Tr U_\Gamma = \Tr U_{\Gamma'} U_\mu(y).
\end{equation}

The quantity $\langle \Tr U_{\Gamma'}\tau^a U_\mu(y) \rangle$, due to the invariance of the measure in the path
integral, must be invariant under the transformation
\begin{equation}
U_\mu(y)\rightarrow(1+i\varepsilon\tau^a)U_\mu(y).
\end{equation}
Collecting the terms linear in $\varepsilon$ one obtains
\begin{align}
\langle \Tr U_{\Gamma'}\tau^a \tau^a U_\mu(y) \rangle &= \frac\beta {2N}\big
\langle \Tr (U_{\Gamma'}\tau^a U_\mu(y))\times\nonumber\\
\ssum_{\rho\neq\mu}\big(&\Tr U_\rho(y) U_\mu(y+\rho) U_\rho(y+\mu) U_\mu^\dagger(y)\tau^a - \nonumber\\
&\Tr \tau^a U_\mu(y) U_\rho(y+\mu) U_\mu^\dagger(y+\rho) U_\rho^\dagger(y) + \nonumber\\
&\Tr U_\rho^\dagger(y-\rho) U_\mu(y-\rho) U_\rho(y+\mu-\rho) U_\mu^\dagger(y)\tau^a- \nonumber\\
&\Tr \tau^a U_\mu(y) U_\rho^\dagger(y+\mu-\rho) U_\mu^\dagger(y-\rho) U_\nu(y-\rho)\big) \big\rangle,
\end{align}
where the traces in the sum consist of all the oriented plaquettes in the action that contain $U_\mu(y)$ or its
hermitian conjugate. One can now use the property of the generators $\tau^a$:
\begin{equation}
\sum_{a=1}^{N^2}\tau^a_{ij}\tau^{a}_{kl} = \tfrac12\delta_{il}\delta_{jk}
\end{equation}
and perform the sum over $a$ to get the loop equation:
\begin{equation}
  \langle W_\Gamma\rangle =
  \frac{1}{g^2N}\sum_{\rho\neq\mu}\left(\langle
  W_{\Gamma'_{+\rho}(y,\mu)}\rangle - \langle
  W_{\Gamma''_{+\rho}(y,\mu)}\rangle + \langle
  W_{\Gamma'_{-\rho}(y,\mu)}\rangle - \langle
  W_{\Gamma''_{-\rho}(y,\mu)}\rangle\right),
\label{eq:ds}
\end{equation}
where the contours are:\\[-42pt]
\begin{adjustwidth}{-0.4cm}{-0.4cm}
\begin{align*}
&\Gamma=(x,\ldots,y,y+\mu,\ldots,x)&\sim \includegraphics{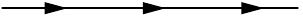}\nonumber\\
&\Gamma'_{+\rho}(y,\mu)=(x,\ldots,y,y+\rho,y+\rho+\mu,y+\mu,\ldots,x)&\sim
\includegraphics{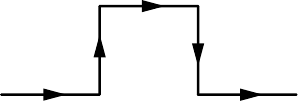}\nonumber\\
&\Gamma''_{+\rho}(y,\mu)=(x,\ldots,y,y+\mu,y+\rho+\mu,y+\rho,y,y+\mu,\ldots,x)\hspace{-0.3cm}&\sim
\includegraphics{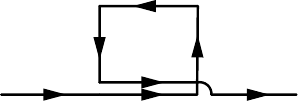}\nonumber\\[12pt]
&\Gamma'_{-\rho}(y,\mu)=(x,\ldots,y,y-\rho,y-\rho+\mu,y+\mu,\ldots,x)&\sim
\raisebox{-24pt}{\includegraphics{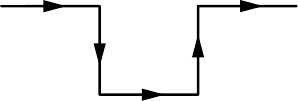}}\nonumber\\
&\Gamma''_{-\rho}(y,\mu)=(x,\ldots,y,y+\mu,y-\rho+\mu,y-\rho,y,y+\mu,\ldots,x)\hspace{-0.3cm}&\sim
\raisebox{-24pt}{\includegraphics{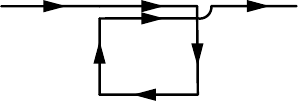}}\nonumber
\end{align*}
\end{adjustwidth}

When the link $U_\mu(y)$ is encountered more than once in the contour we get additional terms in Eq.~\ref{eq:ds}. For
example, when
\begin{equation*}
\Gamma=(x,\ldots,y,y+\mu,\ldots,z,\ldots,y,y+\mu,\ldots,x)
\end{equation*}
(the contour passes through $U_\mu(y)$ twice in the same direction), on the r.h.s.\ of Eq.~\ref{eq:ds} we get an
additional term proportional to $\langle W_{\Gamma_1} W_{\Gamma_2}\rangle$, where $\Gamma_1 =
(y,y+\mu,\ldots,z,\ldots,y)$ and $\Gamma_2 = (y,y+\mu,\ldots,x,\ldots,y)$ sum up to the contour $\Gamma$. One can
systematically add terms from different link repetitions of $U_\mu(y)$ in a similar manner \cite{wad81}.

Let us now consider the same equations in the reduced (EK) model. For every contour
$\Gamma=(x,x+\alpha,x+\alpha+\beta,\ldots,x+\alpha+\beta+\ldots+\omega)$ in the full model one can uniquely assign its
counterpart $\tilde{\Gamma}=(\alpha,\beta,\ldots,\omega)$ in the reduced model.

The only difference in the obtained Dyson-Schwinger equations is the presence of additional terms coming from the fact
that the reduced counterpart of contour $\Gamma=(x,\ldots,y,y+\mu,\ldots,z,z+\mu,\ldots,x)$ has the same link $U_\mu$
repeated even when $y\neq z$. Therefore in this case we get an additional term proportional to $\langle
W_{\tilde{\Gamma}_1} W_{\tilde{\Gamma}_2}\rangle$, where $\tilde{\Gamma}_{1,2}$ correspond to the following (open)
contours in the full model: $\Gamma_1=(y,y+\mu,\ldots,z)$, $\Gamma_2=(z,z+\mu,\ldots,y)$.

Thus, the volume-reduced theory is in general different from the unreduced one. However, in the limit
$N\rightarrow\infty$ we can factorize the additional terms:
\begin{equation}
  \langle W_{\tilde{\Gamma}_1}W_{\tilde{\Gamma}_2}\rangle =
  \langle W_{\tilde{\Gamma}_1}\rangle\langle W_{\tilde{\Gamma}_2}\rangle + \mathcal{O}(1/N).
\end{equation}
Since $\tilde{\Gamma}_1$ and $\tilde{\Gamma}_2$ correspond to open paths, in both of them at least one term $U_\mu$ will
not have a corresponding $U^\dagger_\mu$. We now use the fact that the reduced action possesses a symmetry
(independently for each lattice direction):
\begin{equation}
U_\mu\to e^{i\phi}U_\mu, \quad \text{where} \quad \phi \in \left\{0,\tfrac{2\pi}N,\ldots,\tfrac{2(N-1)\pi}N\right\},
\end{equation}
called the center symmetry -- it is $\zz_N$ for finite $N$ and becomes $U(1)$ in the large-$N$ limit\footnote{The full
symmetry in $d$ spacetime dimensions is $(\mathbb{Z}_N)^d$ and throughout this work we also use the shortened notation
$\mathbb{Z}_N^d$.}. Using it we obtain
\begin{equation}
  \langle W_{\tilde{\Gamma}_1}\rangle = \langle W_{\tilde{\Gamma}_2}\rangle = 0
\end{equation}
and all the unwanted terms disappear. We thus see that Eguchi-Kawai reduction holds iff the center symmetry is not
spontaneously broken.
\vspace{8pt}
\fancybreak*{{* * *}\\[4pt]}

However, the center symmetry in the Eguchi-Kawai model is in fact spontaneously broken at high $\beta$ for $d>2$, as can
be seen both in the Monte Carlo simulations \cite{bhn82,oka82,knn03} and in perturbation theory \cite{bhn82,kmi82} (see
Sec.~\ref{sec:pert}).
This invalidates the Eguchi-Kawai reduction.

Over the years there were several proposals to cure the center-symmetry breaking and obtain a working large-$N$ volume
reduction. Let us quickly describe the most interesting approaches:
\begin{enumerate}
  \item{Force the eigenvalues to satisfy the center symmetry by explicitly fixing them as different elements of the
  $\zz_N$ group (the so-called Quenched Eguchi-Kawai or QEK model) \cite{bhn82,gki82}. This approach is the historically
  first attempt to heal the Eguchi-Kawai reduction and it gained some popularity. However, several years ago it was shown to
  fail \cite{bs08} due to non-trivial correlations between different lattice directions.}
  \item{Use twisted boundary conditions (Twisted Eguchi-Kawai, or TEK model) \cite{go831,go832}. The original
  choice of the twist was shown not to work \cite{bns06,tva07,ahh08} however a different choice that appears to overcome
  the problems was proposed recently \cite{go10,go121}.}
  \item{The center symmetry is intact for physical lattice size larger than some $aL_\text{crit}$ \cite{knn03} -- as long
  as one keeps the lattice size $L>L_\text{crit}$ the volume reduction allows one to perform the calculations as though it
  was infinite. This idea is known as partial reduction or continuum reduction and was studied both in four \cite{knn03}
  and three dimensions \cite{nn03} (it was also the topic of the master's thesis of the author of this work, see
  Ref.~\cite{kor09} for a review of the results). Note that $L_\text{crit}\to\infty$ in the continuum limit.}
  \item{Use adjoint fermions to stabilize the center symmetry (Adjoint Eguchi-Kawai, or AEK model) \cite{kuy07}. This idea
  is inspired by the large-$N$ orbifold equivalences described in Chapter \ref{ch:equiv} and is the basis of this
  work\footnote{There is also a related idea of trace-deformed (or center-stabilized) reduction \cite{uny08,vai11}, that
  however becomes rather complex when reducing more than one lattice direction.}.}
\end{enumerate}

\chapter{Volume reduction as large-$N$ equivalence}
\chaptermark{Reduction as large-$N$ equivalence}
\label{ch:equiv}

\section{Large-$N$ equivalences}
\label{sec:equiv}

The large-$N$ factorization, Eq.~\ref{eq:fact}, not only greatly simplifies the dynamics of the theory but also
resembles the classical limit of quantum mechanics where quantum fluctuations are suppressed as $\hbar\to0$. In fact
this analogy can be made formal, as was shown in Ref.~\cite{yaf82}. The idea of this construction is to find a basis of
coherent states -- in this basis the expectation values of quantum operators become classical observables when
$N\to\infty$, in the full analogy to the $\hbar\to0$ limit. Let us briefly discuss the main ingredients of this
construction (our treatment follows closely that of Ref.~\cite{kuy04}).

We introduce a Lie group $\mathbf{G}$ (called the coherence group) acting on the Hilbert space of the theory
$\mathcal{H}$ via a set of unitary operators $\{\hat{\mathbf{G}}(u)\}$, $u\in\mathbf{G}$. We choose a base
state\footnote{The precise form of the base state is not important for our purposes, see Refs.~\cite{yaf82,kuy04} for a
more comprehensive discussion.} $\ket{0}\in\mathcal{H}$ and generate the coherent states by
\begin{equation}
\ket{u} = \hat{\mathbf{G}}(u)\ket{0}.
\end{equation}

E.g.\ in the case of quantum mechanics of a point particle $\mathbf{G}$ is the Heisenberg group, consisting of space and
momentum translations, while for $U(N)$ hamiltonian lattice gauge theory $\mathbf{G}$ is generated by a Lie algebra
consisting of all hermitian linear combinations of spatial Wilson loops with up to one conjugate momentum or matter
field insertion \cite{yaf82,kuy04}.

The set of coherent states forms an overcomplete basis of the Hilbert space. As $N\to\infty$ the overlaps between
different states $\braket{u}{u'}$ tend exponentially to 0 and the basis becomes orthogonal.

Operators with finite $N\to\infty$ limit of $\frac{\bra{u}\hat{A}\ket{u'}}{\braket{u}{u'}}$ are called classical
operators. We introduce the classical phase space defined as the coadjoint orbit of the coherence group\footnote{See
Section 3 of Ref.~\cite{yaf82} for a thorough discussion. For example, in the case of the point particle the coadjoint
orbit of the Heisenberg group is simply the two-dimensional plane parametrized by the position and momentum of the
particle.}.
For every classical operator we define the classical observable as
\begin{equation}
a(\zeta) \equiv \lim_{N\to\infty}\bra{u}\hat{A}\ket{u},
\end{equation}
where $\zeta$ denotes a point in the classical phase, uniquely determined by $u$ \cite{yaf82}. The following relations
hold for the matrix elements of classical operators:
\begin{align}
\lim_{N\to\infty}\bra{u}\hat{A}\hat{B}\ket{u} &= a(\zeta)b(\zeta),\\
\lim_{N\to\infty}\bra{u}\big[\hat{A},\hat{B}\big]\ket{u} &= \{a(\zeta),b(\zeta)\}_{\text{PB}},
\end{align}
where $\{\cdot,\cdot\}_{\text{PB}}$ is the Poisson bracket (the construction of the classical phase space always allows
the introduction of the Poisson bracket \cite{yaf82}). The classical dynamics is governed by the classical Hamiltonian,
given by the expectation value of the quantum one:
\begin{align}
h_{cl}(\zeta) &\equiv \lim_{N\to\infty}\tfrac1{N^2}\bra{u}\hat{H}\ket{u},\\
\frac{da(\zeta)}{dt} &= \{h_{cl}(\zeta),a(\zeta)\}_{\text{PB}}.
\end{align}
By minimizing the classical Hamiltonian one obtains the ground state of the large-$N$ theory. It is then possible to
systematically add $1/N$ corrections to find the excited states \cite{yaf82}.

Unfortunately, finding the minimum $\zeta_{\text{min}}$ of $h_{cl}$ has not been attained in the case of gauge theories
except for some simple toy models. One can however use the coherent state method to compare different quantum theories
-- if the classical phase spaces and Hamiltonians of the two theories are identical then the theories are equivalent in
the large-$N$ limit. This method has the advantage over the Dyson-Schwinger approach used in Refs.~\cite{ek82,kuy03} in
being completely general -- while the Dyson-Schwinger equations can have multiple solutions and identical equations are
only guaranteed to give coinciding dynamics in the phase continuously connected to strong-coupling large-mass region
\cite{kuy03}.

In the following parts of this section we present two particularly interesting examples of large-$N$ equivalences: the
orbifold and orientifold equivalences. Both these terms originate from string theory but can be described purely in the
QFT language and will be analyzed in this work without mentioning their stringy interpretation.

\subsection{Orbifold equivalence}

Large-$N$ orbifold equivalences were conjectured in Ref.~\cite{str01} as a way of relating non-perturbative aspects of
supersymmetric and non-supersymmetric theories related by orbifold projection. Then they were put in a rigorous
framework for gauge theories with and without matter fields in Refs.~\cite{kuy03,kuy04}. In this section we limit
ourselves to a general discussion of orbifold equivalences, postponing the detailed discussion until
Sec.~\ref{sec:vroe}.

The orbifold projection is based on a discrete symmetry (the so-called projection group $\mathbf{P}$) of a ``parent''
field theory. We project out all degrees of freedom in the parent that are not invariant under this symmetry, yielding a
``daughter'' field theory. Both theories possess the same large-$N$ limit for a class of observables commuting with
$\mathbf{P}$ (called the ``neutral sector''), provided that $\mathbf{P}$ is not spontaneously
broken\footnote{Technically, the equivalence is independent of the realization of $\mathbf{P}$. However, if
$\mathbf{P}$ is spontaneously broken the minimum of the classical Hamiltonian lies outside the neutral sector, thus
preventing the extraction of information about quantum theories of interest and making the problem rather
academic \cite{kuy04}.}.

The projection groups typically used for orbifold equivalences, such as volume reduction, are cyclic groups or products
thereof. To be specific let us choose just one cyclic group $\mathbf{P} = \mathbb{Z}_m$ embedded in a pure-gauge $U(N)$
lattice gauge theory\footnote{The procedure can be extended to include adjoint matter fields in a completely analogous
way \cite{kuy04}, see also Sec.~\ref{sec:vroe}.}, where $N = mN'$ with integer $N'$.

The embedding is chosen so that the gauge fields $U_\mu(x)$ transform under $\mathbf{P}$ as
\cite{kuy04,sch98}:
\begin{equation}
U_\mu(x) \to \gamma U_\mu(x) \gamma^\dagger,
\label{eq:trans_orb}
\end{equation}
where $\gamma = \Omega^{(m)} \times \mathbb{I}_{N'}$, with $\Omega^{(m)}$ defined as the clock matrix:
\begin{equation}
\Omega^{(m)} \equiv \text{diag}[1,\omega,\ldots,\omega^{m-1}],\ \ \text{with}\ \ \omega=e^{2\pi i/m}.
\label{eq:clock}
\end{equation}

The orbifold projection removes the degrees of freedom that are not invariant under the transformation given in
Eq.~\ref{eq:trans_orb}\footnote{The more general form of the constraint is of the form $U_\mu(x)=\gamma U_\mu(x)
\gamma^\dagger e^{2\pi i r/m}$, where $r\in\mathbb{Z}$ is called the charge of the field. Non-zero charge will be used
for the gauge fields in Sec.~\ref{sec:vroe}, cf.\ Eq.~\ref{eq:cond_orb}.}. As a result, the \linebreak $N\times N$
matrices of the gauge fields are left with non-zero entries only in $m$ blocks, each of size $N'\times N'$ (an example
is given in Fig.~\ref{fig:orbifold}). All of these blocks are unitary by construction, thus the daughter theory will
have a $[U(N')]^m$ symmetry group that can be interpreted as $U(N')$ gauge theory with additional internal space (called
``theory space'' $\mathbf{T}$ by the authors of Ref.~\cite{kuy04}) consisting of $m$ independent factors on a
discretized circle (or a torus in the general case of many cyclic groups).

\begin{figure}[tbp!]
  \begin{center}
    \begin{equation*}
    U^{(\text{daughter})} = \begin{pmatrix}
    U_{11} & U_{12} & 0 & 0 & 0 & 0 \\
    U_{21} & U_{22} & 0 & 0 & 0 & 0 \\
    0 & 0 & U_{33} & U_{34} & 0 & 0 \\
    0 & 0 & U_{43} & U_{44} & 0 & 0 \\
    0 & 0 & 0 & 0 & U_{55} & U_{56} \\
    0 & 0 & 0 & 0 & U_{65} & U_{66} \\
    \end{pmatrix}
    \end{equation*}
    \caption{An example result of orbifold projection, with $\mathbf{P}=\mathbb{Z}_3$ acting on a $U(6)$ gauge theory.
    The fields invariant under $\mathbf{P}$ are given by the condition $U=\gamma U \gamma^\dagger$, where $\gamma$
    (in a convenient basis) is given by $\gamma = \Omega^{(3)}\times \mathbb{I}_2 = \text{diag}[1,1,e^{2\pi i/3},
    e^{2\pi i/3},e^{-2\pi i/3},e^{-2\pi i/3}]$. The remaining symmetry of the daughter theory is $[U(2)]^3$.}
    \label{fig:orbifold}
  \end{center}
\end{figure}

There is a bijective mapping of the neutral observables between the daughter and parent theories. For example, for the
Wilson loops:
\begin{equation}
\frac1N \Tr U_\Gamma = \frac1m\sum_{i\in\mathbf{T}}\frac1{N'}\Tr U^{(i)}_\Gamma,
\end{equation}
where the discrete index $i$ is used for averaging over the theory space.

In Ref.~\cite{kuy04} the authors have proven that the subgroups of the coherence groups that define the neutral sectors
are isomorphic in the two theories, thus giving the same classical phase spaces and Hamiltonians -- and making the
theories equivalent in the large-$N$ limit, as discussed in the earlier part of this section. One particularly
interesting example of the orbifold equivalence will be the volume reduction, discussed in Sec.~\ref{sec:vroe}, where
the theory space is identified with the physical spacetime.

\subsection{Orientifold equivalence}

Another example of large-$N$ equivalence that attracted a lot of attention is the orientifold equivalence which relates
large-$N$ limits of QCD(Adj) with $n_f$ adjoint Majorana fermions and QCD(AS) with $N_f$ Dirac fermions, at $n_f=N_f$
\cite{as031,asv05}.

This equivalence was initially investigated in the case of $n_f=1$ where it relates a supersymmetric theory
($\mathcal{N}=1$ SYM) with a non-supersymmetric one ($N_f=1$ QCD(AS)). Another particularly interesting possibility,
which will be used in the following part of this work, is $n_f=2$. QCD(AS) at large number of colors is a very
natural large-$N$ limit of the physical $N=3$ QCD \cite{as032} (called the Corrigan-Ramond limit, see
Sec.~\ref{sec:primer}), especially at $N_f=2$ which describes the 2 lightest physical quarks.

In this section we follow the terminology of Ref.~\cite{uy06}. In this treatment the orientifold equivalence is an
example of ``daughter-daughter'' equivalence in the sense that both theories can be constructed from a common parent by
applying different orientifold projections.

The orientifold projections are based on $\mathbb{Z}_2$ projection groups related with charge conjugation
($\mathcal{C}$) symmetry. The parent theory for both theories under investigation is $SO(2N)$ gauge theory with $n_f$
adjoint Majorana fermions. The two theories result from different $\mathbb{Z}_2$ projections of the parent theory
\cite{uy06}. The QCD(Adj) with $n_f$ Majorana fermions is a result of imposing the constraint
\begin{equation}
U = JUJ^T,\ \psi=J\psi J^T,
\end{equation}
where $J=i\sigma_2 \times \mathbb{I}_N\in SO(2N)$, $U$ symbolically denotes the bosonic degrees of freedom and $\psi$
the fermionic degrees of freedom. On the other hand, QCD(AS) with $N_f(=n_f)$ Dirac fermions is obtained from the parent
theory by the projection with the constraint
\begin{equation}
U = JUJ^T,\ \psi=-J\psi J^T.
\label{eq:cond_orient2}
\end{equation}
The projection in Eq.~\ref{eq:cond_orient2} involves additional factor $(-1)^F$, which multiplies the fermionic fields
by $-1$.

One can show that the neutral sectors in both theories consist of $\mathcal{C}$-even operators (see Ref.~\cite{uy06} for
details). Thus, for the equivalence to be meaningful it is necessary that the charge conjugation symmetry is not
spontaneously broken in either of the theories \cite{uy06,asv07}. For the Wilson loops this requirement can be written
simply as
\begin{equation}
\langle\Tr U_\Gamma\rangle = \langle\Tr U_\Gamma^\dagger\rangle.
\end{equation}

While the $\mathcal{C}$ symmetry is expected to be preserved on $\mathbb{R}^4$ \cite{asv07} it was shown to be broken in
QCD(AS) on $\mathbb{R}^3\times S^1$, with periodic boundary conditions for fermions, for small enough radius of the
circle \cite{uy06}.

\section{Volume reduction as large-$N$ orbifold equivalence}
\label{sec:vroe}
Volume reduction/expansion in the language of orbifold projections was introduced in Refs.~\cite{kuy07,neu03}. For
definiteness and ease of notation we consider the following two $d$-dimensional lattice gauge theories:
\begin{itemize}
	\item {Theory 1: $U(N')$ gauge theory on a periodic lattice of volume $\Lambda = L^d$, 	with or without adjoint matter
	fields,}
	\item {Theory 2: $U(N)$ single-site model with the same matter content as theory 1.}
\end{itemize}
The generalization to arbitrary volume of theory 2 and anisotropic lattices is also possible (the only difference is a
somewhat more complicated notation, see the Appendix of Ref.~\cite{kuy07} for details).

\subsection{Theory 1 $\to$ Theory 2 (volume reduction)}
The volume reduction from theory 1 to 2 is implemented by discarding all fields carrying non-zero momentum \cite{kuy07}.
This can be described in the language of orbifold projections with theories 1 and 2 being the parent and daughter
theories respectively, with $N'=N$.

Due to the periodic boundary conditions theory 1 has a $\mathbb{Z}_L^d$ translational invariance. We choose the
projection group $\mathbf{P}=\mathbb{Z}_L^d$ and eliminate all the fields that are not invariant under the translations.
The invariant fields are manifestly visible in the momentum space -- consider a lattice Fourier transform for some
generic field $\Phi$:
\begin{equation}
\Phi_{\text{Th.1}}(x) = \sum_{n\in\mathbb{Z}_L^d} \tilde{\Phi}_n e^{2\pi i n\cdot x/L} \xrightarrow{\,n=0\,}
\Phi_{\text{Th.2}} = \tilde{\Phi}_0.
\end{equation}
As a result of the projection all the components except the (constant in space) zero-momentum mode are discarded.

The projection defines a one-to-one mapping between the Wilson loops in the parent theory, averaged over spacetime, and
the Wilson loops in the daughter theory\footnote{With winding numbers being integer multiples of $L$ (this includes all
the ``ordinary'' contractible Wilson loops with zero winding number).}. For example, the Wilson action of pure gauge
theory is cast to the Eguchi-Kawai action times $L^d$ (the factor of volume accounts for the ratio of discarded degrees
of freedom of the parent theory). This result also applies to the Wilson loops with arbitrary number of adjoint matter
field insertions along the loop (called ``single-trace observables'' by the authors of Ref.~\cite{kuy07}).

\subsection{Theory 2 $\to$ Theory 1 (volume expansion)}
\label{sec:vroe_exp}

We start with pure gauge theory and choose $N=L^d N'$. The single-site model is now the parent theory whereas the
``big'' lattice of theory 1 is the daughter. We choose $\mathbf{P}=\mathbb{Z}_L^d$ which is a subgroup of the
$\zz_N^4$ center symmetry of theory 2.

As usual, the orbifold projection eliminates all degrees of freedom not invariant under $\mathbf{P}$ -- the proper
embedding of the projection group in the full symmetry of the theory will allow us to identify the subblocks of gauge
matrices in the parent with different points in the spacetime of the daughter theory.

The projection is equivalent to imposing a following set of constraints on the gauge fields \cite{kuy07}:
\begin{equation}
U_\mu=\left\{
  \begin{array}{l l}
    \gamma_\nu U_\mu \gamma_\nu^\dagger\,e^{2\pi i/L}, & \quad \mu=\nu\\
    \gamma_\nu U_\mu \gamma_\nu^\dagger\,, & \quad \mu\neq\nu\\
  \end{array} \right.
  \label{eq:cond_orb}
\end{equation}
Here $\gamma_\nu$ are defined as
\begin{equation}
\gamma_\nu = \underbrace{\mathbb{I}_L\times\ldots}_{\nu-1}\times\,\Omega^{(L)}\times
\underbrace{\mathbb{I}_L\times\ldots}_{d-\nu} \times\mathbb{I}_{N'},
\end{equation}
with $\Omega^{(L)}$ being the clock matrix, cf.\ Eq.~\ref{eq:clock}.

As a result of the projection, in every $U_\mu$ there are only $L^d$ non-zero blocks of size $N'\times N'$ left. Each of
the $N'\times N'$ blocks is a unitary matrix by itself, and can be naturally associated with a single link in the
``big'' lattice by inspection of how it couples to other blocks in the projected Eguchi-Kawai action\footnote{The phase
factor $e^{2\pi i / L}$ in Eq.~\ref{eq:cond_orb} is chosen so that the correct coupling of blocks to the nearest
neighbors is obtained -- see the Appendix of Ref.~\cite{kuy07} for a comprehensive discussion.}.

Under the same mapping the Wilson loops of the parent theory are associated one-to-one with Wilson loops in the daughter
theory averaged over spacetime volume. Likewise the action becomes the standard big lattice action up to a constant
factor that ensures the equality of 't Hooft couplings in the two theories.

Addition of adjoint matter fields  (both scalars and fermions) is straightforward as these fields transform in the same
way as the gauge fields, they also preserve center symmetry. Thus the effect of the projection is simply
\begin{equation}
\Phi = \gamma_\nu \Phi \gamma_\nu
\end{equation}
where $\Phi$ is the matter field matrix. There is a one-to-one mapping between the single-trace observables, just as in
the pure-gauge case.

\subsection{Large-$N$ equivalence}

As discussed earlier, the large-$N$ dynamics of parent and daughter theories related by orbifold projections coincide in
the neutral sectors. However, for the ground states (and thus the physical properties) of the theories to coincide, the
symmetries defining the projections must not be spontaneously broken. Of these symmetries, the one that is the most
non-trivial to satisfy is the center symmetry of the small-volume model -- it is broken in the pure-gauge case, thus
invalidating Eguchi-Kawai reduction. In the subsequent chapter we will analyze the introduction of adjoint fermions in
order to keep the center symmetry intact.

\subsection{Effective system size at finite $N$}
\label{sec:l_eff}

The orbifold equivalence is demonstrated by taking $N\to\infty$. However, since in computer simulations we are always
dealing with finite $N$ it is useful to consider, at least qualitatively, the effective size $L_\text{eff}$ of the
volume-reduced lattice (or, equivalently, the effective volume $V_\text{eff}=L_\text{eff}^4$) and its finite-$N$
dependence.

If the large-$N$ equivalence holds, we expect that the theory on a single site with $N$ colors gives the same physical
results as the theory on a volume $V_\text{eff}$ with $N_\text{eff}$ colors -- up to corrections suppressed by powers of
$1/N_\text{eff}$. There is a trade-off between increasing $L_\text{eff}$ and $N_\text{eff}$ and the value of
$N_\text{eff}$ must be large enough so that the finite-$N_\text{eff}$ corrections to the quantities of interest are not
too large. We choose $N_\text{eff}$ fixed and ask what is the dependence $L_\text{eff}(N)$.

The orbifold projection of presented in Sec.~\ref{sec:vroe_exp} gives an explicit prescription of packaging the matrices
in different spacetime points into a larger gauge matrix. The $N\times N$ link matrices are partitioned into blocks of
size $N_\text{eff}\times N_\text{eff}$, with $N_\text{eff} = N/L_\text{eff}^4$. If we fix $N_\text{eff}$ to some
constant value (e.g.\ 3) we obtain the effective-size scaling
\begin{equation}
L_\text{eff}(N)\propto N^{1/4}.
\label{eq:l_eff_orb}
\end{equation}

\chapter{Volume reduction with adjoint fermions}
\chaptermark{Volume reduction w/ adjoint fermions}
\label{ch:aek}

\section{Definition of the Adjoint Eguchi-Kawai model}
\label{sec:def}
Addition of adjoint fermions was proposed by the authors of Ref.~\cite{kuy07} as a way to stabilize center symmetry.
They have shown that the massless adjoint fermions with periodic boundary conditions give a repulsive contribution to
the one-loop potential of Polyakov loop eigenvalues that allows the center symmetry to be preserved. There are also
reasons to believe that the center symmetry may be preserved even with heavy adjoint fermions (see Sec.~\ref{sec:pert}).

In this section we define the model that will be the main topic of this work, the Adjoint Eguchi-Kawai (AEK) model,
i.e.\ a single-site lattice theory with $SU(N)$ gauge group and $N_f$ adjoint Dirac fermions. The generating functional
of the theory is
\begin{equation}
Z_\text{AEK}=\int \prod_\mu [dU]\,\exp\left(-S_\text{gauge}[U] + \ln(\det D_W[U])^{N_f} \right),
\label{eq:Z_AEK}
\end{equation}
where the gauge part is the Eguchi-Kawai action (cf.\ Eq.~\ref{eq:z_ek}):
\begin{equation}
S_\text{gauge}=-2Nb \sum_{\mu<\nu} \text{Re}\Tr U_\mu U_\nu U_\mu^\dagger U_\nu^\dagger + \text{constant}.
\label{eq:S_gauge}
\end{equation}
The constant is independent of the gauge configuration and we neglect it in the numerical simulations. We use Wilson
fermions with periodic boundary conditions in all directions:
\begin{equation}
D_W =
1 - \kappa \left[\sum_{\mu=1}^4 \left( 1 - \gamma_\mu\right) 
U^\text{adj}_\mu
+ \left(1 + \gamma_\mu\right)U^{\dag \text{adj}}_{\mu}\right].
\label{eq:D_W}
\end{equation}

The Wilson discretization is chosen due to its simplicity (especially when working with fermions that are not very
light), following Refs.~\cite{bks11,bs09} (see also Ref.~\cite{cgu10} for a related calculation on a $2^4$ lattice).
Overlap discretization was also used in the literature \cite{hn09,hn11}.

The bare quark mass is zero at $\kappa=1/8$. However, since Wilson fermions do not preserve chiral symmetry at finite
$a$, the physical quark masses are additively renormalized. We thus define $\kappa_c(b)$ as the value of $\kappa$ at
which the physical quark mass becomes 0. The value of $\kappa_c$ goes to 1/8 as $a\to0$ but it is in general different
(larger) than that at finite lattice spacing, and the physical quark mass becomes:
\begin{equation}
m_\text{phys} = \frac1a\left(\frac{1}{2\kappa}-\frac{1}{2\kappa_c}\right).
\end{equation}

The gauge theory with adjoint fermions is asymptotically free if $N_f<N_f^I=11/4$, independently on the value of $N$. It
is argued by a range of analytic methods \cite{ds07,san09} that (also independently on $N$) there exists a value
$N^*_f<N_f^I$ above which the massless theory loses its confining character and develops an infrared fixed point
(becomes conformal). The range $N_f\in[N_f^*,N_f^I)$ is called the ``conformal window''. $N_f^*$ is estimated by various
methods to be in $N_f^*\in[1\tfrac1{16},2\tfrac3{40}]$ \cite{ds07}. However, only non-perturbative studies can
give a definite answer whether the theory with given $N_f$ lies in the conformal window or not\footnote{This is
especially interesting in the case $N_f=2$ for which the analytic methods are not unanimous. Several lattice studies
have been performed for this theory (see below).}.

\begin{figure}[tbp!]
  \begin{center}
    \includegraphics[width=13cm]{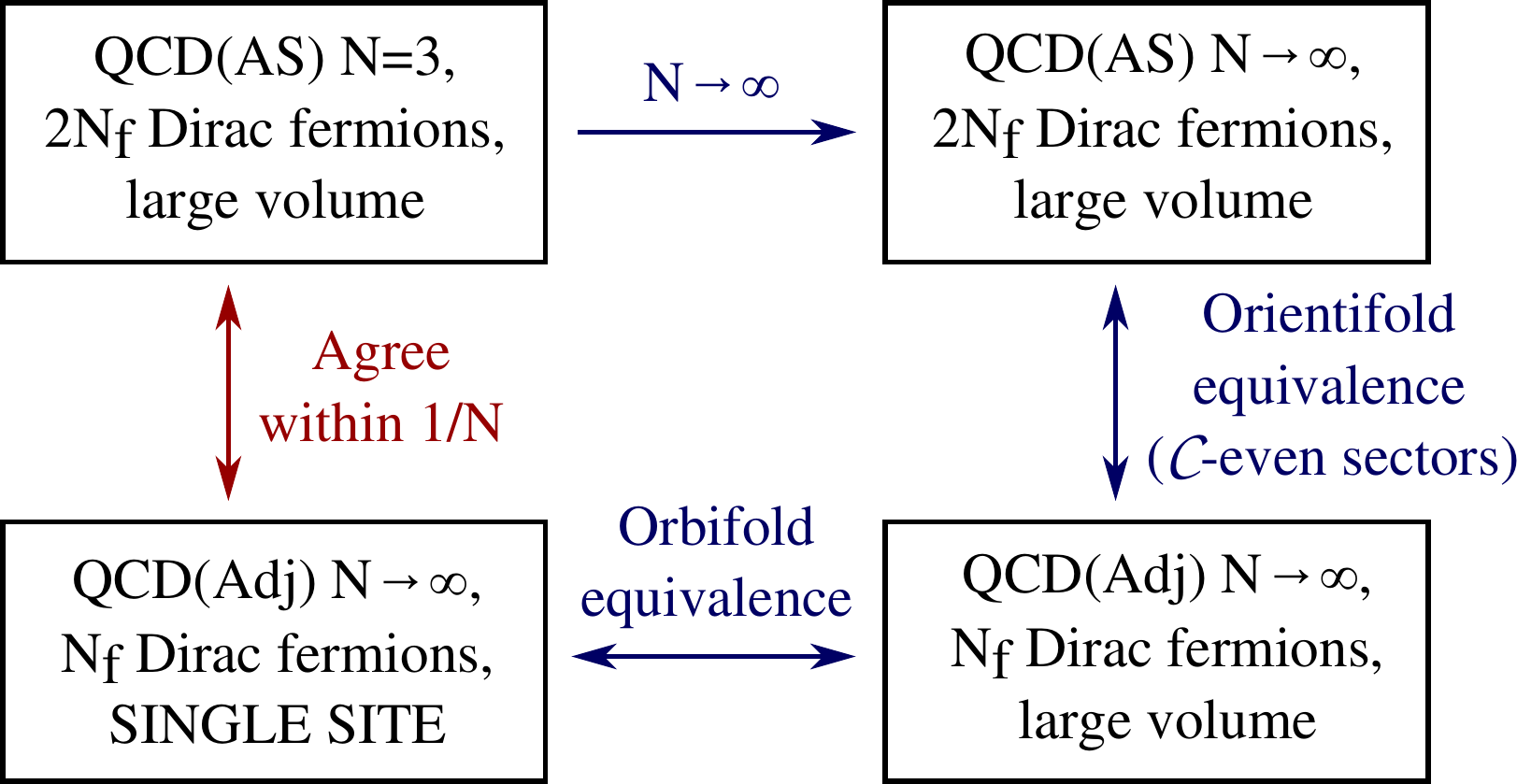}
    \caption{The chain of orbifold-orientifold equivalences connecting the QCD with $N=3$ (in the antisymmetric
    representation which is equivalent to the fundamental representation at this value of $N$) with $2N_f$ flavours and the
    AEK model with $N_f$ flavours.}
    \label{fig:qcd_aek}
  \end{center}
\end{figure}

Thus, the AEK models with $N_f=1/2$, $N_f=1$ and $N_f=2$ are all interesting, for different reasons. Let us briefly
review the putative large-volume equivalents of the three theories:
\begin{enumerate}
  \item{$N_f=1/2$ (single Majorana fermion): this corresponds, in the massless case, to the large-$N$ limit of
  $\mathcal{N}=1$ SYM. This theory has been extensively studied, also using lattice methods, although this is somewhat
  difficult due to the so-called sign problem \cite{gbc08,end09,dff10}. The lattice regularization also breaks the
  supersymmetry and a set of specific methods has to be used to analyze this theory efficiently. We do not attempt to
  analyze the supersymmetric case in this work.}
  \item{$N_f=1$: as discussed in Sec.~\ref{sec:equiv} the large-$N$ orientifold equivalence connects the theory with
  $N_f$ adjoint Dirac fermions to the theory with $2N_f$ Dirac fermions in the antisymmetric representation. On the
  other hand, large-$N$ QCD(AS) with two flavours is the Corrigan-Ramond limit of the physical QCD with 2 lightest
  quarks. Thus there exists a chain of orbifold-orientifold equivalences, pictured in Fig.~\ref{fig:qcd_aek}, that connects
  $N_f=1$ AEK to $N_f=2$ QCD, up to $1/N$ corrections \cite{kuy07}! The large-volume theory, $N_f=1$ QCD(Adj), is expected
  to be confining and to show spontaneous breaking of chiral symmetry \cite{ds07,san09}.}
 	\item{$N_f=2$: the theory with two flavours of massless adjoint fermions is expected to lie in the conformal window or
 	close to it \cite{ds07,san09}. The best analyzed case is $N=2$ due to its use in the walking technicolor theory -- it
 	is now rather well established that the theory is conformal \cite{cgs08,hrt09,bdh11}. This result is expected to
 	persist for all $N$ \cite{ds07,san09} -- e.g.\ note that the gluonic and fermionic degrees of freedom scale with $N$
 	in the same way and that the Gell-Mann--Low	$\beta$ function of the theory is independent on $N$ up to two loops
 	(although there exists an $N$-dependence in the fourth-order correction to the $\beta$ function \cite{cza04}). Also
 	note that the theories inside the conformal window are perfectly feasible to analyze using volume reduction
 	\cite{uy10}.}
\end{enumerate}

Also, if the center symmetry is preserved for heavy quarks we expect that in this region the AEK model well approximates
the dynamics of the large-volume pure-gauge model, regardless of $N_f$ \cite{bs09}. That would be a realization of a
working Eguchi-Kawai reduction.

\section{Perturbative calculation}
\label{sec:pert}

In this section we investigate the perturbative properties of the volume-reduced systems. We first consider the pure
gluonic model and show the emergence of the center symmetry breaking that invalidates the volume reduction. Then we
analyze the impact of adjoint fermions, both in the massless and in the massive case.

\subsection{Pure-gauge case}
\label{sub:pert_pg}
The perturbative calculation of the one-loop potential in the (four-dim\-ensional) Eguchi-Kawai model can be found in
Refs.~\cite{bhn82,kmi82}\footnote{In this derivation we use the timelike gauge following Ref.~\cite{kmi82} however we do
not use the smart parametrization trick used in that Ref. This (arguably) makes the derivation simpler in our approach,
at the cost of harder generalization to arbitrary dimensionality.}. The action of the model can be rewritten, up to a
constant factor, as
\begin{equation}
S_\text{gauge} = Nb \sum_{\mu<\nu} \Tr\!\left([U_\mu, U_\nu][U_\mu, U_\nu]^\dagger\right).
\label{eq:comm}
\end{equation}
The minimum of the action is obtained when $[U_\mu,U_\nu]=0\,$ i.e.\ when the link matrices can be simultaneously
diagonalized. Thus it is convenient to parametrize:
\begin{equation}
U_\mu = V_\mu D_\mu V_\mu^\dagger, \ \mu = 1,\ldots,4
\label{eq:u_vdv}
\end{equation}
where $D_\mu=\text{diag}\,[e^{i\vartheta_\mu^1},\ldots,e^{i\vartheta_\mu^N}]$ and $V_\mu$ is a unitary matrix.

At large $b$ the effective potential can be found by calculating the partition function in the vicinity of the diagonal
link matrices. We change the integration variables to $\vartheta$ and $V$ obtaining
\begin{align}
Z_\text{EK} &= \mathcal{N} \int\! \big(\sprod_\mu \sprod_i d\vartheta_\mu^i\big) \sprod_\mu \sprod_{i<j}
\sin^2\tfrac{\vartheta_\mu^i-\vartheta_\mu^j}2 Z'(\vartheta),\\
Z'(\vartheta) &= \int\! \big(\sprod_\mu dV_\mu\big) \exp\!\big(Nb\ssum_{\mu\neq\nu}\Tr(V_\mu D_\mu V_\mu^\dagger
V_\nu D_\nu V_\nu^\dagger V_\mu D_\mu^* V_\mu^\dagger V_\nu D_\nu^* V_\nu^\dagger)\big)
\end{align}

We may now fix the gauge -- we choose the ``timelike'' gauge $V_1=I$, eliminating one of the integrals. When $b$ is
large the remaining link matrices are close to being diagonal, thus we can write
\begin{equation}
V_\mu = \exp(i A_\mu), \ \mu = 2,\ldots,4
\end{equation}
and expand in the hermitian matrices $A_\mu$. Note that after choosing the timelike gauge there still exists a residual
gauge freedom -- Eq.~\ref{eq:u_vdv} is invariant with respect to the transformation:
\begin{equation}
V_\mu\to V_\mu \Lambda_\mu,
\end{equation}
where $\Lambda_\mu$ is an arbitrary unitary diagonal matrix. To remove this freedom, we require (following
Ref.~\cite{kmi82}) that matrices $A_\mu$ have vanishing entries on the diagonal.

The first order of the expansion in $A_\mu$ disappears and in the second order we obtain:
\begin{align}
Z'(\vartheta)=\int\!\big(\!\sprod_{\mu>1}\sprod_{i>j}d^{\,2\!}A_\mu^{ij}\big)
\exp\Big(&\!-\!16Nb\ssum_{i>j}\ssum_{\mu\neq\nu}\sin^2\tfrac{\vartheta_\mu^i-\vartheta_\mu^j}2
\sin^2\tfrac{\vartheta_\nu^i-\vartheta_\nu^j}2\label{eq:z_pert_long}\\
&\times\big(|A_\mu^{ij}|^2+|A_\nu^{ij}|^2-A_\mu^{ij}A_\nu^{ji}-A_\mu^{ji}A_\nu^{ij}\big)\Big),\nonumber
\end{align}
with $A_1^{ij}\equiv0$. For each given $i,j$ the integral in Eq.~\ref{eq:z_pert_long} is a $3$-dimensional complex
Gaussian integral. We evaluate the determinants and obtain:
\begin{align}
&Z_\text{EK} = \mathcal{N}'\!\int\!\big(\sprod_{\mu,i} d\vartheta_\mu^i\big)
\exp\!\big(\!-\!V_\text{1-loop}(\vartheta)\big),\\
&V_\text{1-loop}(\vartheta) = 2\ssum_{i>j}\log
\big(\ssum_\mu\sin^2\tfrac{\vartheta_\mu^i-\vartheta_\mu^j}2\big),
\end{align}
where $V_\text{1-loop}(\vartheta)$ is called the one-loop effective potential (or effective action). In
Refs.~\cite{bhn82,kmi82} the effective potential was calculated for arbitrary lattice dimensionality:
\begin{equation}
V_\text{1-loop}(\vartheta) = (d-2)\ssum_{i>j}
\log\big(\ssum_\mu\sin^2\tfrac{\vartheta_\mu^i-\vartheta_\mu^j}2\big)
\end{equation}

The validity of the large-$N$ volume reduction in $d=2$ is well-known by other methods \cite{ek82,gw80}. However when
$d>2$ the effective potential for the phases $\vartheta_\mu^i$ is attractive and favours a peaked distribution of
eigenvalues, signalling spontaneous breaking of the center symmetry. This phenomenon is in fact seen in the Monte Carlo
simulations \cite{bhn82,oka82,knn03,nn03,kor09} and it invalidates the Eguchi-Kawai volume reduction.

\subsection{The effect of adjoint fermions}
\label{sec:pert_ferm}
We now add $N_f$ adjoint Wilson fermions with periodic boundary conditions to the theory. In one-loop perturbation
theory the Wilson Dirac operator, Eq.~\ref{eq:D_W}, is diagonal in color space \cite{br091} and one can easily calculate
the one-loop potential (up to a $\vartheta$-independent term) \cite{ahu10}:
\begin{equation}
V_\text{1-loop}(\vartheta)= 2\ssum_{i>j} \log\big(\ssum_\mu\sin^2\tfrac{\vartheta_\mu^i-\vartheta_\mu^j}2\big)
- 4N_f \ssum_{i>j} \log\big(\ssum_\mu\sin^2(\vartheta_\mu^i-\vartheta_\mu^j)+m_W^2(\vartheta)\big),
\label{eq:1loop}
\end{equation}
where the first term is the contribution of the gauge fields and the second term is the fermionic part, with $m_W$
being the contribution from the bare mass and the Wilson term:
\begin{equation}
m_W(\vartheta) = am_0+2\ssum_\mu\sin^2\tfrac{\vartheta_\mu^i-\vartheta_\mu^j}2.
\end{equation}

The fermionic term in the potential has the opposite sign to the gauge part (the fermions give a repulsive contribution
to the potential) and the analysis which term dominates is more involved in this case. In particular, the
singularities for coinciding eigenvalues can now lead to incorrect conclusions if not analyzed with proper care
\cite{ahu10}.

The situation is simpler in the case of only one compactified direction. Ref.~\cite{kuy07} contains the one-loop result
for massless fermions in the $\mathbb{R}^3\times S^1$ case (in the continuum):
\begin{equation}
V_\text{1-loop}(\Omega) = \big(N_f-\tfrac12\big)\frac1{\pi^2 L^4}\sum_{r=1}^\infty\frac1{n^4}|\Tr \Omega^r|^2,
\end{equation}
where $\Omega$ is the Polyakov loop matrix in the compactified direction. For $N_f>1/2$ this potential prefers the
vanishing traces of $\Tr\Omega^r$ and thus it is repulsive for the phases of eigenvalues, resulting in the preservation
of the center symmetry\footnote{For $N_f=1/2$ (the supersymmetric case, see Sec.~\ref{sec:def}) the one-loop potential
vanishes and different methods were used to show that the center symmetry is preserved in this case, see
Ref.~\cite{kuy07}.}.

The continuum analysis has been extended to the massive case in Refs.\ \cite{mo09,hm09}. A corresponding lattice
analysis with Wilson fermions was presented in Refs.~\cite{br091,br092}. Both these approaches show that as we increase
the mass from zero there is a cascade of transitions breaking $\zz_N$ symmetry to its $\zz_K$ subgroups with $K$
decreasing from large values at very small mass to 1 at very large mass\footnote{This silently assumes that $N$ is
divisible by $K$ -- however, as we will see in Chapter~\ref{ch:ph_diag}, there exist phases with only approximate
$\zz_K$ symmetry, e.g.\ $\zz_N$ at odd $N$ can break into approximate $\zz_2$ with bunches of eigenvalues differing by
$\mathcal{O}(1/N)$  -- this is a subleading effect at large $N$ and we neglect this subtlety in the analysis of this
chapter.}.

This effect can be understood as follows \cite{br092}. The one-loop effective potential can be written as
\begin{equation}
V_\text{1-loop}(\Omega) = \sum_{r=1}^{\infty}V_r |\Tr \Omega^r|^2 + \text{Const.},
\end{equation}
where the coefficients $V_r$ are of the form\footnote{The explicit expressions are not important for our purposes, see
Ref.~\cite{br092} for details.}:
\begin{equation}
V_r= N_f V_r^\text{(ferm.)} - V_r^\text{(gauge)}\quad \text{where}\quad V_r^\text{(ferm.)}, V_r^\text{(gauge)} > 0\,.
\label{eq:coeffs}
\end{equation}

The sign of $V_r$ determines the realization of the center symmetry (at the one-loop level). If all $V_r\geq0$ for
$1\leq r < N$ (and at least one of them is greater than zero) then the center symmetry is unbroken. It is shown in
Ref.~\cite{br092} that this is the case for massless fermions with $N_f>1/2$. On the other hand, if for some $K<N$ we
have $V_K<0$ then the symmetry is broken to the $\zz_K$ subgroup. Ref.~\cite{br092} numerically shows that this is the
case for the massive fermions, with $K\sim 1/am$.

This result is easy to understand intuitively \cite{br092} -- in the reduction language $r$ corresponds to euclidean
distance ($\Tr \Omega^r$ wraps $r$ times around the compactified direction) in the corresponding volume-expanded theory.
If the fermions have a mass then their range is smaller than that of the massless gluons and at some $r$ the fermionic
contribution, that dominates at small distance, must become smaller than the gluonic one.

Thus the one-loop analysis with one compact direction allows reduction only with the fermions of mass of order
$\mathcal{O}(1/aN)$, which vanishes in the large-$N$ limit.

However, as pointed in Refs.~\cite{uy10,ahu10}, this picture is far from being complete, especially when compactifying
multiple directions. For example, the breaking of the symmetry in the single-site model causes the eigenvalues to
coincide which results in an IR singularity in Eq.~\ref{eq:1loop}. This is a result of integrating out massless modes
that are necessary for the correct description of the long-distance behaviour of the theory \cite{ahu10}.

The authors of Ref.~\cite{ahu10} give a semi-quantitative description of the neglected modes by introducing a matrix
model from which they infer the non-perturbative fluctuation scale at which the one-loop analysis breaks down. They
estimate the size of the eigenvalue separation to be of order $\sim b^{-1/4}$.

Therefore, when the perturbation theory suggests that the center symmetry is broken to a $\zz_K$ subgroup with $K\gg1$,
the separation of the bunches of eigenvalues may be in fact smaller than their width and the resulting phase is
indistinguishable from the completely unbroken phase, thus leading to a working large-$N$ volume reduction. The final
picture can however only be resolved by non-perturbative calculations, such as the one presented in the subsequent
chapters.

\chapter{Monte Carlo simulation of the AEK model}
\chaptermark{AEK model: Monte Carlo simulation}
\label{ch:mc}

In this chapter we present the numerical methods to analyze the Adjoint Eguchi-Kawai model by means of Monte Carlo
simulations. We analyze both the $N_f=1$ and $N_f=2$ cases. We use Hybrid Monte Carlo algorithm \cite{dkp87} to
generate the ensembles, equipped with the rational approximation in the case of the single fermionic flavour
\cite{hks99,cke07}.

\section{Hybrid Monte Carlo -- general idea}
\label{sec:hmc_gen}

Hybrid Monte Carlo (HMC) is a standard algorithm used in lattice gauge theories with dynamical fermions. Let us first
review the basic concepts of the algorithm (for a more in-depth discussion see e.g.\ \cite{lus10,gat09}).

Consider a general bosonic field $A$ with action $S[A]$. For simplicity we employ matrix notation and omit the indices
-- in lattice gauge theory one has $A\equiv A^a_\mu(x)$.

HMC is based on introduction of auxiliary Gaussian-distributed momenta conjugate to $A$. We can symbolically write the
resulting Hamiltonian as
\begin{equation}
  H[A,P] = \tfrac{1}{2}\Tr P^2 + S[A]\,.
\end{equation}
Note that one can easily integrate out the momenta and recover the original expectation values:
\begin{equation}
  \langle O\rangle_{A,P} =
  \frac{\displaystyle{\int}\mathcal{D}[A]\mathcal{D}[P]\;\!e^{-H[A,P]}\;\!O[A]}
  {\displaystyle{\int}\mathcal{D}[A]\mathcal{D}[P]\;\!e^{-H[A,P]}} =
  \frac{\displaystyle{\int}\mathcal{D}[A]\;\!e^{-S[A]}\;\!O[A]}
  {\displaystyle{\int}\mathcal{D}[A]\;\!e^{-S[A]}} = \langle O\rangle_{A}\,.
\end{equation}
The Hamiltonian gives the classical equations of motion, called Molecular Dynamics (MD) equations, that leave $H$ unchanged
and thus lead to exact microcanonical evolution of the system in additional ``computer'' time $\tau$:
\begin{equation}
  \dot{A} = \frac{\partial H}{\partial P} = P\,,\;\;\dot{P} = - \frac{\partial
  H}{\partial A} = - \frac{\partial S}{\partial A}\,.
  \label{eq:md_gen}
\end{equation}

In computer practice we integrate the MD equations numerically, introducing a discrete step size
$\varepsilon=\Delta\tau$. This method introduces systematic errors. To balance this effect the algorithm utilizes a
Metropolis accept/reject step with acceptance probability
\begin{equation}
  P_{acc}(A\to A',P\to P')=\min\left\{1,\,\exp(H[A,P]-H[A',P'])\right\}
  \label{eq:metro}
\end{equation}
after integrating the equations from $\tau=0$ to $\tau=\tau_{fin}$ (which is most often set to 1).

One can show that this corrects the errors and satisfies the detailed balance condition for $A$ provided that the
Molecular Dynamics integration is reversible
\begin{equation}
  P_{MD}(A\to A', P\to P') = P_{MD}(A'\to A,-P'\to -P)
\end{equation}
and preserves the integration measure $\mathcal{D}[A]\mathcal{D}[P]$. The simplest and most commonly used integrator
that satisfies these conditions \cite{lus10} is the leap-frog integrator:
\begin{equation}
  \mathcal{I}(\varepsilon, \tau_{fin}=\varepsilon N_{MD}) =
  \left(\mathcal{P}_{\tfrac{\varepsilon}{2}\,}
  \mathcal{A}_{\:\displaystyle{\varepsilon}\,}
  \mathcal{P}_{\tfrac{\varepsilon}{2}}\right)^{N_{MD}},
\end{equation}
where
\begin{align}
  \mathcal{P}_{\displaystyle{\varepsilon}}: \{A(\tau_A),P(\tau_P)\} &\to
  \left\{A(\tau_A),\,P(\tau_P+\varepsilon) = P(\tau_P) -
  \varepsilon\,\tfrac{\partial S}{\partial A}\big|_{A(\tau_A)} \right\}, \\[6pt]
  \mathcal{A}_{\,\displaystyle{\varepsilon}}: \{A(\tau_A),P(\tau_P)\} &\to
  \left\{A(\tau_A + \varepsilon) = A(\tau_A) + \varepsilon
  P(\tau_P),\,P(\tau_P)\right\}.
\end{align}

\begin{figure}[tbp!]
  \begin{center}
    \includegraphics[width=10cm]{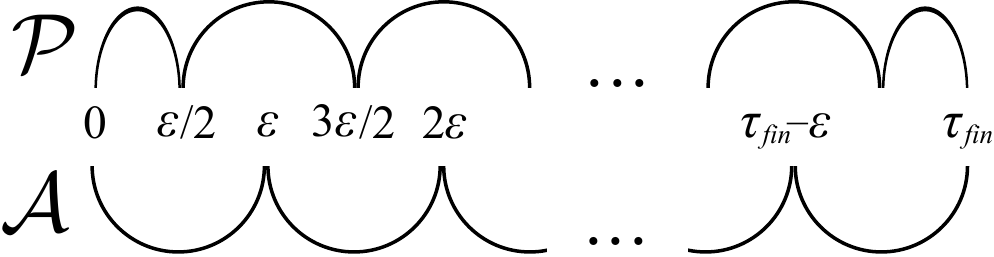}
    \caption{Schematic picture of a multi-step leap-frog evolution of $A$ and $P$ fields.}
    \label{fig:leapfrog}
  \end{center}
\end{figure}

Fig.~\ref{fig:leapfrog} shows a pictorial explanation of leap-frog's action on $A$ and $P$. Leap-frog integrator
introduces errors of order $\mathcal{O}(\varepsilon^2)$. They are corrected by the Metropolis step, however if the
errors are too big the acceptance will become poor. This is why it is important to properly choose $\varepsilon$. In our
calculations $\varepsilon$ is set so that the acceptance rates are approximately between 0.7 and 0.85.

Most of the calculations presented in this work were done using the leap-frog algorithm. There exist however more
advanced integrators, systematically studied in Ref.~\cite{ome03} by Omelyan et al.\ (hence the commonly used name
``Omelyan integrators''). Recently, we have implemented the second order minimum-norm (2MN) integrator
\cite{ome03,tdf05}. It requires two calculations of $\,\tfrac{\partial S}{\partial A}$ (``force calculations'') per MD
step, i.e.\ it is approximately twice as costly as the leap-frog, however the resulting difference between the final and
the initial hamiltonian ($\Delta H$) is an order of magnitude smaller in the case of 2MN (for the same $\varepsilon)$.
This gives a significant speed-up of the algorithm and 2MN is the integrator-of-choice in our current
calculations\footnote{There is also a possibility to use the higher-order integrators analyzed in Ref.~\cite{ome03},
e.g.\ the ones that introduce $\mathcal{O}(\varepsilon^4)$ errors. We have implemented the fourth order integrator
4MN4FP (we follow the naming convention of Ref.~\cite{tdf05}) that requires 4 force calculations per MD step. It gives
$\Delta H$ much smaller than 2MN, however in practice we found its efficiency to be inferior to 2MN at the system sizes
we currently use -- its practicality should increase as one goes to larger systems than the ones we can currently
achieve.}.

\section{Hybrid Monte Carlo -- application to $N_f=2$ AEK}
\label{sec:hmc_nf2}

\subsection{Introduction}

The action for the $N_f=2$ AEK model is
\begin{equation}
S[U] = S_\text{gauge}[U] - \ln(\det D[U])^2\,,
\end{equation}
where $S_\text{gauge}$ is the Wilson plaquette action defined in Eq.~\ref{eq:S_gauge}, and the (Wilson) Dirac operator
is defined in Eq.~\ref{eq:D_W}.

The explicit calculation of the fermionic determinant is very costly and the way to avoid it is to introduce
pseudofermion fields. The determinant is real due to the $\gamma_5$-hermiticity of the Dirac operator:
\begin{equation}
\gamma_5 D \gamma_5 = D^\dagger.
\end{equation}
Thus we can write:
\begin{equation}
(\det D)^2 = \det D \det D^\dagger = \det(DD^\dagger).
\end{equation}
Next we note that one can interpret the determinant as a result of a bosonic integral:
\begin{equation}
  \det(DD^\dagger) = \frac{1}{\det(D D^\dagger)^{-1}} = Const. \int
  \mathcal{D}[\phi]\text{e}^{-\phi^\dagger (D D^{\dagger})^{-1}\phi}\,,
\end{equation}
where $\phi$ is a complex bosonic field with the same indices as the fermionic fields (hence the name pseudofermions).
The last observation is that the action is indifferent to the substitution
\begin{equation}
  D \to Q = D\gamma_5\,,\;\;Q=Q^\dagger.
  \label{eq:Q}
\end{equation}

We are now ready to write the HMC Hamiltonian for the $N_f=2$ AEK model:
\begin{equation}
H = \frac12 \sum_\mu \Tr (P_\mu^2) - \frac{Nb}2
\sum_{\mu\ne \nu} (\Tr U^\square_{\mu\nu} + h.c.)
+ \phi^\dagger Q^{-2} \phi\,.
\end{equation}
$P_\mu$ are momenta conjugate to the gauge fields $A_\mu$ (the link variables $U_\mu = \exp(iA_\mu)$) -- they are
traceless hermitian $N\times N$ matrices, while the pseudofermion $\phi$ is a complex (bosonic) Dirac field in the adjoint
representation (thus having $4(N^2-1)$ complex components).

The standard HMC algorithm produces momenta, gauge fields and pseudofermions distributed according to the probability
density $e^{-H}$. This is done using the following steps, undertaken with a given ``starting'' set of $U_\mu$:
\begin{itemize}
\item{New momenta $P_\mu$ are drawn directly from the Gaussian distribution $\exp(-\Tr P_\mu^2/2)$.}
\item {A new pseudofermion $\phi$ is obtained by first drawing a random \linebreak pseudofermion field $\psi$ from a Gaussian
distribution, with weight \linebreak $\exp\{-\Tr(\psi^\dagger\psi)\}$, and then setting
\begin{equation}
  \phi = Q \psi \,.
\end{equation}
}
\item {The initial Hamiltonian is evaluated. Note that the pseudofermion term in the action can be easily obtained from the
Gaussian fields $\psi$.}
\item {The Molecular Dynamics equations are then solved numerically using the leap-frog algorithm (or
some more sophisticated integrator). The MD equation for the gauge field is
\begin{equation}
\dot{U_\mu} = i P_\mu U_\mu \,,
\end{equation}
while that for $\dot{P_\mu}$ must be determined by enforcing that $\dot{H}=0$ (the specific calculations for AEK model
are presented in the next subsection). The pseudofermion field $\phi$ is unchanged during the evolution.}
\item {At the end of the MD trajectory the new fields $U_\mu'$ and $P_\mu'$ are obtained and the final Hamiltonian is
evaluated using these fields. Finally, the Metropolis accept/reject step is performed, i.e.\ the new gauge configuration
is accepted with probability Eq.~\ref{eq:metro}.}
\end{itemize}
One ends up with a (possibly) new set of $U_\mu$ and then repeats the steps.

\subsection{Calculation of the HMC force}
The right hand side of the Molecular Dynamics equation for $\dot{P}$ (called force in analogy with classical mechanics)
corresponds to a change of the action with respect to an infinitesimal change in the gauge field (see
Eq.~\ref{eq:md_gen}):
\begin{equation}
  U \to U\text{e}^{i\omega}\,,
\end{equation}
where $\omega$ is an infinitesimal traceless hermitian matrix. The force separates into the gluonic and fermionic part:
\begin{equation}
\dot{P_\mu} = \dot{P_\mu}^{U} + \dot{P_\mu}^{\phi}
\,.
\end{equation}

The former is
\begin{equation}
\dot{P_\mu}^{U}
=
i N b \sum_{\nu\ne\mu} U_\mu \left[
U_\nu U_\mu^\dagger U_\nu^\dagger
+
U_\nu^\dagger U_\mu U_\nu \right] + {h.c.}
\,.
\end{equation}
Note that this result is automatically traceless and that it has the same structure as the large-volume HMC result (see
e.g.\ Ref.~\cite{lus10}), despite different derivation -- the large-volume gauge action is linear in $U_\mu(x)$ while the
EK action is quadratic in $U_\mu$.

To calculate the fermionic part we first express the variation of the pseudofermion action in terms of variation of
$Q[U]$:
\begin{equation}
  \phi^\dagger\delta(Q^{-2})\phi = -2 \phi^\dagger \mathrm{Re} \{Q^{-2} \delta Q Q^{-1}\} \phi =
  -2\,\mathrm{Re}\{\chi^\dagger \delta Q \psi\}\,
\end{equation}
where we have introduced
\begin{equation}
\chi = Q^{-2} \phi
\ \ \text{and}\ \
\psi = Q \chi
\,.
\label{eq:chi_psi_nf2}
\end{equation}
Next we use the explicit form of the Dirac operator, Eq.~\ref{eq:D_W}, and obtain the final result for the fermionic force:
\begin{align}
\dot{P_\mu}^{\phi}
=&\;
i (-\kappa) \Big\{
(\gamma_5-\gamma_\mu\gamma_5)_{\alpha\beta}
\left[
U_\mu \psi_\beta U_\mu^\dagger \chi_\alpha^\dagger
-
\chi_\alpha^\dagger U_\mu \psi_\beta U_\mu^\dagger \right]
\nonumber\\
&-
(\gamma_5+\gamma_\mu\gamma_5)_{\alpha\beta}
\left[
\psi_\beta U_\mu \chi_\alpha^\dagger U_\mu^\dagger
-
U_\mu \chi_\alpha^\dagger U_\mu^\dagger \psi_\beta \right]
\Big\}
+ {h.c.} \,.
\label{eq:nf2_force}
\end{align}
Again, the tracelessness of $P_\mu$ is maintained.

\subsection{Some technical details}

The most computationally expensive operation in HMC is the inversion of the Dirac operator. In the MD equation, the only
place where we need the inversion of $Q$ is the calculation of $\chi$ in Eq.~\ref{eq:chi_psi_nf2}. We use the fact that
$Q^2$ is a hermitian positive definite operator and use Conjugate Gradients algorithm to obtain an iterative
approximation to $\chi$. In this way we never need to explicitly calculate the Dirac operator $Q$ -- we only need to
calculate the action of $Q^2$ on a vector.

This greatly reduces the memory consumption of the algorithm and also allows significant CPU-time
reduction\footnote{That is, unless the number of CG iterations $N_{CG}$ grows proportionally to $N$ -- a possibility
that cannot be easily excluded in the volume-reduced case where $Q$ is a dense matrix. The actual scaling, however,
ranges from $\mathcal{O}(N^0)$ to $\mathcal{O}(N^{1/2})$ as will be discussed further in this section.
The matter is more obvious in the large-volume simulations where $Q$ is a sparse matrix or at least most of its values
are very close to 0. The lack of a simple zero-structure in the volume-reduced case also greatly hampers the use of
CG-preconditioners commonly used in the large-volume case. No preconditioning was used in the calculations presented in
this work.}. It also allows us to avoid the explicit construction of the adjoint matrices $U_\mu^\text{adj}$. Instead we
represent the pseudofermion fields $\phi$ in the color space as a traceless hermitian matrix on which $U^\text{adj}_\mu$
acts as
\begin{equation}
U_\mu^\text{adj}\phi \to U_\mu \phi\, U_\mu^\dagger. 
\end{equation}
In this way the action of $Q$ on a vector only requires the multiplication of $N\times N$ matrices so it has the time
scaling $\mathcal{O}(N^3)$.

We calculate the pseudofermion part of the final Hamiltonian in analogous manner. The only difference is a stronger
stopping criterion of the CG. For the MD we require that the residue
\begin{equation}
r \equiv \phi-Q^2 \chi
\end{equation}
satisfies $|r|/|\phi|<10^{-5}$. The accept-reject step compensates for any errors introduced due to the truncation of
the CG so we only need to take care that the lower precision does not affect the acceptance rate too much. In the
accept-reject step, on the other hand, we need to assure that the precision is good enough so we use a stopping
criterion $|r|/|\phi|<10^{-15}$ which is comparable to the numerical precision of the exact inverter.

\begin{figure}[tbp!]
\centering
\includegraphics[width=12cm]{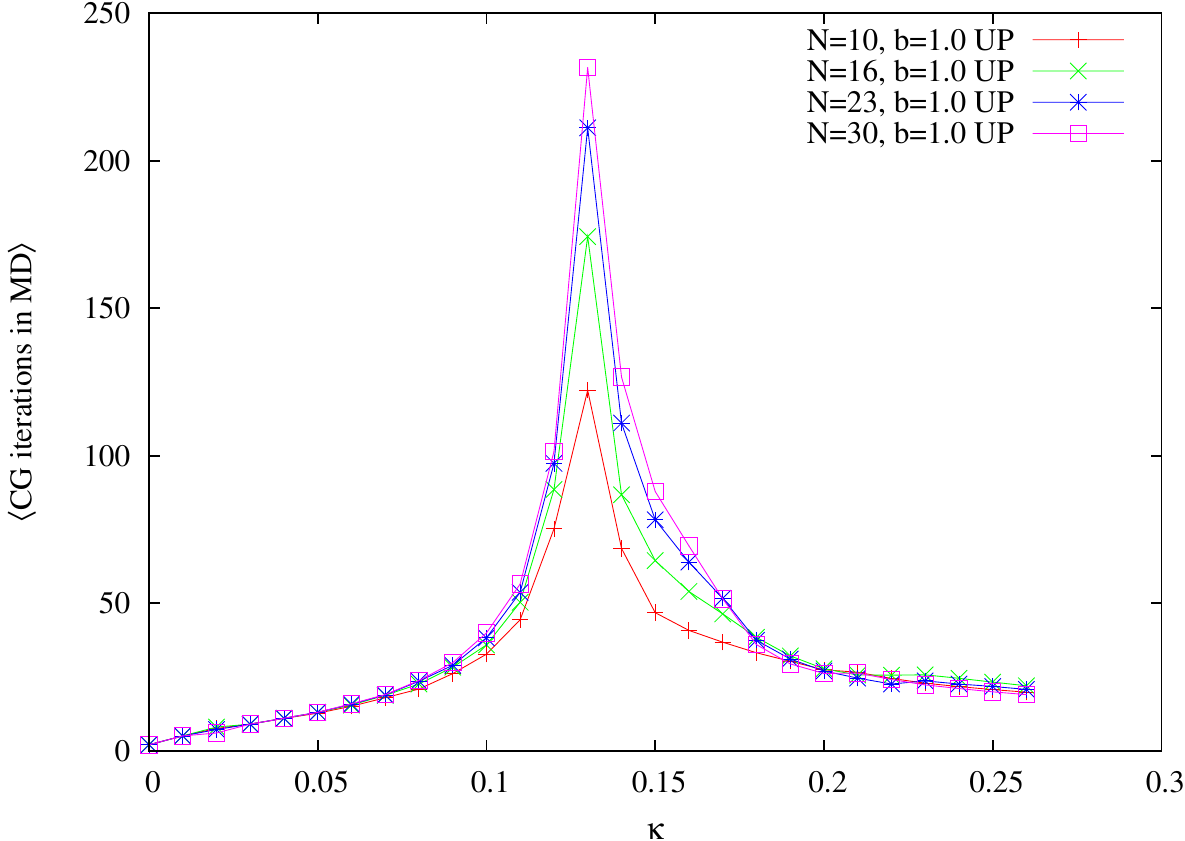}
\caption{Average number of CG iterations in the MD updates for various $N$ as a function of $\kappa$ at $b=1.0$,
$\kappa_c\approx0.13$.}
\label{fig:ncg}
\end{figure}

Next, we want to estimate the CPU-time scaling of the algorithm with $N$. To do that we need to know how the number of CG
iterations, $N_{CG}$ depends on $N$ -- we found that this depends on the quark mass. For the heavy quarks, away from the
critical line $\kappa_c$, $N_{CG}$ is independent of $N$ (for a given stopping criterion) while for the light quarks
(close to $\kappa_c$) it grows approximately as $N^{1/2}$. An illustration of this is given in Fig.~\ref{fig:ncg}.

The last ingredient of the time-scaling behaviour of the HMC algorithm is the number of MD steps per trajectory (for a
given acceptance rate). We find it to grow approximately linearly with $N$. Thus the final CPU time scaling ranges from
$\mathcal{O}(N^4)$ for heavy quarks to $\mathcal{O}(N^{4.5})$ for light quarks.

It is a common practice in large-volume simulations to use larger time steps for the fermionic force than for the
gluonic force. This is based on the fact that the gluonic part of the force is typically much larger (i.e.\ the fields
change faster under its influence and have to be evaluated with larger accuracy). In our case we may expect that, since
fermions play crucial role in the center symmetry restoration, their impact on the dynamics (and the corresponding size
of the force) can be comparable to the gluonic one. We found that this is in fact the case in the Monte Carlo
simulations (see Ref.~\cite{bks11} for details) so we conclude that using different time steps is not practical in our
case.

Finally, almost all simulations in this work were done using serial code working on a single CPU core. Recently, we have
implemented a parallel simulation code that can be efficiently executed on ($16\times Volume$) cores. This allows
working with 16 cores on a single-site and 256 cores on $2^4$ lattice. In this work, only one result obtained with the
new code is presented -- see Sec.~\ref{sec:funnel_n}.

\section{Simulation of $N_f=1$ AEK -- Rational Hybrid Monte Carlo}
\label{sec:rhmc}

\subsection{The rational approximation}

When trying to construct the HMC algorithm for odd number of flavours we encounter several problems of both conceptual
and technical nature. First, we note that although the fermionic determinant in the models we analyze is always positive
\cite{bs09}, there may exist gauge configurations where some eigenvalues of the Wilson Dirac operator will have a
negative real part. This invalidates the concept of the pseudofermion integral and to protect from that, we replace
$\det D$ with
\begin{equation}
  \det|D| = \det\sqrt{DD^\dagger} = \det|Q| = Const. \int
  \mathcal{D}[\phi]\text{e}^{-\phi^\dagger |Q|^{-1}\phi}\,,
\end{equation}
where $Q$ is the hermitian operator defined in Eq.~\ref{eq:Q}.

Explicit calculation of $|Q|^{-1}$ is, however, very costly and one has to rely on some sort of approximation -- we
choose the Zolotarev optimal rational approximation:
\begin{equation}
\mathcal{R}(x) = A \prod_{i=1}^{n}\frac{(x+c_{2i-1})}{(x+c_{2i})} =  
A\left(1+\sum_{i=1}^{n}\frac{r_{i}}{x+a_{i}}\right)
\xrightarrow[n\to\infty]{} 1/\sqrt{x},
\label{eq:R}
\end{equation}
where he coefficients $A,a_i,r_i$ are set to minimize the error (see e.g.\ Ref.~\cite{lus10} for an accessible review or
Ref.~\cite{ach92} for a more in-depth treatment):
\begin{equation}
 \delta = \max_{\varepsilon\leq x \leq 1}|1-\sqrt{x}\mathcal{R}(x)|.
\label{eq:err}
\end{equation}
If the spectrum of $Q^2$ is contained in $[\varepsilon M^2,M^2]$ then the approximation we need is
\begin{equation}
|Q|^{-1} \simeq \frac{1}{M}\mathcal{R}(Q^2/M^2) \equiv R\,. 
\end{equation}

To protect ourselves from any imperfections of the approximation we introduce additional correcting pseudofermion field
$\phi_{corr}$ and write the determinant as:
\begin{equation}
  \det|Q| = Const. \int
  \mathcal{D}[\phi,\,\phi_{corr}]\,\exp\left\{-\phi^\dagger R\,\phi
  -\phi_{corr}^\dagger (|Q|R)^{-1} \phi_{corr} \right\}.
\end{equation}
If the quality of the approximation is good, then the operator $|Q|R\,$ is close to identity and has a very weak
dependence on the gauge configuration. Thus using only $\phi$ in the Molecular Dynamics should not affect the acceptance
rate very much.

\subsection{Changes compared to $N_f=2$}

Compared to the $N_f=2$ case, there are three parts of the algorithm that need to be changed for the Rational Hybrid
Monte Carlo (RHMC):
\begin{enumerate}
\item {Generation of the pseudofermion fields: we generate fields $\psi$ and $\psi_{corr}$ with Gaussian distribution
and calculate
\begin{equation}
\phi = C\psi\,,\;\;\phi_{corr} = B\psi_{corr}\,,
\end{equation}
where $C^\dagger C = R^{-1}$ and $B^\dagger B = |Q|R$.

Operator $C$ that fulfills the above condition can be easily found from Eq.~\ref{eq:R}:
\begin{equation}
C =  \sqrt{\frac{M}{A}}\prod_{i=1}^{n}
\frac{(Q/M+i\sqrt{c_{2i-1}})}{(Q/M+i\sqrt{c_{2i}})}.
\label{eq:rat_c}
\end{equation}
To find $B$ note that the operator
\begin{equation}
Z = Q^2 R^2 - 1
\end{equation}
is very small (of order $\delta$) so one only needs the first few terms of the power series:
\begin{equation}
B = (1 + Z)^{1/4} = 1 + \tfrac14 Z - \tfrac{3}{32} Z^2 + \ldots
\end{equation}
to compute $B$ up to the machine precision \cite{lus10}.}
\item {Calculation of the pseudofermion contribution to the Hamiltonian for the Metropolis step. The initial Hamiltonian
can be calculated from $\psi$ and $\psi_{corr}$. The final Hamiltonian however requires calculation of $R\phi$ and
$(|Q|R)^{-1}\phi_{corr}$. The latter can be found as a power series in $Z$.}
\item {Calculation of the fermionic force in the Molecular Dynamics equations. The variation of the $\phi^\dagger
R\,\phi$ part is (remember that we do not include $\phi_{corr}$ in the MD):
\begin{align}
\phi^\dagger\delta R\,\phi &=
\frac{A}{M}\sum_{i=1}^{n}r_i\phi^\dagger\delta\left((Q^2/M^2+a_i)^{-1}
\right)\phi = \nonumber\\
&=-\frac{2A}{M^3}\sum_{i=1}^{n}r_i\phi^\dagger\mathrm{Re}\{(Q^2/M^2+a_i)^{-1}\delta
Q\,Q\,(Q^2/M^2+a_i)^{-1}\}\phi= \nonumber\\
&=-\frac{2A}{M^3}\sum_{i=1}^{n}r_i\,\mathrm{Re}\{\chi_i^\dagger \delta Q\psi_i\},
\label{eq:f1}
\end{align}
where
\begin{equation}
\chi_i = (Q^2/M^2+a_i)^{-1} \phi
\ \ \text{and}\ \
\psi_i = Q \chi_i
\,.
\label{eq:chi_psi_nf1}
\end{equation}
From Eq.~\ref{eq:f1} we see that the force is nothing but a sum of terms equivalent to the force in the $N_f=2$ HMC
(compare Eq.~\ref{eq:chi_psi_nf2}).
}
\end{enumerate}

Note that the shifted structure of Eqs.~\ref{eq:chi_psi_nf1} allows their simultaneous solution at the cost of solving
the single (most expensive) equation, using the multi-shift Krylov solver CG-M \cite{jeg96}. Similar calculations can be
applied to other parts where the approximation is used.

Calculating $C$ in the pseudofermion generation, Eq.~\ref{eq:rat_c}, is the only place where we need to invert a matrix
that is not hermitian positive definite -- we do this using CGNE-M algorithm (which is a combination of standard CGNE
with the multi-shift solver CG-M -- the normal equations preserve the shifted structure).

\begin{figure}[tbp!]
\centering
\includegraphics[width=12cm]{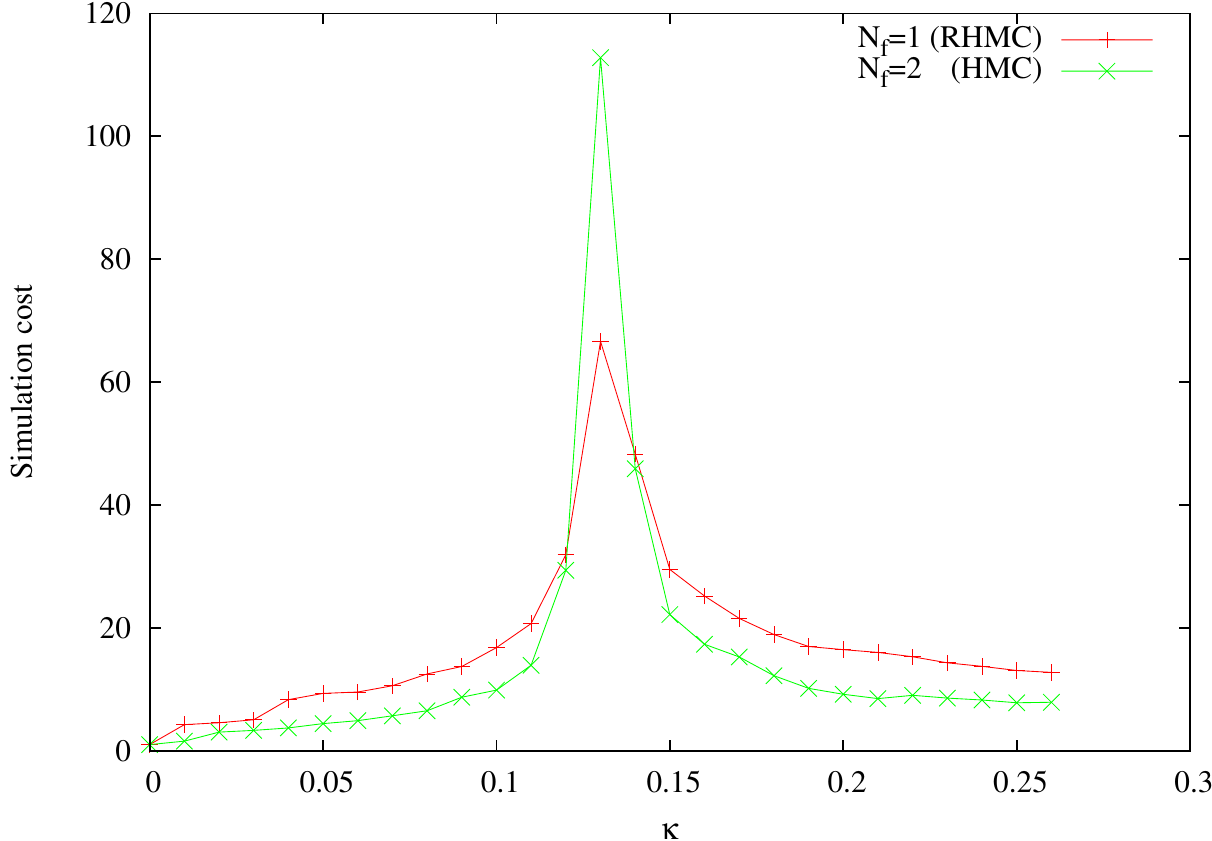}
\caption{Example comparison of simulation cost of $N_f=1$ and $N_f=2$ simulations for a representative scan. We define the
simulation cost as the CPU time per HMC trajectory divided by the average acceptance rate and we normalize it to the
pure-gauge value (set to one). The results are presented as a function of $\kappa$ at $N=16$, $b=1.0\,$.}
\label{fig:cost}
\end{figure}

It is interesting to compare the cost of the $N_f=1$ and $N_f=2$ simulations. An example comparison is presented in
Fig.~\ref{fig:cost}. The first observation is that in both cases the simulation becomes much more costly as one
approaches $\kappa_c$ (even by two orders of magnitude). In the analyzed case the cost of the simulation for heavy
fermions is roughly two times bigger for the RHMC algorithm than for the ordinary HMC. As we get closer to $\kappa_c$
the additional overhead due to the rational approximation becomes less significant and at $\kappa=0.13$ we can see that
the result for $N_f=2$ is in fact much larger. This is most likely caused by a smaller physical quark mass which requires
more CG iterations in the two-flavour case ($\kappa_c$ have different values for the two models due to a different additive
renormalization).

\chapter{Phase diagram of the AEK model}
\chaptermark{AEK model: Phase diagram}
\label{ch:ph_diag}

\begin{figure}[btp!]
\centering
\includegraphics[width=12cm]{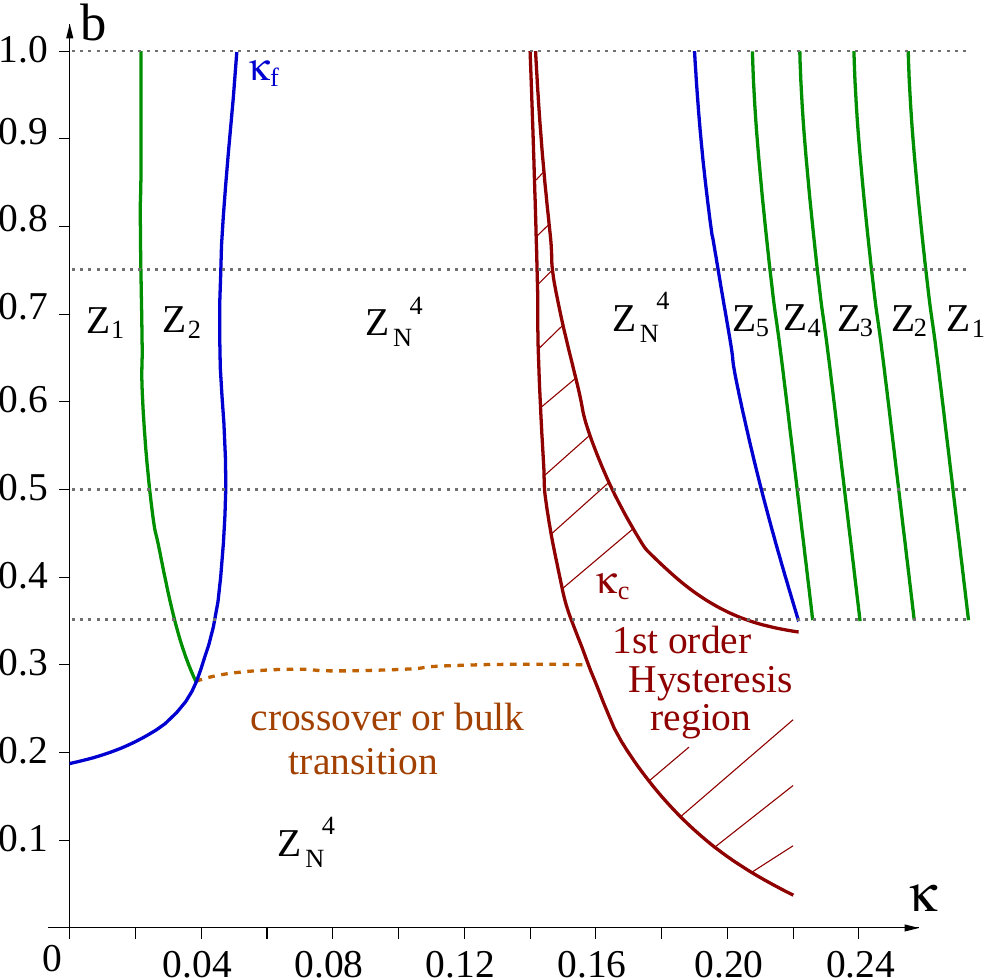}
\caption{Sketch of the phase diagram for the $N_f=2$ AEK model with $N\approx30$ (a completely analogous picture is
valid for $N_f=1$). The region named $\zz_N^4$ is the center-symmetric funnel in which the volume reduction is valid.
Surrounding it, there are several phases named $\zz_1$-$\zz_5$ after the pattern of partial breaking of the center
symmetry (the phases are labeled $\zz_K$, not $\zz_K^4$ because every time we observe the center symmetry breaking there
are substantial correlations between different lattice directions, see Sec.~\ref{sec:scans}). $\zz_1$ is the name for
the region where the center symmetry is completely broken (as in the pure-gauge EK model, that is represented by the
$\kappa=0$ line). The grey dotted lines denote the values of $b$ used for scans in Sec.~\ref{sec:scans}.}
\label{fig:phase_diag}
\end{figure}

The ultimate goal of the volume reduction technique is to use the single-site model to extract the properties of
physical (large-volume) systems. To do that we must first determine the values of parameters of the AEK model for which
the center symmetry is unbroken. In this section we present the methods and the numerical results used to establish
these values. The sketch of the deduced phase diagram in the $\kappa-b$ plane (cf.~Eqs.~\ref{eq:kappa}, \ref{eq:b}) is
presented in Figure~\ref{fig:phase_diag}.

The main feature of the phase diagram is the presence of a broad region of parameters in which the $\zz_N^4$ center
symmetry is intact. Due to its funnel-like shape we call this region the center-symmetric ``funnel''. In this region the
large-$N$ volume reduction holds and the measurements of observables can be mapped to the large-volume ones. The
smallest value of $\kappa$ (for given $b$) at which the center-symmetry is unbroken is denoted as $\kappa_f$.

This chapter mostly presents the results for $N_f=2$ contained in Ref.~\cite{bks11} with some minor extensions. We find
that the results in $N_f=1$ are very similar and thus we only show a limited number of plots for this case -- the only
substantial difference is the width of the center-symmetric funnel and we present the results for both cases in
Sec.~\ref{sec:funnel_n}.

\section{Measured quantities}

Observables used to detect the center symmetry breaking are the general open loops:
\begin{equation}
K_{n} = \tfrac1{N}\Tr\, U^{n_1}_1\, U^{n_2}_2\, U^{n_3}_3\, U^{n_4}_4,
\quad \text{with}\ \ n_\mu =0,\pm 1, \pm 2, \dots
\label{eq:op_loops}
\end{equation}
where $U^{-n}\equiv U^{\dagger n}$. These loops transform non-trivially under the center symmetry (unless all $n_\mu$
are integer multiples of $N$) and are thus very general order parameters for the center-symmetry breaking. We put the
most focus on the simplest examples of such loops, which are the 4 Polyakov loops:
\begin{equation}
P_\mu=\tfrac1N \Tr U_\mu
\end{equation}
and the 12 ``corner variables'':
\begin{equation}
M_{\mu\nu}=\tfrac1N\Tr U_\mu U_\nu \quad \text{and} \quad M_{\mu,-\nu}=\tfrac1N\Tr U_\mu U_\nu^\dagger
\end{equation}
with $\mu\neq\nu$. The corner variables were found to be particularly helpful in finding the center symmetry breaking in
the Quenched Eguchi-Kawai model \cite{bs08} because of their sensitivity to partial $\zz_N^4$ breakings\footnote{The
partial breaking of $\zz_N\!$ to $\zz_K$ ($1<K<N$) is observed when all the loops $\Tr[(U_\mu)^L]$ with $0<L<K$ vanish
but $\Tr[(U_\mu)^K]$ is non-zero.}, also the ones including correlations between different lattice directions.

We also analyze the more complicated open loops, although to keep the quantity of data manageable, we limit ourselves to
$-5\leq n_\mu \leq 5$. In this case the loops are not sensitive to more complicated patterns of partial symmetry
breaking, e.g.\ $\zz_N^4\to \zz_{10}$. To be able to observe such patterns we analyze the eigenvalues of link matrices.
As already noted in Sec.~\ref{sec:pert} each link can be represented as
\begin{equation}
U_\mu = V_\mu D_\mu V_\mu^\dagger, \quad \text{with}\ \ 
D_\mu=\text{diag}[e^{i\vartheta_\mu^1},e^{i\vartheta_\mu^2},\ldots,e^{i\vartheta_\mu^N}]
\end{equation}
The single-site gauge transformation $U_\mu\to\Omega U_\mu \Omega^\dagger$ leaves the set of eigenvalues unchanged. In
the center-symmetric phase one expects the distribution of phases $\vartheta_\mu^a$ of link eigenvalues to be invariant
under translations by $2\pi n/N$. The partial symmetry breaking can be detected when only a subgroup of the
translational symmetry is unbroken.

If the coupling is not very strong one also expects that the partition function is dominated by the link matrices
that are close to being simultaneously diagonalizable (see Sec.~\ref{sec:pert}). One can then use the gauge freedom to
set the $U_1$ to be diagonal (analogously to what was done in the perturbative calculation of Sec.~\ref{sub:pert_pg})
and analyze the elements of the other link matrices. In particular, the phases of the diagonal elements are expected to
be close to the phases of the link eigenvalues and the correlations between them (and the exact eigenvalues of $U_1$)
give extra input to the realization of the center symmetry (see Sec.~\ref{sec:scans}).

Apart from the aforementioned observables we also use the average plaquette, described in Chapter~\ref{ch:obs}, which is
very helpful to map the gross features of the phase diagram.

\section{Scans of $\kappa-b$ plane}
\label{sec:scans}

To establish the center symmetry realization in the model we performed a series of scans in the $\kappa-b$ plane. The
gross features of the phase diagram were analyzed with the scans in $\kappa$, at fixed $b$ (``horizontal scans''). In
these runs we used $N\leq30$. The gathered information was later supplemented by looking at the selected points with
greater precision and larger values of $N$ (up to 53, and in one case up to 60). We also made several ``vertical scans''
(changing $b$ at fixed $\kappa$).

We mostly focused on the range of $b\in[0.35,1]$. This reaches from the non-perturbative regime above the bulk transition
to a weakly coupled regime that can be compared to the perturbative calculations -- e.g.\ for $N=3$ this corresponds to
the range of $\beta$ between 6.3 and 18.

For the $N_f=2$ case we also made several scans extending to the unphysical strongly-coupled phase (where the center
symmetry is intact also in the pure-gauge case) as well as to extremely weakly coupled theory (up to $b=200$). The
values of $\kappa$ were mostly in the range between 0 and 0.26 but we also made some runs extending as far as
$\kappa=0.6$.

In particular, for both $N_f=1$ and $N_f=2$ we performed detailed scans at $b=0.35, 0.5, 0.75, 1.0$ with $N=10,16,23,30$
for $\kappa\in[0,0.26]$ measured every 0.01, both increasing the value of $\kappa$ (called ``UP'' scans in the
following) and decreasing $\kappa$ (called ``DOWN'' or ``DN'' scans). 

\begin{figure}[tbp!]
\begin{adjustwidth}{-1.2cm}{-1.2cm}
\centering
\subbottom[$N_f=2$] {\label{fig:poly_scan_nf2} \includegraphics[width=7.5cm]{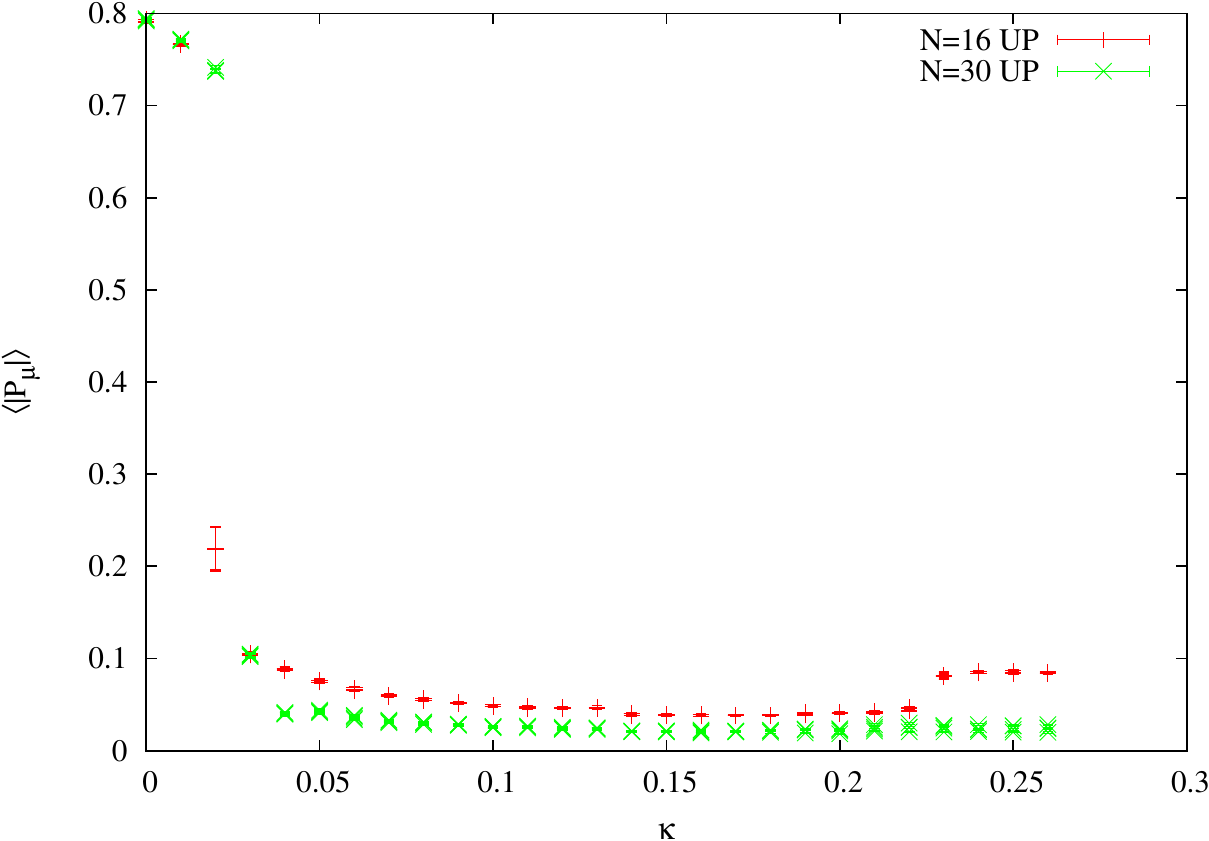}}
\hspace{0.2cm}
\subbottom[$N_f=1$] {\label{fig:poly_scan_nf1} \includegraphics[width=7.5cm]{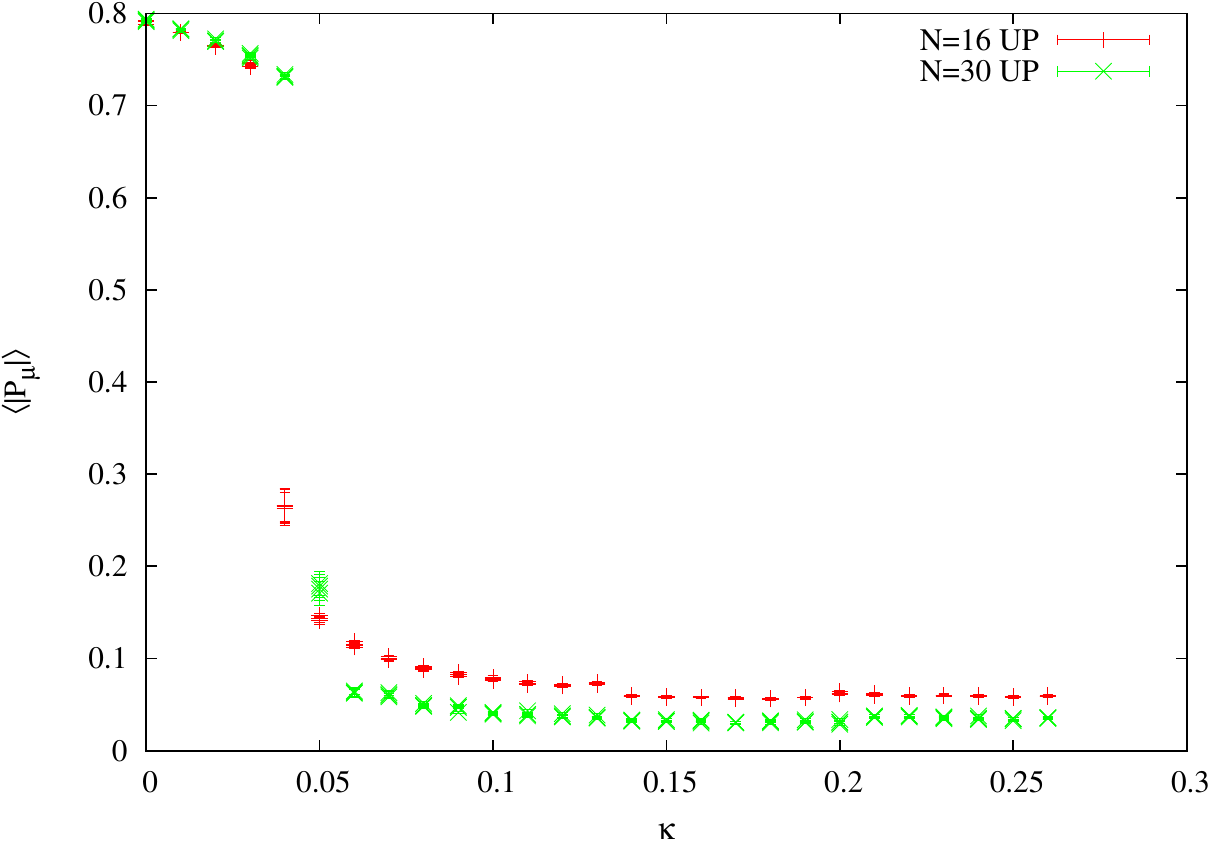}}
\caption{Absolute values of the Polyakov loops for $N=16$ and 30, $b=1.0$. All 4 directions are shown (in both panels).}
\label{fig:poly_scan}
\end{adjustwidth}
\end{figure}

In Fig.~\ref{fig:poly_scan} we present the example results of Polyakov loops measured in the runs at $b=1.0$ for both
values of $N_f$ (for the sake of clarity, only the UP scans at $N=16$ and 30 are plotted). At pure gauge, $\kappa=0$,
the absolute values of the Polyakov loops are clearly different from 0 signalling a spontaneously broken center
symmetry, as expected. For $N_f=2$, Fig.~\ref{fig:poly_scan_nf2}, at around $\kappa=0.02-0.04$ there is a jump to
significantly smaller values which suggest that the center symmetry is not broken, up to finite-$N$
fluctuations\footnote{Note that, strictly speaking, one can only discuss phase transitions and spontaneous symmetry
breaking in infinite systems so the terminology we use is only adequate for $N\to\infty$. In practice, however,
spontaneous symmetry breaking is effectively observed in simulations at finite but large values of $N$ and we keep using
the language of phase transitions throughout this work.}. The situation does not change throughout all higher values of
$\kappa$ up to $\kappa_c$, around $0.13$, and beyond. There may be some concerns perhaps about the $N=16$ run for
$\kappa>0.22$. For $N_f=1$, Fig.~\ref{fig:poly_scan_nf2}, the situation is similar, however the transition to smaller
absolute values of Polyakov loops takes place at higher values of $\kappa=0.03-0.05$.

\begin{figure}[tbp!]
\begin{adjustwidth}{-1.2cm}{-1.2cm}
\centering
\subbottom[$N=16,\, b=1.0,\, \kappa=0$]
{\label{fig:sc_broken}
\includegraphics[width=7.5cm]{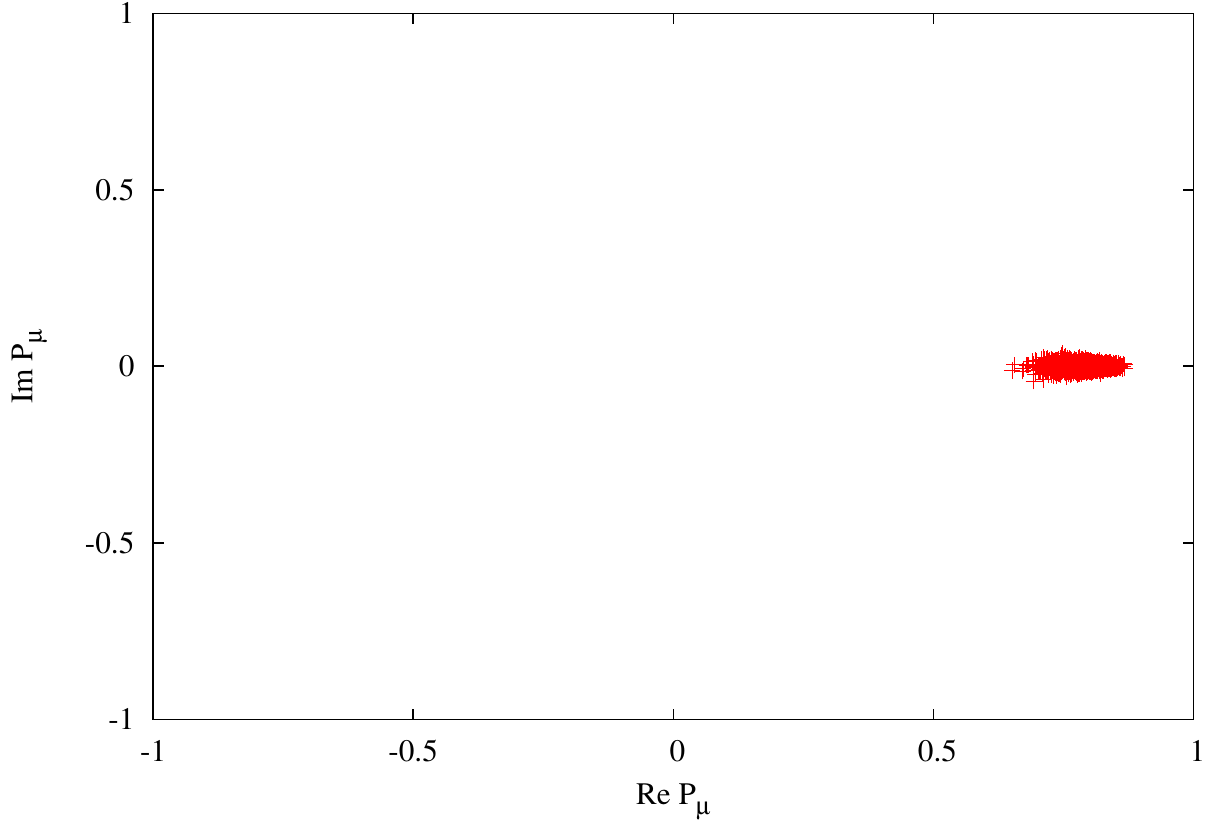}}
\hspace{0.2cm}
\subbottom[$N=10,\, b=1.0,\, \kappa=0.01$]
{\label{fig:sc_mult_vac}
\includegraphics[width=7.5cm]{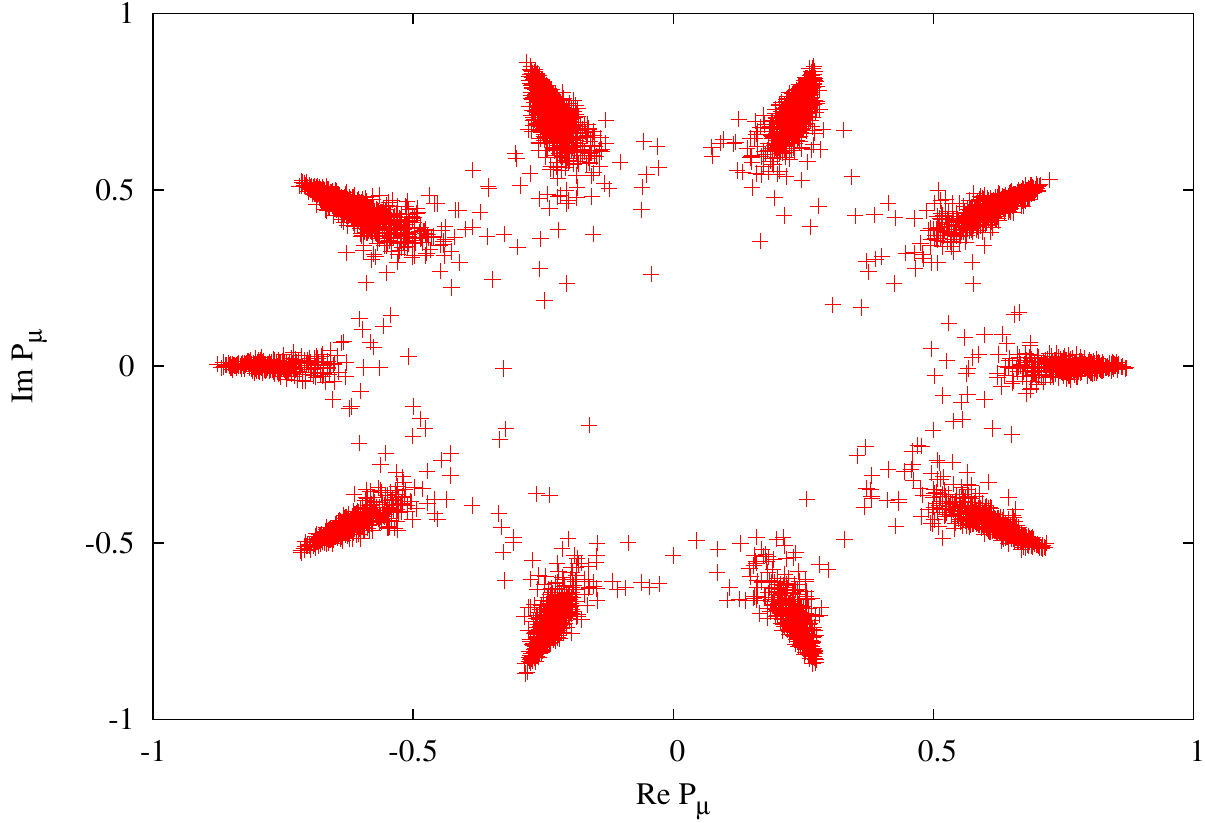}}\\[6pt]
\subbottom[$N=16,\, b=1.0,\, \kappa=0.09$]
{\label{fig:sc_unbroken}
\includegraphics[width=7.5cm]{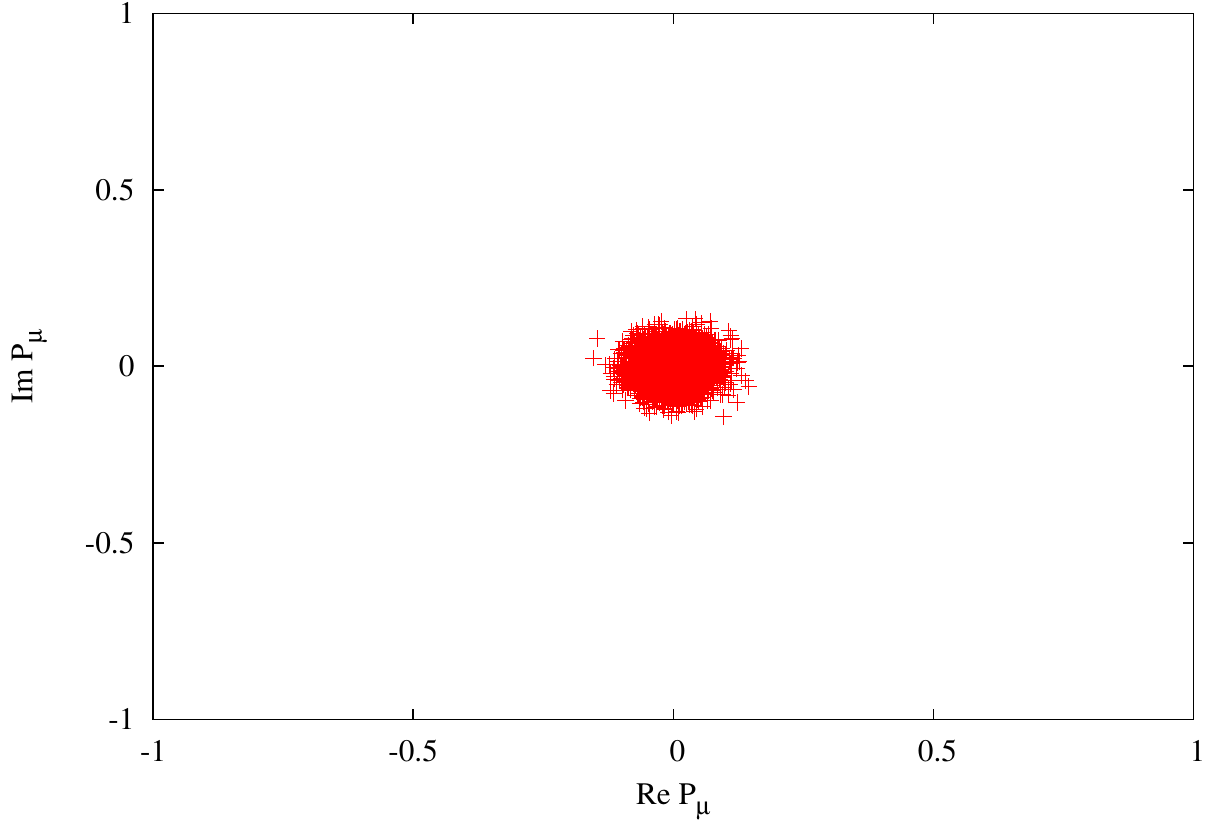}}
\hspace{0.2cm}
\subbottom[$N=16,\, b=1.0,\, \kappa=0.25$]
{\label{fig:sc_part_broken}
\includegraphics[width=7.5cm]{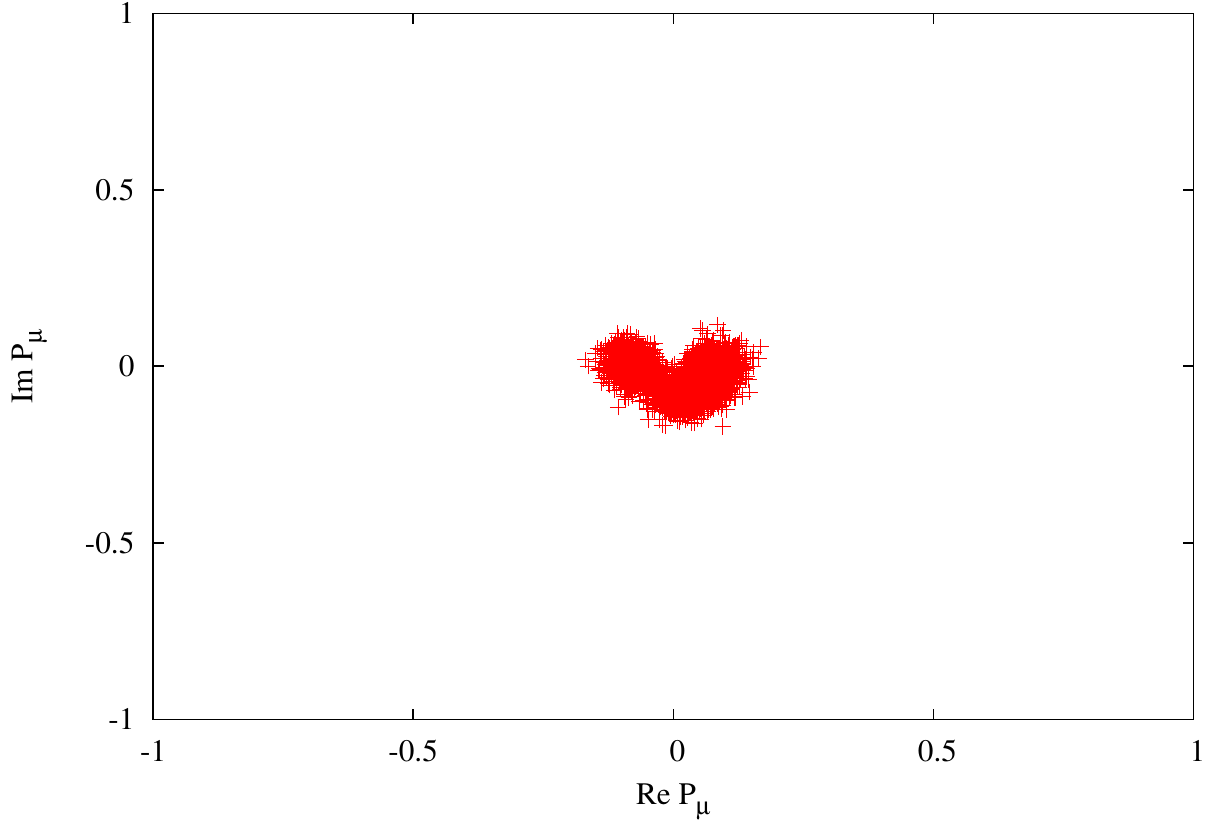}}
\caption{Scatter plots of Polyakov loops in the $N_f=2$ model. All 4 directions are pictured together.}
\label{fig:pol_scatter}
\end{adjustwidth}
\end{figure}

The picture is confirmed by looking at the scatter plots of the Polyakov loops in the complex plane, presented in
Fig.~\ref{fig:pol_scatter} (for the $N_f=2$ case). In the pure-gauge model, Fig.~\ref{fig:sc_broken}, the values of the
Polyakov loops are close to 1, clearly forming a state with broken center symmetry. With smaller $N$ one can observe
tunneling between the different vacua with broken center symmetry, see Fig.~\ref{fig:sc_mult_vac}. The tunneling is
however suppressed as the size of the system becomes larger. The run with $\kappa=0.09$, Fig.~\ref{fig:sc_unbroken}, is
on the other hand uniformly distributed around 0, signalling that the center symmetry is likely to be unbroken. The last
plot, Fig.~\ref{fig:sc_part_broken}, is also distributed close to zero but the shape of the blob is rather asymmetric,
signalling that there may occur a partial breaking of the $\zz_N^4$ symmetry.

\begin{figure}[tbp!]
\begin{adjustwidth}{-1.2cm}{-1.2cm}
\centering
\subbottom[$N_f=2$] {\label{fig:corner_scan_nf2} \includegraphics[width=7.5cm]{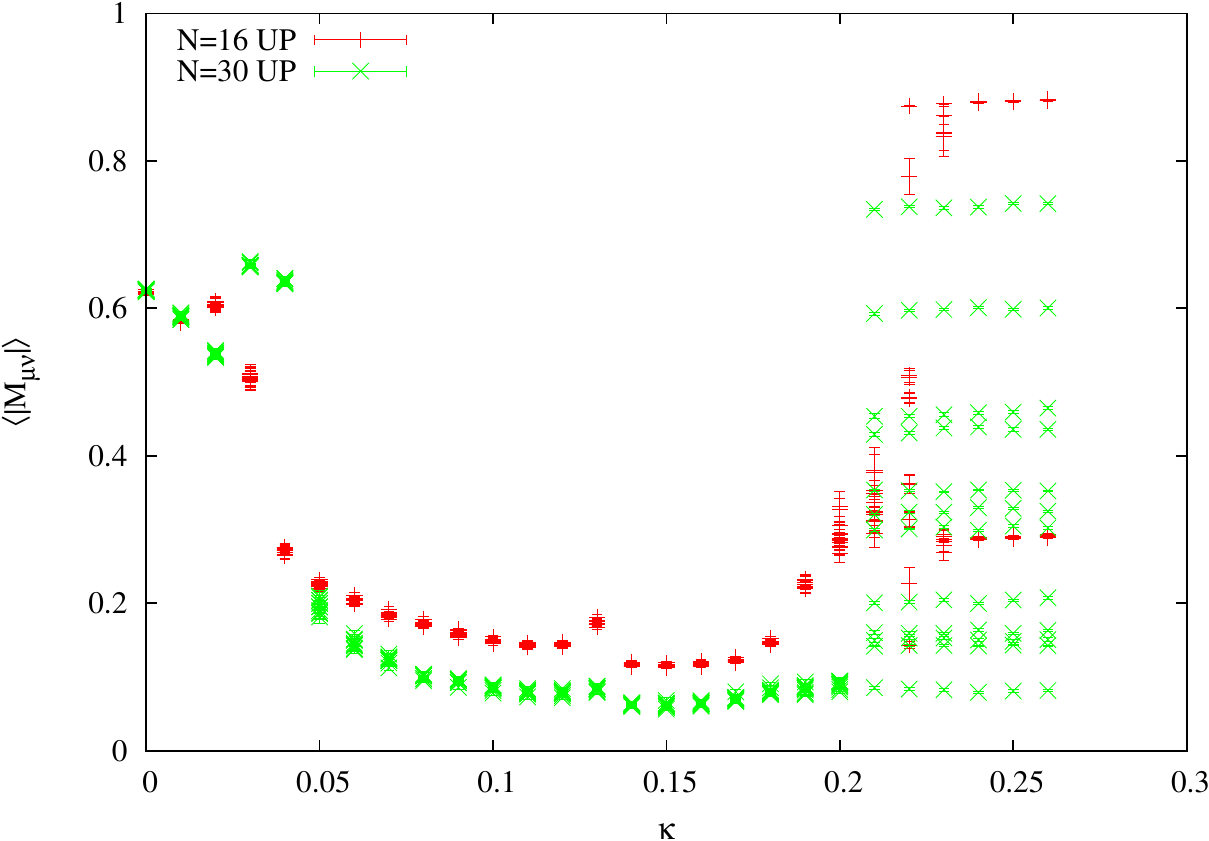}}
\hspace{0.2cm}
\subbottom[$N_f=1$] {\label{fig:corner_scan_nf1} \includegraphics[width=7.5cm]{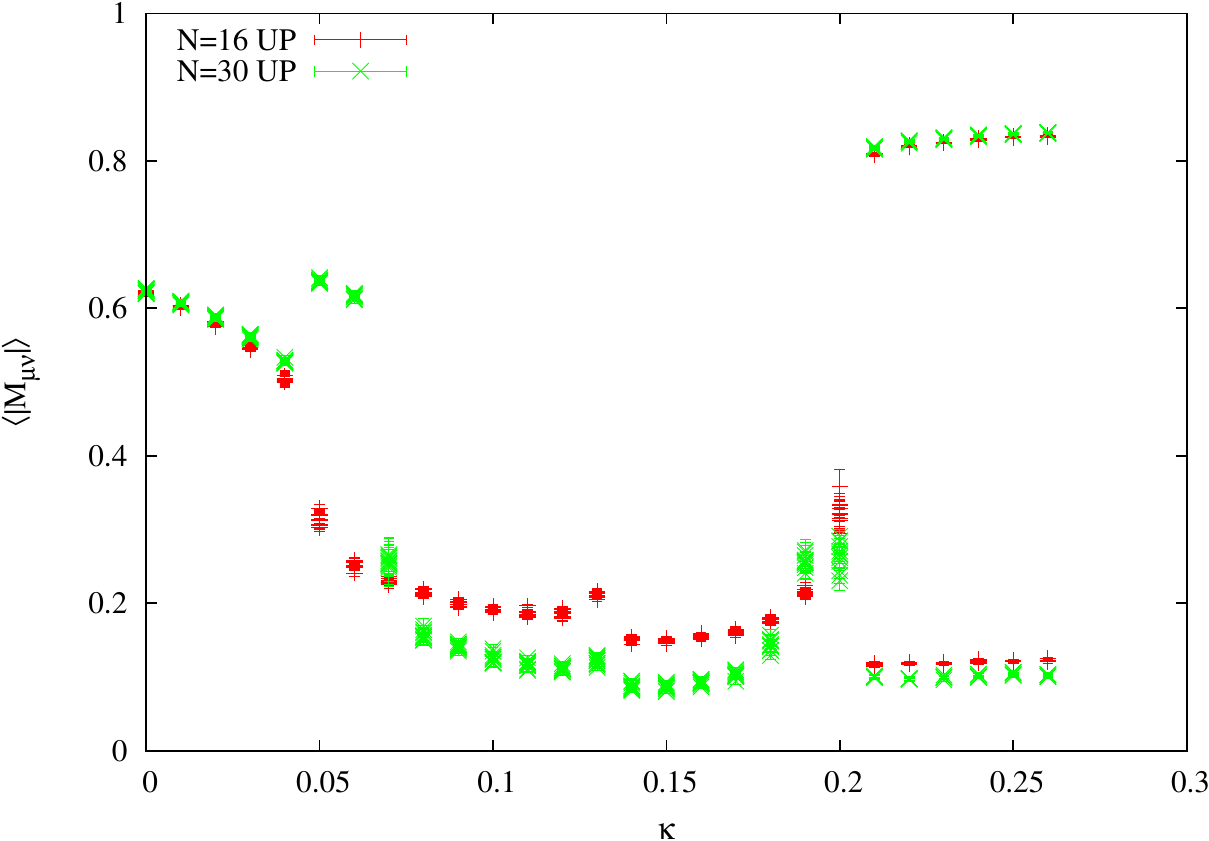}}
\caption{Absolute values of the corner variables for $N=16$ and 30, $b=1.0$. All 12 directions are shown (in both
panels).}
\label{fig:corner_scan}
\end{adjustwidth}
\end{figure}

We can further analyze the situation by looking at the corner variables $M_{\mu\nu}$, Fig.~\ref{fig:corner_scan}. They
are sensitive to partial symmetry breaking, especially in the case where correlations between different directions are
involved. We can see that the jump to $|M_{\mu\nu}|$ close to zero occurs at larger values of $\kappa$ than in the case
of Polyakov loops. Also, in the region $\kappa\gtrsim0.2$ the corner variables are clearly non-zero (both for $N_f=1$
and $N_f=2)$. We thus conjecture that the center-symmetric phase is separated from the completely broken phases by
regions with partially broken symmetry.

This conjecture can be further verified by looking directly at the histograms of eigenvalues of $U_\mu$. The link
eigenvalues provide a more thorough test of the realization of the center symmetry than $P_\mu$ and $M_{\mu\nu}$ alone
as they are sensitive also to patterns of symmetry breaking in which both of these loops vanish. From now on we focus on
the $N_f=2$ model. All the results presented in this section are analogous for the $N_f=1$ case, except the larger
$\kappa_f$ (which will be analyzed for both values separately in Sec.~\ref{sec:funnel_n}) and slightly larger finite-$N$
corrections.

\begin{figure}[tbp!]
\begin{adjustwidth}{-1.2cm}{-1.2cm}
\centering
\subbottom[$\ N=30$, $b=0.35$, $\kappa=0.09$, \mbox{600 configs, all links}]
{\label{fig:hist_zn}
\includegraphics[width=7cm]{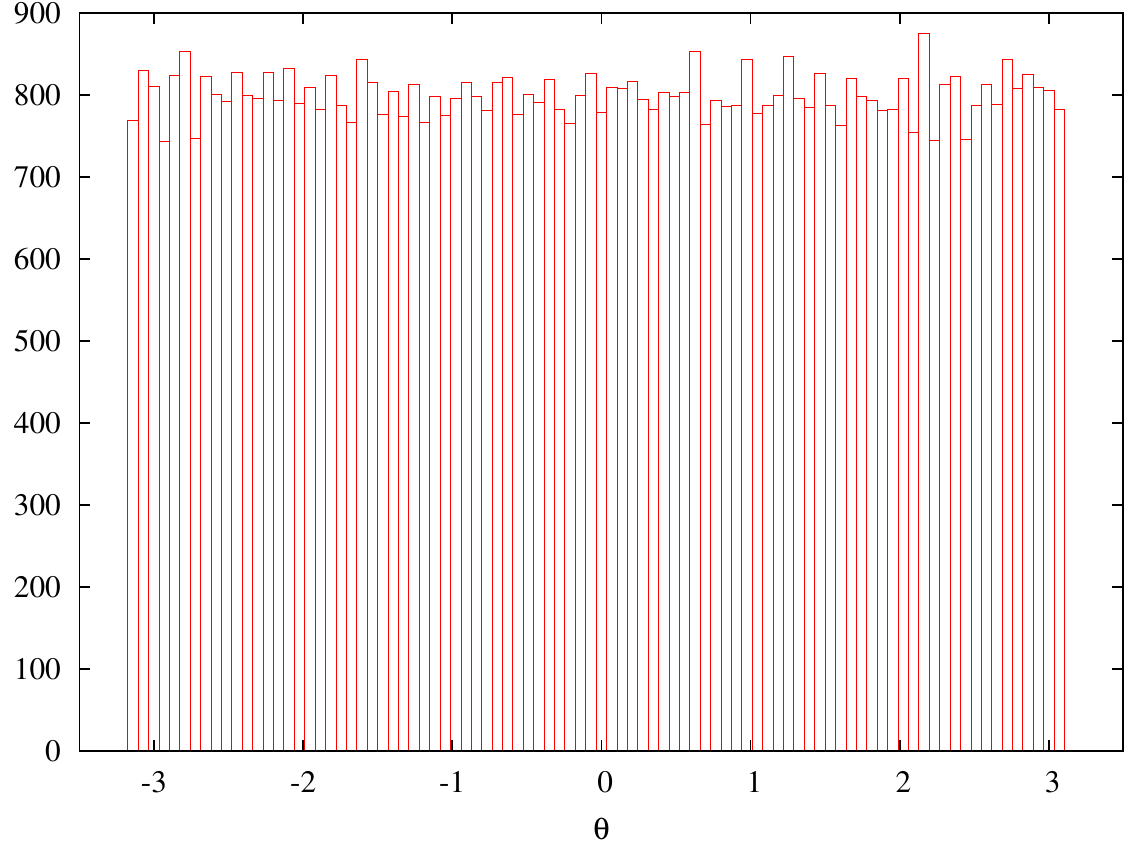}}
\hspace{0.5cm}
\subbottom[$\ N=23$, $b=1.0$, $\kappa=0.01$, \mbox{2000 configs, all links}]
{\label{fig:hist_z1}
\includegraphics[width=7cm]{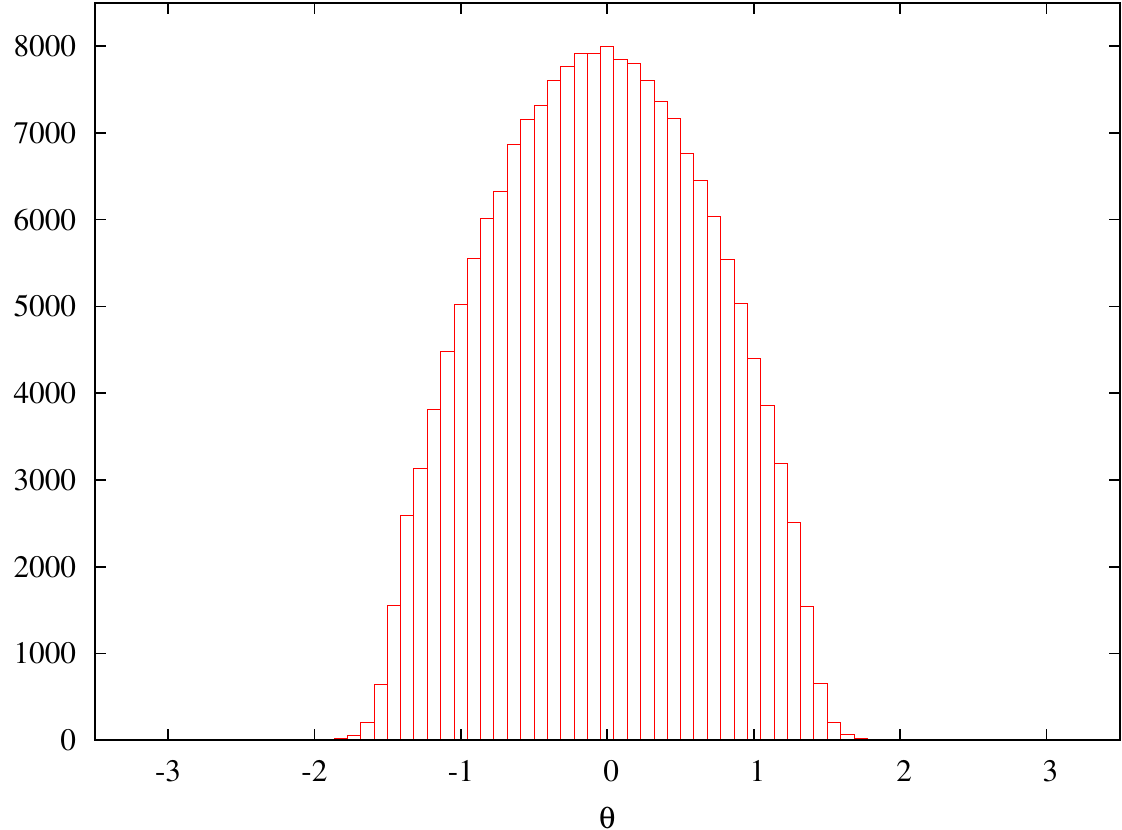}}\\[6pt]
\subbottom[$\ N=23$, $b=1.0$, $\kappa=0.03$, \mbox{2000 configs, $U_2$ only}]
{\label{fig:hist_z2}
\includegraphics[width=7cm]{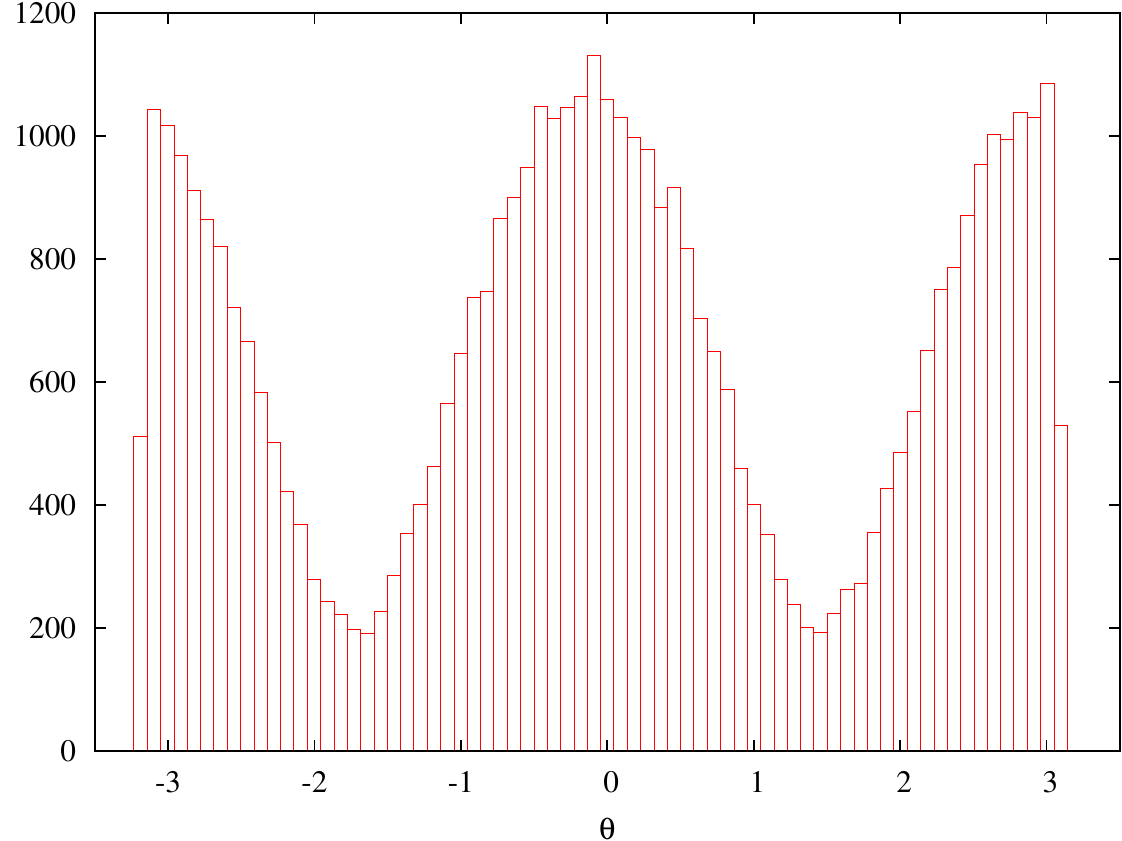}}
\hspace{0.5cm}
\subbottom[$\ N=16$, $b=1.0$, $\kappa=0.24$, \mbox{5000 configs, $U_3$ only}]
{\label{fig:hist_z3}
\includegraphics[width=7cm]{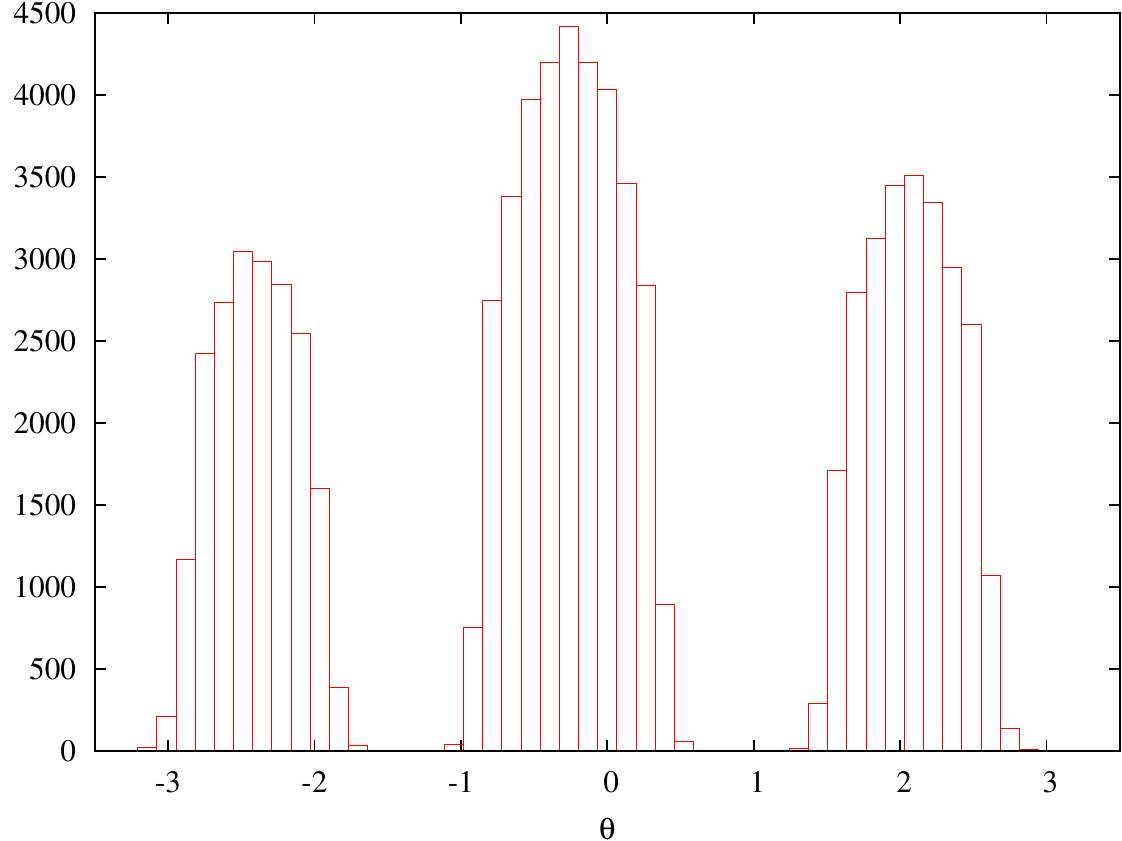}}\\[6pt]
\subbottom[$\ N=16$, $b=0.35$, $\kappa=0.22$, \mbox{3000 configs, $U_3$ only}]
{\label{fig:hist_z4}
\includegraphics[width=7cm]{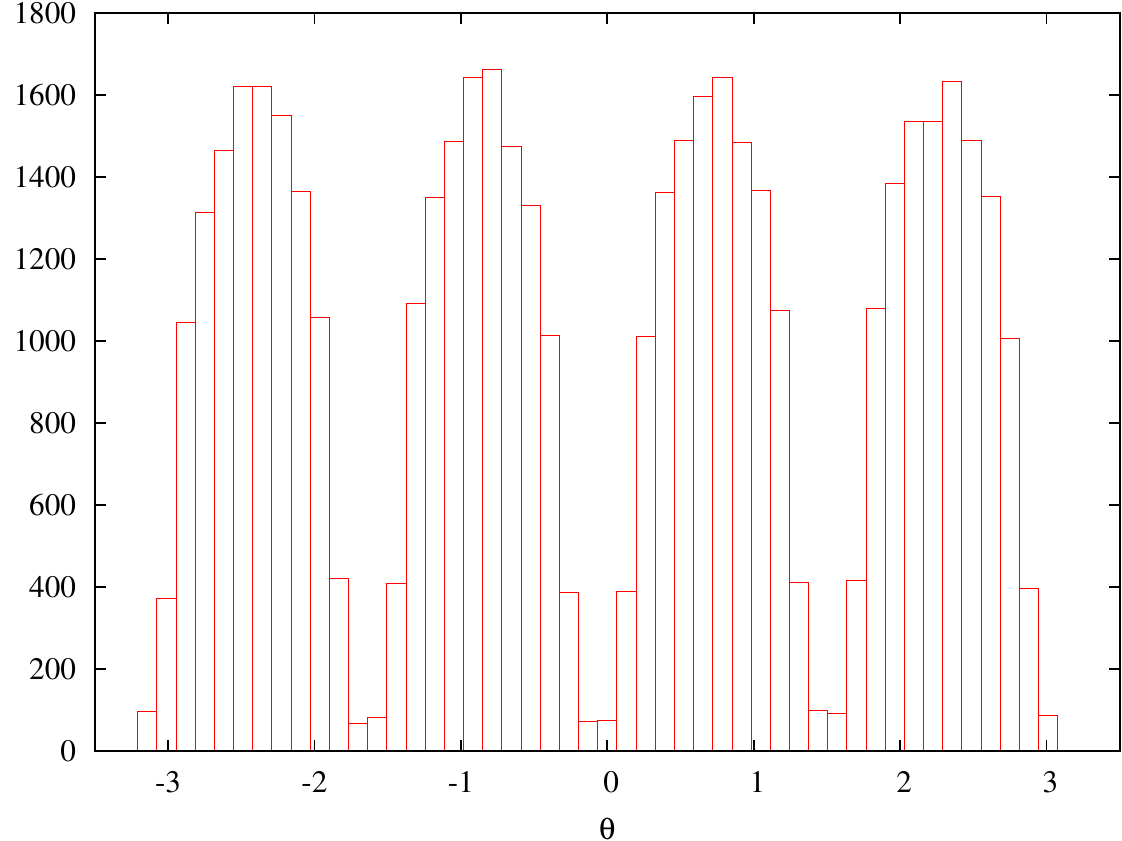}}
\hspace{0.5cm}
\subbottom[$\ N=30$, $b=1.0$, $\kappa=0.23$, \mbox{1000 configs, $U_1$ only}]
{\label{fig:hist_z5}
\includegraphics[width=7cm]{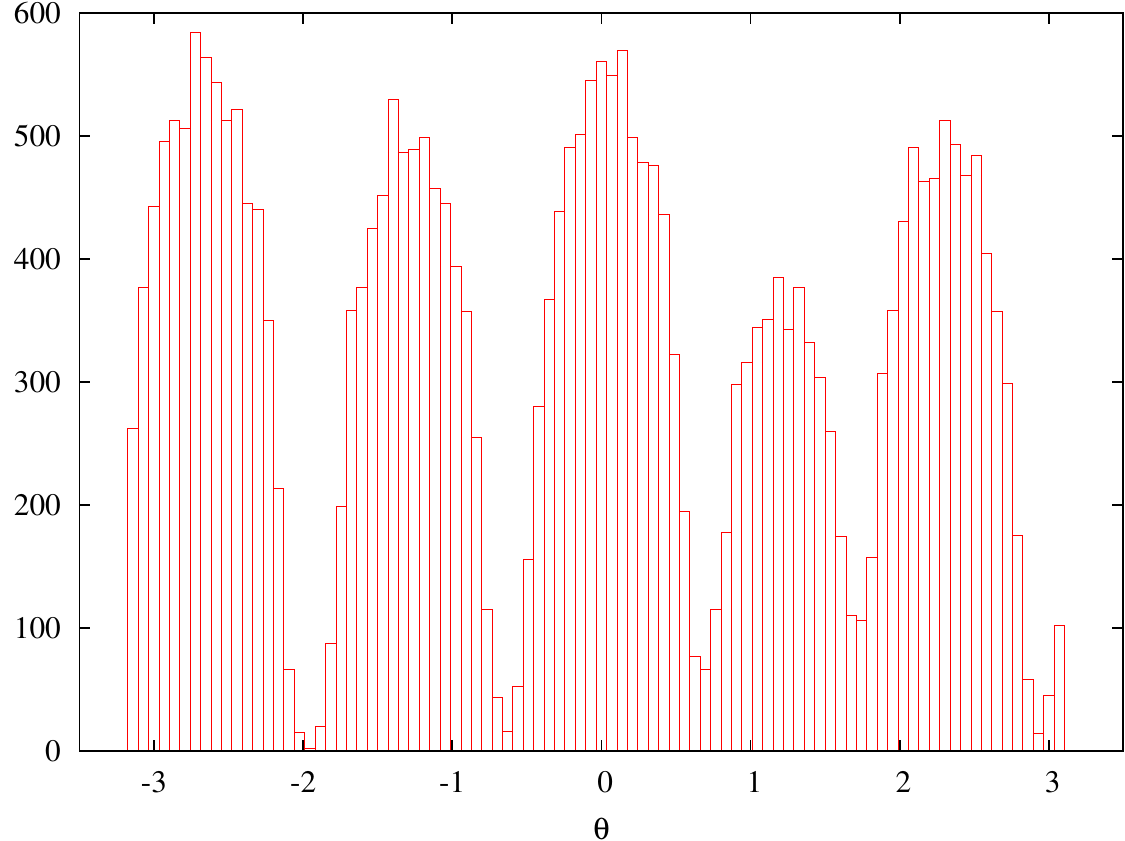}}
\caption{The (unnormalized) histograms of the phases $\vartheta_\mu^a$ of the link eigenvalues. The uncertainties
in each bin are of the order of $\sqrt{Number\ of\ counts\:}$. More details of the binning are discussed in the text.}
\label{fig:hist_leig}
\end{adjustwidth}
\end{figure}

In Fig.~\ref{fig:hist_leig} we present examples of the results that allow us to fill in the details of the phase diagram
in Fig.~\ref{fig:phase_diag}. The phases of eigenvalues (defined in the range $\vartheta^{a}_\mu\in(-\pi,\pi]\:\!$)
presented in each histogram are collected in $3N$ bins of width $2\pi/3N$ each. The $\zz_N$ symmetry implies that the
histogram should be periodic under translation by multiples of 3 bins (up to the errors, that are proportional to the
square root of the number of counts in each bin).

An example histogram inside the funnel is presented in Fig.~\ref{fig:hist_zn}. The distribution of $\vartheta^a_\mu$ is
invariant under $\zz_N$ translations and within errors it is consistent with being uniform. This is, in fact, a little
surprising as the pure-gauge model is known to reproduce the Haar-measure-like probability density distribution in the
unbroken phase \cite{knn03}:
\begin{equation}
p(\vartheta^a_\mu) = \left(\frac1{2\pi} - \frac{(-1)^N}{N\pi} \cos(N\vartheta^a_\mu)\right),
\end{equation}
which has visible oscillations in computer simulations (that vanish like $1/N$). As a check on our code, we have
confirmed that the simulation reproduces this equation at $\kappa=0$ with $b$ below the bulk transition.

Outside the funnel the attraction between eigenvalues leads to the breaking of the center symmetry that manifests itself
by a formation of groups (``clumps'') of eigenvalues on the unit circle. By counting the number of clumps we can
identify the approximate remnant symmetry ($\zz_k$-symmetric histogram has $k$ clumps).

In Fig.~\ref{fig:hist_leig} we show some examples of the clumping patterns -- Figs.~\ref{fig:hist_z1} and
\ref{fig:hist_z2} show $\zz_1$ and $\zz_2$ phases on the l.h.s of the funnel (small $\kappa$ region) while
Figs.~\ref{fig:hist_z3}, \ref{fig:hist_z4} and \ref{fig:hist_z5} show $\zz_3$, $\zz_4$ and $\zz_5$ phases on the r.h.s.
of the funnel (large $\kappa$ region).

Note that the remnant symmetry is not always exact. For example, in Fig.~\ref{fig:hist_z2} we find two clumps for $N=23$
and in Fig.~\ref{fig:hist_z3} three clumps for $N=16$ -- the eigenvalues cannot be equally distributed between the
clumps and the symmetry is only approximate. Also in the case of Fig.~\ref{fig:hist_z5} where there are five clumps for
$N=30$ we see that the clumps are not even and correspond to 7,6,7,6,4 eigenvalues respectively. We also find that
different runs can have different patterns of clumping, e.g.\ 7,7,6,6,4 vs.\ 7,6,6,6,5, but that it is rare for the
clumping to change during a run. Thus it appears that there are different competing ``vacua'' that are only
approximately related by center symmetry transformations.

We can now understand the results given by the Polyakov loops and the corner variables. The Polyakov loops are
practically insensitive to the partial symmetry breaking while the corner variables are giving various, sometimes
complex, patterns in that case (cf.\ Fig.~\ref{fig:corner_scan}) -- also in the case of more complicated patterns
such as $\zz_5$ where we expect the ``squared'' Polyakov loop $\Tr U_\mu^2$ to be insensitive. Thus, as $M_{\mu\nu}$
involve links in different directions, we may expect that these complex patterns mean that there exist substantial
correlations between the different links.

\begin{figure}[tbp!]
\begin{adjustwidth}{-1.2cm}{-1.2cm}
\centering
\subbottom[$U_1$]
{\label{fig:elems_u1}
\includegraphics[width=7.5cm]{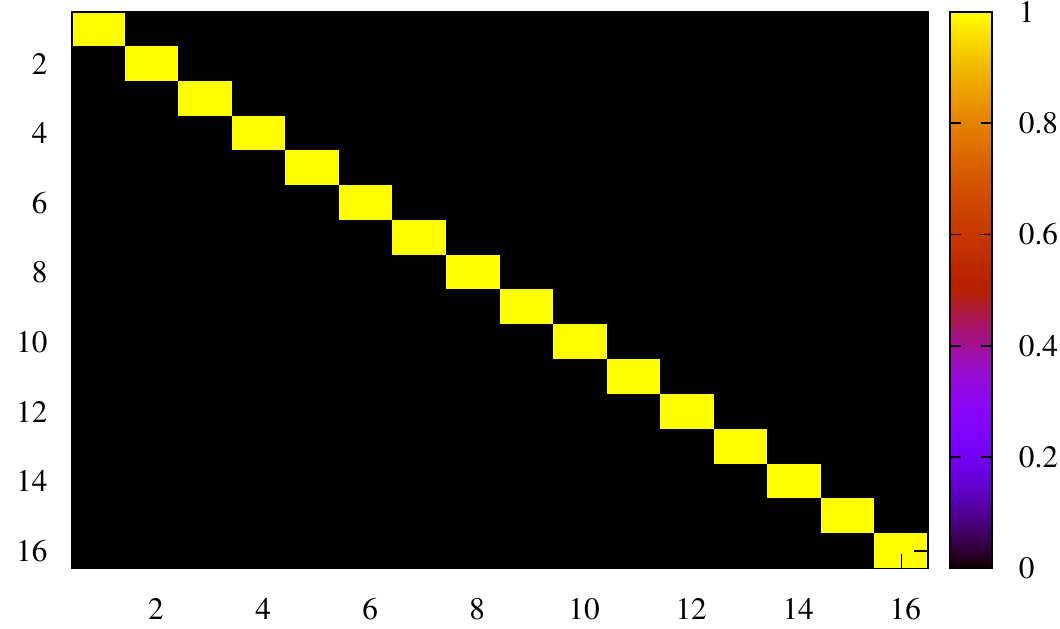}}
\hspace{0.2cm}
\subbottom[$U_2$]
{\label{fig:elems_u2}
\includegraphics[width=7.5cm]{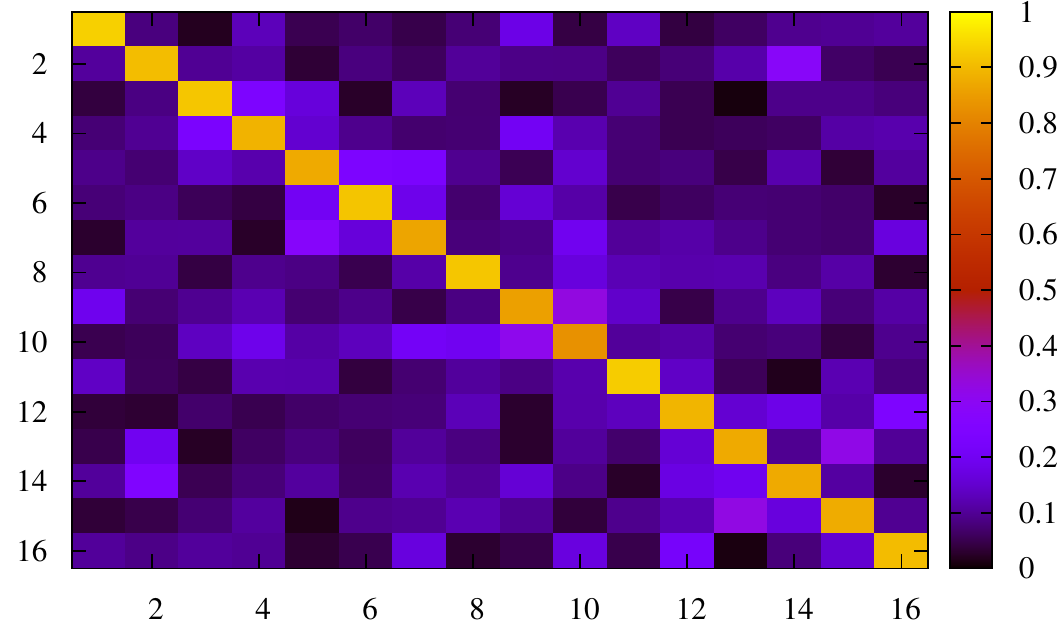}}\\[6pt]
\subbottom[$U_3$]
{\label{fig:elems_u3}
\includegraphics[width=7.5cm]{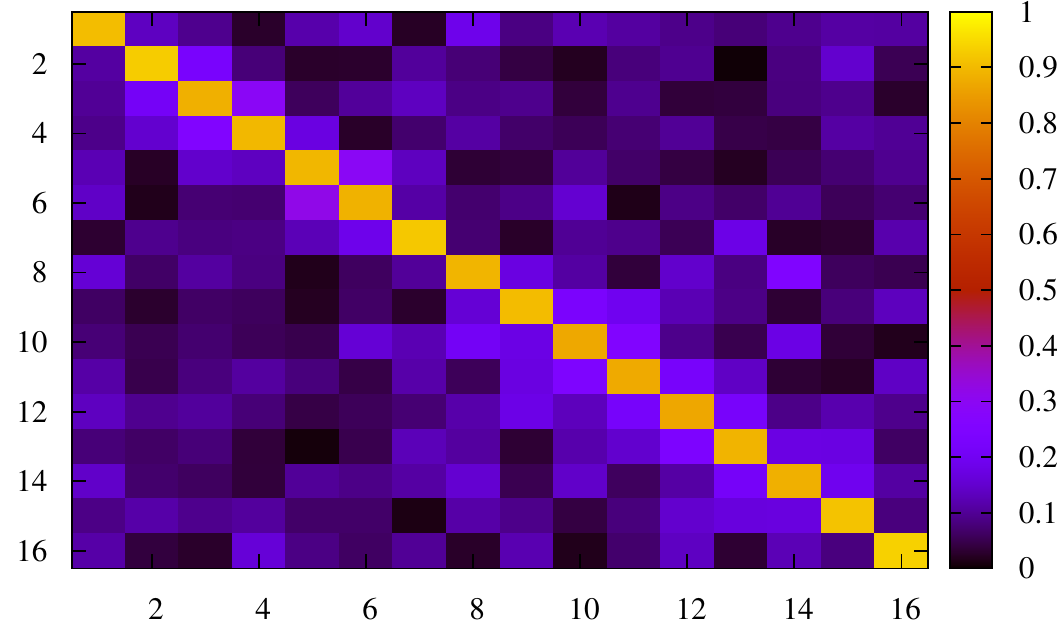}}
\hspace{0.2cm}
\subbottom[$U_4$]
{\label{fig:elems_u4}
\includegraphics[width=7.5cm]{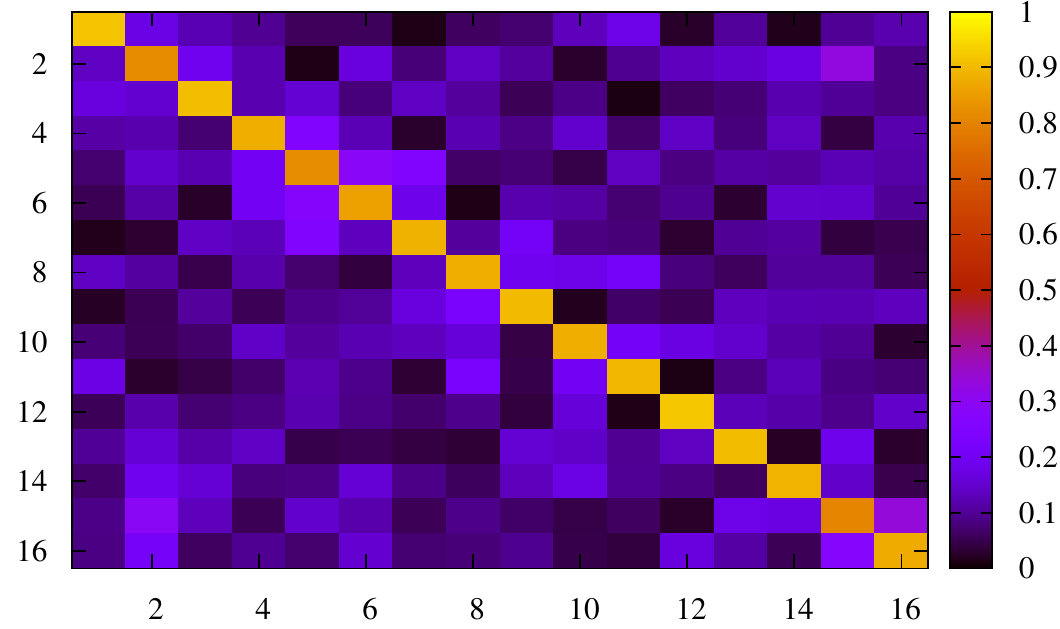}}
\caption{Absolute values of link elements in the ``timelike'' gauge -- an example configuration at $N_f=2$, $N=16$, $b=0.35$,
$\kappa=0.23$. The diagonal elements are clearly the largest ones.}
\label{fig:link_elems}
\end{adjustwidth}
\end{figure}

To measure this, at least at relatively weak coupling, we use the ``timelike'' gauge in which $U_1$ is diagonal, in a
similar manner as in Section \ref{sec:pert}. At large $b$ we expect that in this gauge the remaining links are also 
close to diagonal (cf.\ Eq.~\ref{eq:comm}) and thus we can treat their diagonal elements ``almost'' as eigenvalues. In
Fig.~\ref{fig:link_elems} we see an example configuration which shows that even at $b$ as low as 0.35 the diagonal
dominance is very clear (at least for $\kappa>\kappa_c$; for $\kappa<\kappa_c$ the diagonal dominance is only clear for
$b\gtrsim1.0$).

\begin{figure}[tbp!]
\centering
\includegraphics[width=12cm]{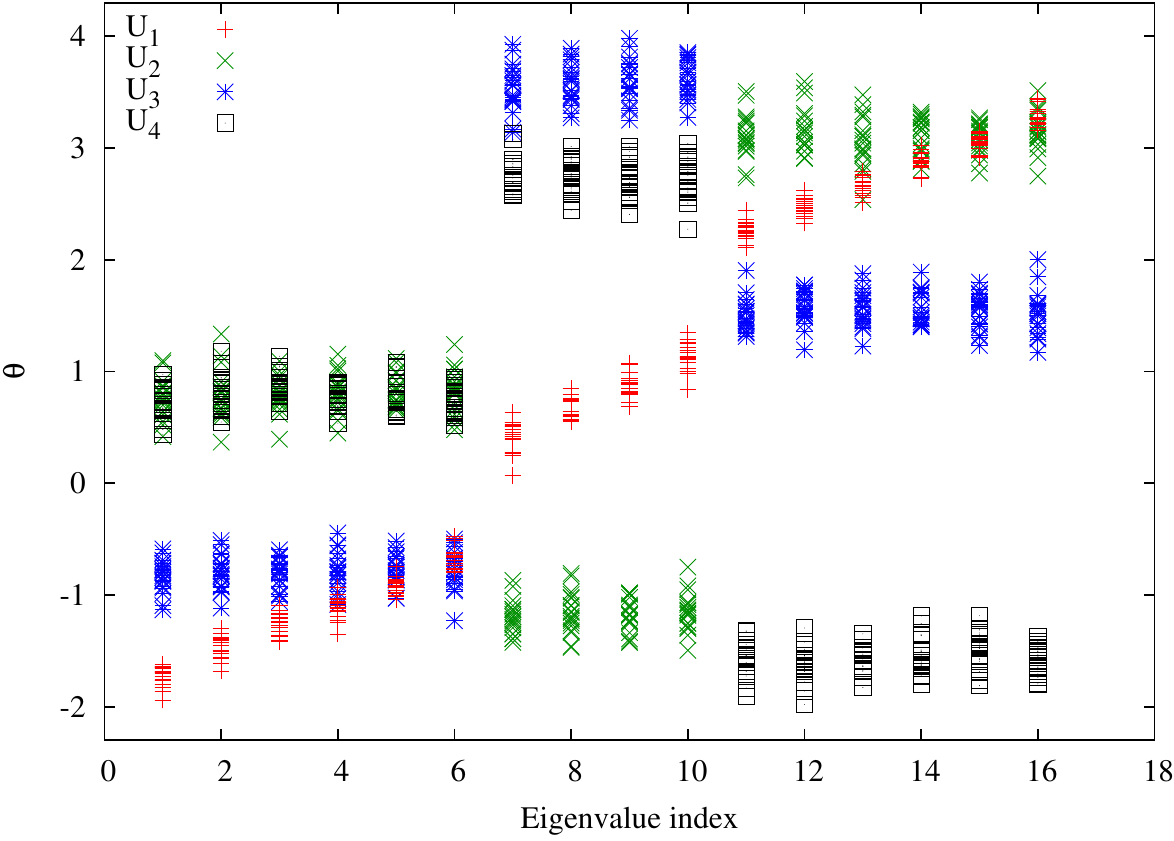}
\caption{Phases of the diagonal elements of $U_\mu$, in the gauge where $U_1$ is diagonal and its phases are ordered, for
20 thermalized configurations at $N_f=2$, $N=16$, $b=0.35$, $\kappa=0.23$. The phase is $\zz_3$ and the range of the phases
was adjusted so as to avoid the cut through one of the three clumps.}
\label{fig:link_phases}
\end{figure}

Fig.~\ref{fig:link_phases} presents the phases of $U_\mu$ in 20 configurations collected in a run with the same
parameters as in Fig.~\ref{fig:link_elems}. We set the gauge in such a way that the phases of eigenvalues of $U_1$ are
ordered -- note that no ambiguity exists in the ordering of the remaining diagonal elements once we specify the order
for $U_1$. We see three clumps of sizes 6,4,6 that, while being positioned at different angles, are almost completely
correlated between all four links, and do not change during the Monte Carlo evolution. Because of these correlations,
seen every time when the center symmetry is broken, we infer that the approximate remnant symmetry of
Fig.~\ref{fig:link_elems} is $\zz_3$, not $\zz^4_3$.

Ref.~\cite{ahu10} argues that one should expect that the number of clumps should become smaller as one gets further away
from the funnel (i.e.\ as the quark mass gets higher). That is indeed what we observe. On the large-$\kappa$ side we
extended some scans up to $\kappa=0.6$. We find that $\vartheta^a_\mu$ in the UP scans form less and less clumps after
leaving the funnel, until they end up in a two clump state. The DOWN scans, started from an ordered start, begin with a
single clump and as $\kappa$ is decreased have a transition to two clumps. The transition appears to occur in stages
where the eigenvalues gradually ``peel off'' from the original clump. As $\kappa$ is further decreased, there appear
more and more clumps until we enter the funnel and the eigenvalues become uniformly distributed on the unit circle. The
largest number of clumps depends on $N$, and the largest we have observed is five, as shown in Fig.~\ref{fig:hist_z5}.

One expects a similar phenomenon on the small-$\kappa$ side. In fact, as we increase $\kappa$ (thus reducing the quark
mass) from 0 we encounter similar transitions although the maximum number of clumps before entering the funnel is
significantly smaller in this region. For $N<23$ we only observe the $\zz_1$ phase, for $23\leq N<47$ we see $\zz_1$ and
$\zz_2$, while for $N=47$ and 53 we find $\zz_1$, $\zz_2$ and $\zz_3$ phases. The arguments of Ref.~\cite{ahu10} imply
that the maximum number of clumps should increase with $b$. Indeed, that is what we observe -- for $b>1$ the phase
$\zz_3$ appears at smaller values of $N$.

We have also checked the results given by the higher-order open loops defined in Eq.~\ref{eq:op_loops}, following the
method of Ref.~\cite{bs09}, as well as performed several vertical scans in the $\kappa-b$ plane to supplement our
knowledge of the phase diagram. We found that these calculations confirm the results presented earlier in this section
and, for the sake of brevity, we do not present them here.

\section{The width of the center-symmetric funnel}

\subsection{The $N$-scaling of $P_\mu$ and $M_{\mu\nu}$}
\label{sec:scaling}

The encouraging results of the previous section lead to the crucial question -- what happens to the center-symmetric
funnel as $N\to\infty$? In order to study the large-$N$ limit we have extended the calculations to larger values of $N$
in several points of the $\kappa-b$ plane.

Let us begin with the large-$N$ extrapolations of $\langle|P_\mu|^2\rangle$ and $\langle|M_{\mu\nu}|^2\rangle$. In the
$N\to\infty$ limit they are equal to $|\langle P_\mu\rangle|^2$ and $|\langle M_{\mu\nu}\rangle|^2$ (cf.\
Eq.~\ref{eq:fact}) -- which are both equal to zero if the center symmetry is unbroken. Thus an important check of the
proposed phase diagram, presented in Fig.~\ref{fig:phase_diag}, is that both these values extrapolate to 0 in the
tentative center-symmetric funnel.

\begin{figure}[tbp!]
\centering
\includegraphics[width=12cm]{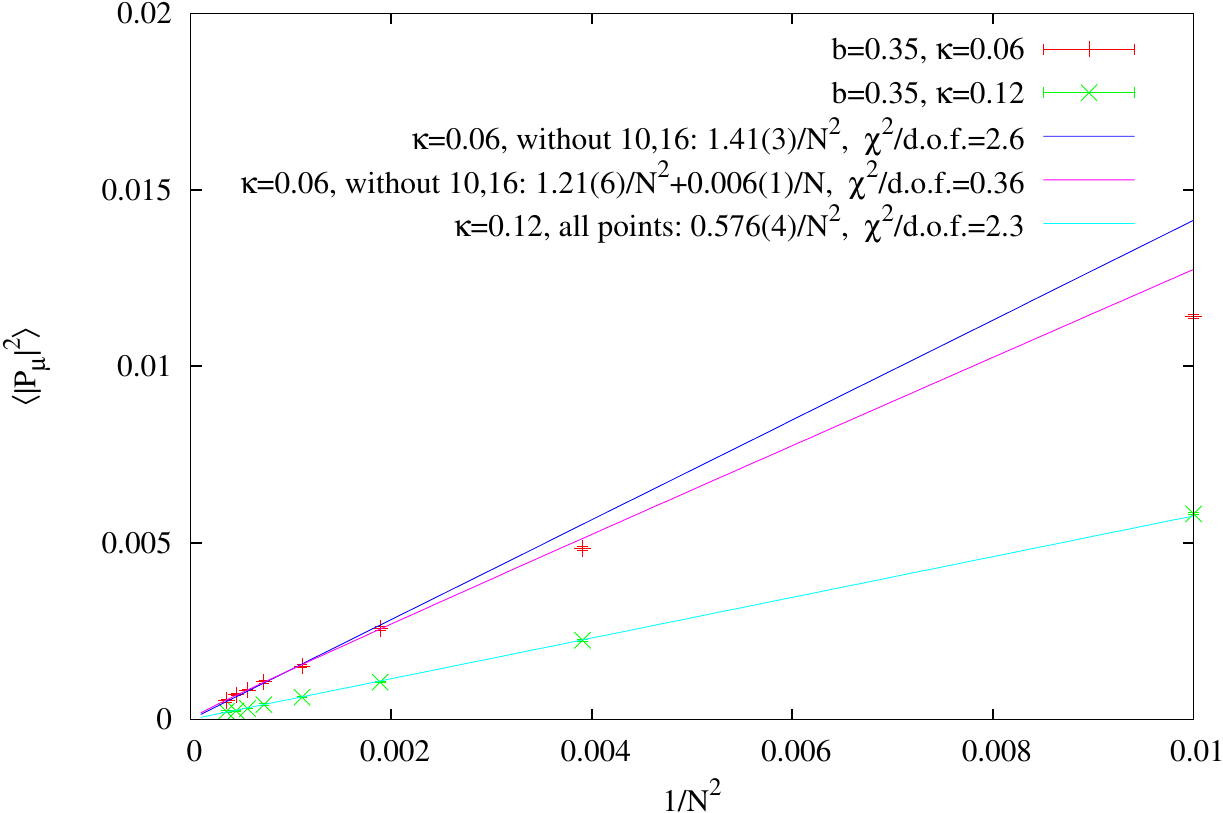}
\caption{Large-$N$ extrapolations of $\langle|P_1|^2\rangle$ for $N_f=2$, $b=0.35$ and two values of $\kappa$: 0.06
and 0.12. The values of $N$ are 10, 16, 23, 30, 37, 42, 47 and 53.}
\label{fig:poly_extr}
\end{figure}

Ordinarily finite-$N$ effects are $\mathcal{O}(1/N^2)$ but in the volume-reduced models one can also encounter
$\mathcal{O}(1/N)$ corrections (see Sec.~\ref{sec:plaq}). In Fig.~\ref{fig:poly_extr} we show $\langle|P_1|^2\rangle$ as
a function of $1/N^2$ for two representative points in the $\kappa-b$ plane. Although the $1/N^2$ term is clearly the
largest finite-$N$ contribution in many cases we found that pure $1/N^2$ fits have large values of
$\chi^2/\text{d.o.f.}$. Addition of $1/N$ term as well as dropping the 2 lowest values of $N$ (10 and 16) lead to
satisfactory fits in all analyzed cases -- the values of the fit parameters are collected in Table~\ref{tab:loops_extr}.
The coefficients of the $1/N$ term are small in all cases, and in fact consistent up to $\sim 3\sigma$ with zero except
for $b=0.35$, $\kappa=0.06$. We also show the results of fits to $1/N^2$ plus a constant term -- the fits are of similar
quality and the constant term is consistent with 0 except, again, the point $b=0.35$, $\kappa=0.06$. We conclude that
(except this one point near the edge of the tentative funnel) the behaviour of Polyakov loops is consistent with
hypothesis that the center symmetry is intact in the funnel.

\begin{table}[tbp!]
\footnotesize
\renewcommand{\arraystretch}{1.5}
\begin{adjustwidth}{-1.2cm}{-1.2cm}
\centering
\setlength{\tabcolsep}{3mm}
\begin{tabular}{ccccccccc}
\hline\hline
Qty&$b$ & $\kappa$ &
$c_1$&$c_2$&$\frac{\chi^2}{\text{d.o.f.}}$ &$c'_0$&$c'_2$&$
\frac{\chi'^2}{\text{d.o.f.}}$ 
\\ \hline
$\langle |P_{1}|^2\rangle$& 0.35 & 0.06 & 
0.006(1) & 1.21(6) & 0.36 & $9(1)\times 10^{-5}$ & 1.31(1) & 0.39
\\
$\langle |P_{1}|^2\rangle$& 0.35 & 0.09 & 
0.0014(9) & 0.73(3) & 0.67 & $2(1)\times 10^{-5}$& 0.76(2) & 0.73
\\
$\langle |P_{1}|^2\rangle$& 0.35 & 0.12 & 
0.001(3) & 0.57(3) & 0.82 & $0(2)\times 10^{-5}$ & 0.56(1) & 0.84
\\ \hline
$\langle |M_{12}|^2\rangle$& 0.35 & 0.09 & 
0.152(5) & 2.9(1) & 0.42  & 0.0023(3) & 5.3(4) & 2.57
\\
$\langle |M_{12}|^2\rangle$& 0.35 & 0.12 & 
0.036(6) & 3.5(2) & 1.0 & $5(1)\times 10^{-4}$ & 4.1(2) & 1.2
\\ \hline
$\langle |P_{1}|^2\rangle$& 1.0 & 0.06 & 
$-0.0001(3)$ & 1.17(1) & 0.025 & $-1(3)\times 10^{-6}$ & 1.17(1) &
0.025
\\
$\langle |P_{1}|^2\rangle$& 1.0 & 0.09 & 
$-0.0003(4)$ & 0.70(1) & 1.2 & $-5(6)\times 10^{-6}$ & 0.70(1) & 1.2
\\
$\langle |P_{1}|^2\rangle$& 1.0 & 0.12 & 
$-0.0010(3)$ & 0.55(1) & 0.60 & $-1.4(4)\times 10^{-5}$ & 0.54(1) & 0.58
\\
\hline
$\langle |M_{12}|^2\rangle$& 1.0 & 0.06 & 
0.69(3) & -1.6(7) & 0.58 & 0.010(1) & 9(2) & 2.7
\\
$\langle |M_{12}|^2\rangle$& 1.0 & 0.09 & 
0.0053(7) & 6.1(2) & 0.99  & $8(2)\times 10^{-4}$ & 6.9(3) & 1.7
\\
$\langle |M_{12}|^2\rangle$& 1.0 & 0.12 & 
$0.01(1)$ & 5.5(4) & 0.4 & $0(1)\times 10^{-4}$ & 5.4(2) & 0.41
\\
\hline
\end{tabular}
\caption{Results from fits to the large-$N$ behavior of $\langle |P_{1}|^2\rangle$ and $\langle|M_{12}|^2\rangle$ in the
$N_f=2$ case \cite{bks11}. Fits are to $N=23$, $30$, $37$, $42$, $47$ and $53$ using $f_1(N)=c_1/N + c_2/N^2$ and
$f_2(N)=c'_0 + c'_2/N^2$, showing statistical errors.}
\label{tab:loops_extr}
\end{adjustwidth}
\end{table}

\begin{figure}[tbp!]
\centering
\includegraphics[width=12cm]{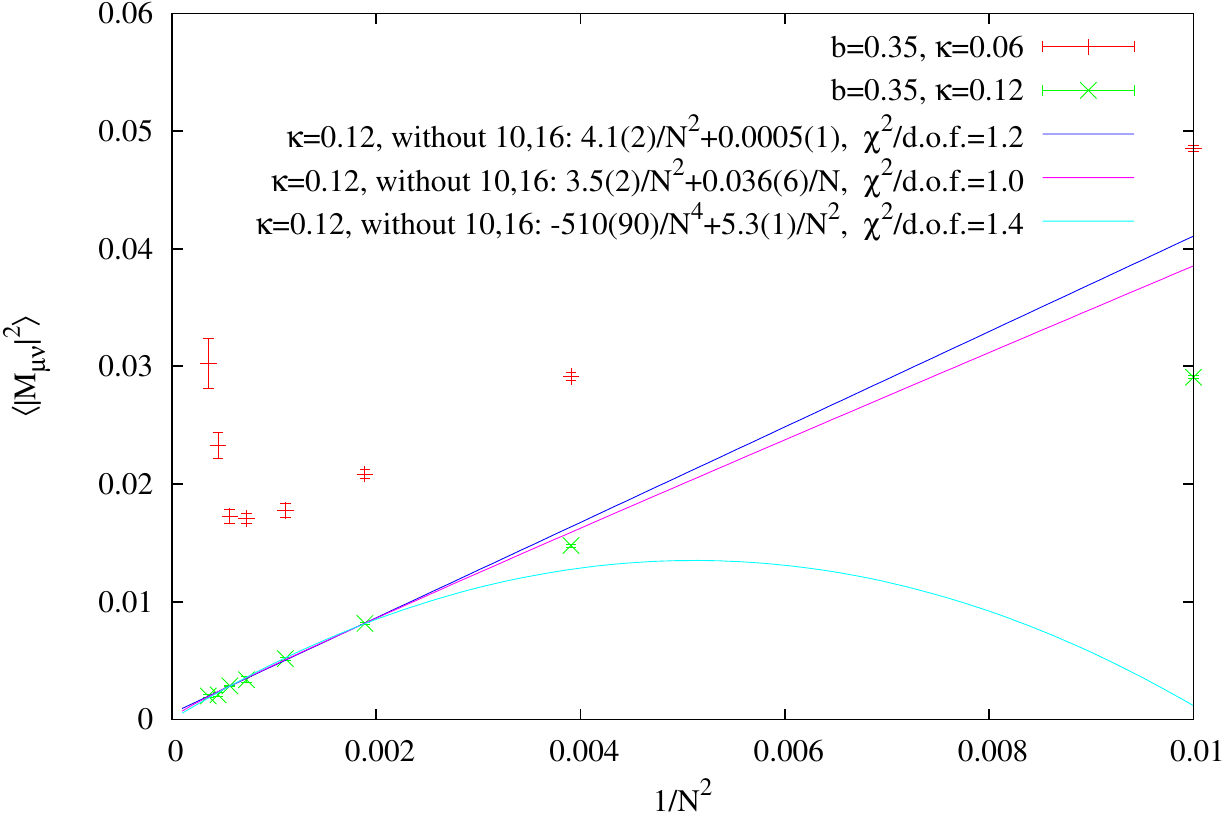}
\caption{Large-$N$ extrapolations of $\langle|M_{12}|^2\rangle$ for $N_f=2$, $b=0.35$ and two values of $\kappa$: 0.06
and 0.12.}
\label{fig:corner_extr}
\end{figure}

Example data for the corner variables are presented in Fig.~\ref{fig:corner_extr}. The particularly striking is the data
at $b=0.35$, $\kappa=0.06$ -- the pictured $\langle|M_{12}|^2\rangle$ begins to rise for $N\gtrsim40$ and clearly does
not extrapolate to zero. This result is rather surprising as otherwise there is no clear sign of breaking of the center
symmetry -- for all analyzed values of $N$, $M_{\mu\nu}$ as well as other open loops from Eq.~\ref{eq:op_loops} are
distributed approximately near the origin and the distribution of link eigenvalues is consistent with uniform.

Our interpretation is that the increase in $\langle|M_{\mu\nu}|^2\rangle$ with $N$ is due to the lower edge of the
funnel, $\kappa_f$, increasing with $N$ (possibly reaching $\kappa_f>0.06$ in the large-$N$ limit)\footnote{This effect
can be also observed (for $b=1.0$) in Fig.~\ref{fig:corner_scan} where the jump in the corner variables for $N=30$ is at
visibly higher values of $\kappa$ than for $N=16$.}. In that case, we might be able to observe clear signal of $\zz_N^4$
symmetry breaking at higher values of $N$ and the hints of the transition are already visible in the values we use due
to the vicinity of increasing $\kappa_f$. The results for $M_{\mu\nu}$ also explain why the $P_\mu$ fits were less
satisfactory in this point.

For all other analyzed points we find that the corner variables decrease monotonically with $N$. As for the Polyakov
loops, the $1/N^2$ fits are rather poor but the addition of $1/N$ term leads to acceptable fits in all cases (note that
contrary to the $P_\mu$ case the $1/N$ terms differ significantly from 0). The fits to $1/N^2$ plus constant lead to
small but often non-zero constant values but they also usually have higher $\chi^2/\text{d.o.f.}$ than the
$c_2/N^2+c_1/N$ fits -- we thus conclude that the latter fits are the most reasonable ones\footnote{We have also
experimented with fits to $c_2/N^2+c_4/N^4$ (an example is shown in Fig.~\ref{fig:corner_extr}) but the fits require
very large coefficients with opposite sign to give a satisfying quality -- we consider such fine tuning as very unlikely
to be the correct description.}.

\subsection{Funnel width as $N\to\infty$}
\label{sec:funnel_n}

The presented results suggest that the center symmetry is in fact unbroken in a sizeable region of parameters, but there
remains some anxiety whether $\kappa_f$ does not in fact converge to $\kappa_c$ in the large-$N$ limit -- in that case
the reduction with massive fermions may only be apparent for some low enough values of $N$ (as in the $b=0.35,
\kappa=0.06$ case).

\begin{figure}[tbp!]
\centering
\includegraphics[width=12cm]{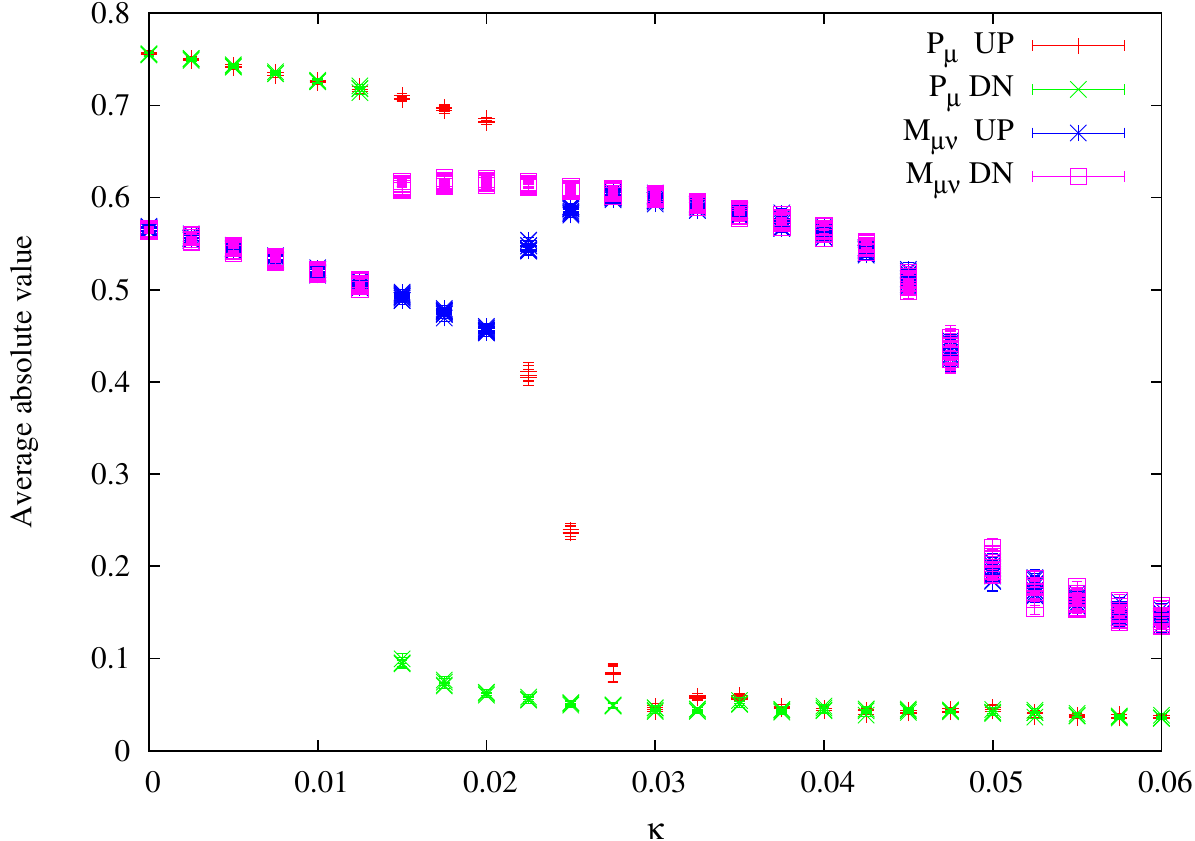}
\caption{Absolute values of all Polyakov loops and corner variables in the low $\kappa$ region, $N_f=2$, $N=30$, $b=0.75$.}
\label{fig:low_kappa}
\end{figure}

To resolve this matter it is important to study the function $\kappa_f(N)$ and to perform its large-$N$ extrapolation.
To investigate this we have performed a series of fine scans of the small $\kappa$ region. An example is pictured in
Fig.~\ref{fig:low_kappa}. There are two phases with broken center symmetry before we enter the funnel: the system is in
the $\zz_1$ phase in the region $0\leq\kappa\lesssim0.02$ and in the $\zz_2$ phase in $0.02\lesssim\kappa\lesssim0.05$.
The transition between the first and the second phase shows a significant hysteresis while that between the second phase
and the funnel does not -- we will discuss the possible order of this phase transition in the following part of this
section.

A reliable large-$N$ extrapolation of $\kappa_f(N)$ requires very costly analysis and with our resources we were only
able to perform it at a single value of coupling, $b=1.0$, albeit for both $N_f=2$ and $N_f=1$. In both cases we have
done very fine scans near the edge of the funnel. $N\leq53$ was used for $N_f=2$ and $N\leq60$ for $N_f=1$. The corner
variables were found to be the most useful in determining the transition (although we performed careful checks if the
other observables do not show any signs of some more complicated pattern of the symmetry breaking that the corner
variables could miss). We were able to pin down the transition, conservatively, to about $\delta\kappa=\pm0.001$.

\begin{figure}[tbp!]
\centering
\subbottom[$N_f=2$] {\label{fig:funnel2} \includegraphics[width=12cm]{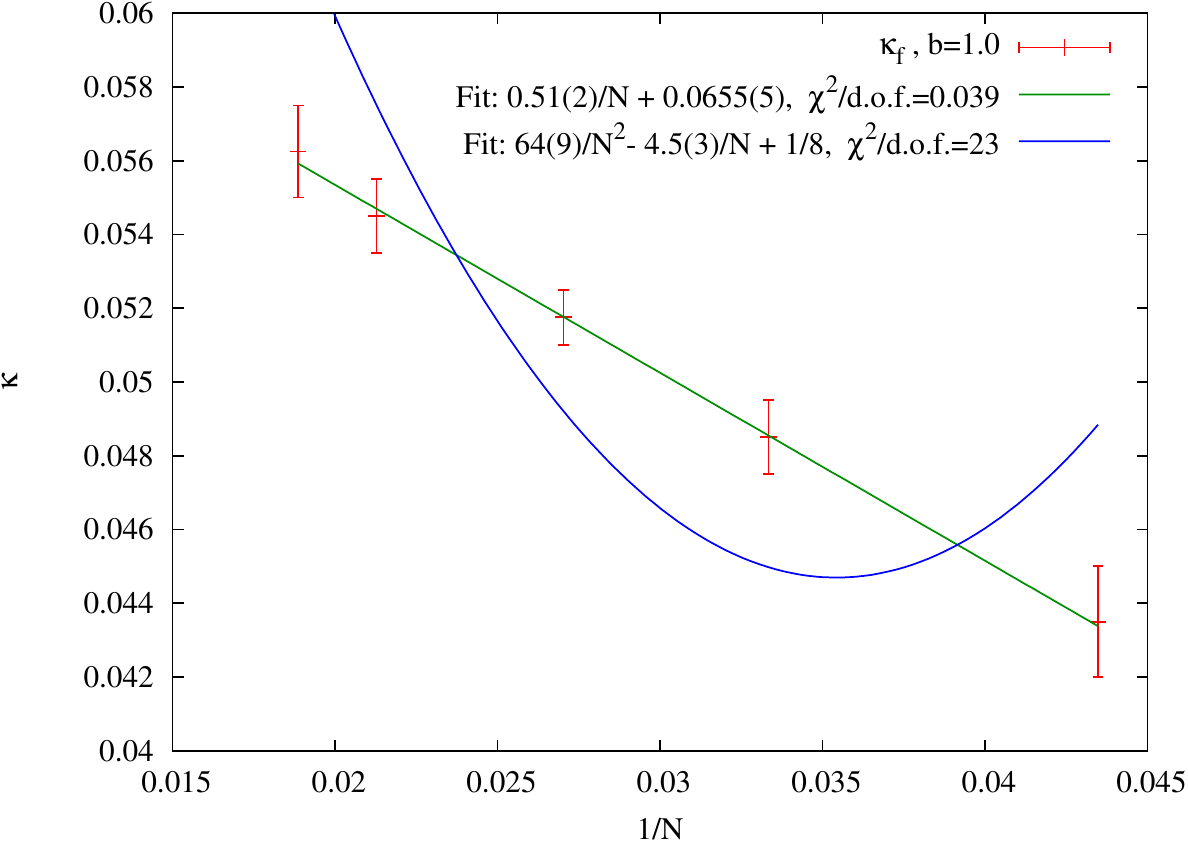}}\\[6pt]
\subbottom[$N_f=1$] {\label{fig:funnel1} \includegraphics[width=12cm]{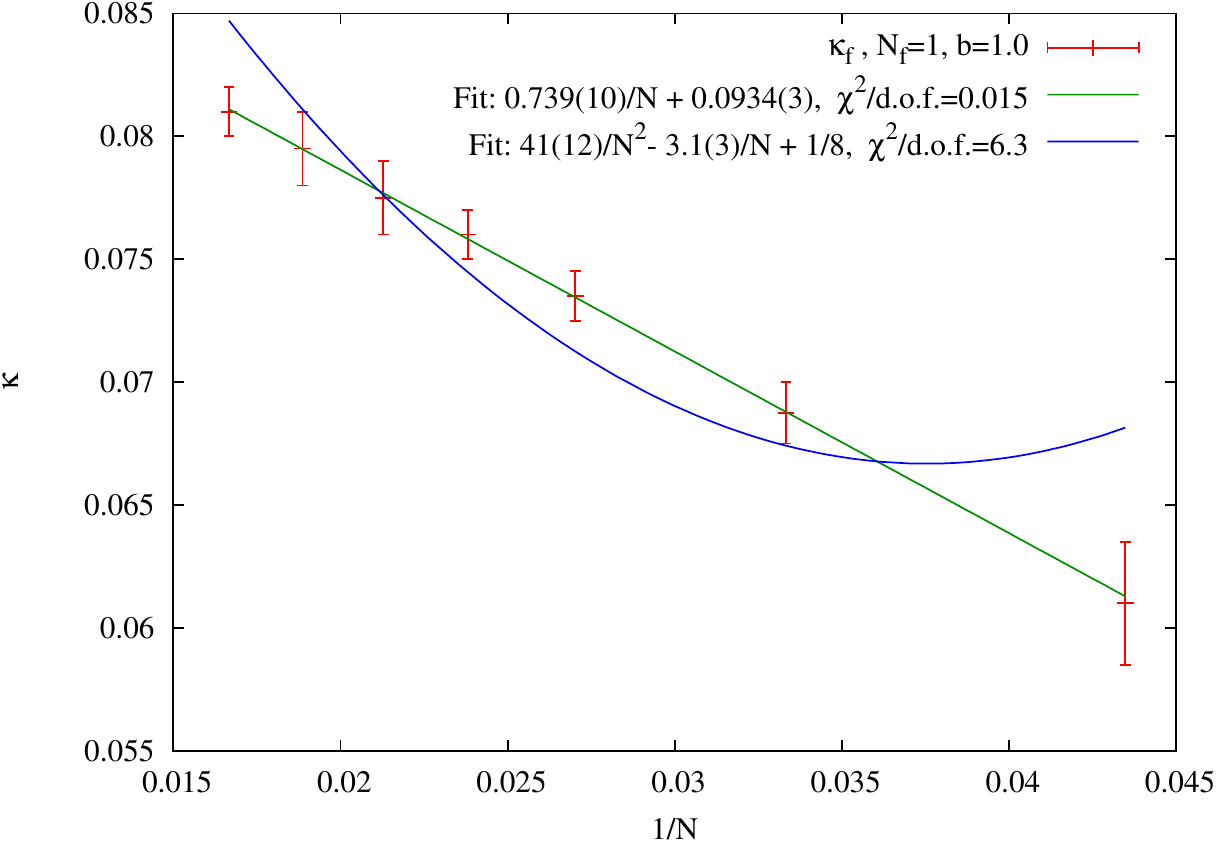}}\\[6pt]
\caption{Large-$N$ extrapolations of $\kappa_f$. Note that the errors are rather conservative as the main goal
was the exclusion of the ``closed funnel'' hypothesis. The point $N=60$ for $N_f=1$ was calculated using the new parallel
code on the \texttt{Deszno} supercomputer.}
\label{fig:funnel}
\end{figure}

The results are presented in Fig.~\ref{fig:funnel} along with two fits each. The first fits are of the form $c_0+c_1/N$
-- they have a very low $\chi^2/\text{d.o.f.}$ (signalling that the uncertainties are perhaps overly conservative) and
they look very reasonable. The extrapolated large-$N$ limits are:
\begin{align}
\kappa_f(N_f=2,N=\infty,b=1) = 0.0655(5),\\
\kappa_f(N_f=1,N=\infty,b=1) = 0.0934(3),
\end{align}
both of which are values substantially lower than $\kappa_c$ (which is $\approx0.13-0.14$ at this value of $b$). Note
that the funnel is significantly narrower for $N_f=1$. This is understandable in the context of the perturbative
calculations presented in Sec.~\ref{sec:pert_ferm} (cf.\ Eq.~\ref{eq:coeffs}).

The second fits are to the function $c_1/N+c_2/N^2+1/8$ where the constant term is the lower limit on
$\kappa_c$\footnote{That is, without taking into account the additive renormalization of the mass. The precise
determination of $\kappa_c$ is very costly and we have limited ourselves to finding the approximate value. Therefore, we
choose the most conservative approach in the fits.}. In both cases the fit is extremely poor ($\chi^2/\text{d.o.f.}=23$
for $N_f=2$ and $\chi^2/\text{d.o.f.}=6.3$ for $N_f=1$), excluding the possibility of the closed funnel -- particularly
taking into account that taking $\kappa_c=0.13-0.14$ makes the fits even worse. We conclude that, at least for $b=1$,
the funnel has finite width in the large-$N$ limit and the reduction holds also for rather heavy fermions.

\subsection{Funnel width as a function of $b$}
\label{sec:funnel_b}

Another interesting property of the funnel is its behaviour as a function of $b$. Ref.~\cite{ahu10} predicts that at
large $b$ the width of the funnel in $am$ should be proportional to $b^{-1/4}$. As was already mentioned in the previous
section, it also predicts that at the edge of the funnel there should be multiple phases with partially broken symmetry
-- the width of each eigenvalue clump is proportional to $b^{-1/4}$ so the maximum number of clumps should increase with
$b$ (as long as $N$ is large enough).

To investigate this we have performed a series of scans in $\kappa$ for $N=10$ (and, in some cases, for $N=30$) at
b=2.5, 5, 10, 50 and 200 (the presented analysis uses $N_f=2$ data only; we have no reasons to suspect a qualitatively
different behaviour of $N_f=1$ in this case). We find that the simulation algorithm performs surprisingly well even at
such extremely high values of $b$. No indication of autocorrelation times that are close to the number of trajectories
used for measurements (7500 in the $N=10$ case) was found\footnote{However the MD step size has to be reduced,
approximately as $\sqrt{b}$, and we find that for $b=200$ the 450 trajectories used for thermalization is sometimes
slightly too little for the system to thermalize completely.}.

We analyze $\kappa_f$ as a function of $b$ to check whether
\begin{equation}
am_f\equiv\frac{1}{2\kappa_f}-\frac{1}{2\kappa_c}\propto b^{-1/4}.
\end{equation} 

It would be of great benefit to repeat the study of the previous section for all values of $b$ -- that is however out of
our numerical possibilities. Also, we find that it is harder to pin down the precise $\kappa_f$ for $b$ away from unity.
At $b=0.35$ the transition to the center-symmetric phase is very smooth, up to the point where locating it precisely is
very hard and in fact it may be a crossover -- one sign of such difficulties was already seen in Sec.~\ref{sec:scaling}
in the case $b=0.35$, $\kappa=0.06$. Also, Ref.~\cite{go122} has triggered a discussion whether the funnel has at all
finite width at this value of $b$. At $b=0.5$ and above the transition becomes well-located and with growing $b$ it
becomes stronger and stronger. At very high $b$ it becomes a strong-first order transition with significant hysteresis
(see also the discussion of the plaquette in Sec.~\ref{sec:plaq}) that makes it difficult to precisely determine
$\kappa_f$ without using more advanced methods. The situation is thus not entirely explained away from $b=1$ and would
definitely benefit from further study.

\begin{figure}[tb!]
\centering
\includegraphics[width=12cm]{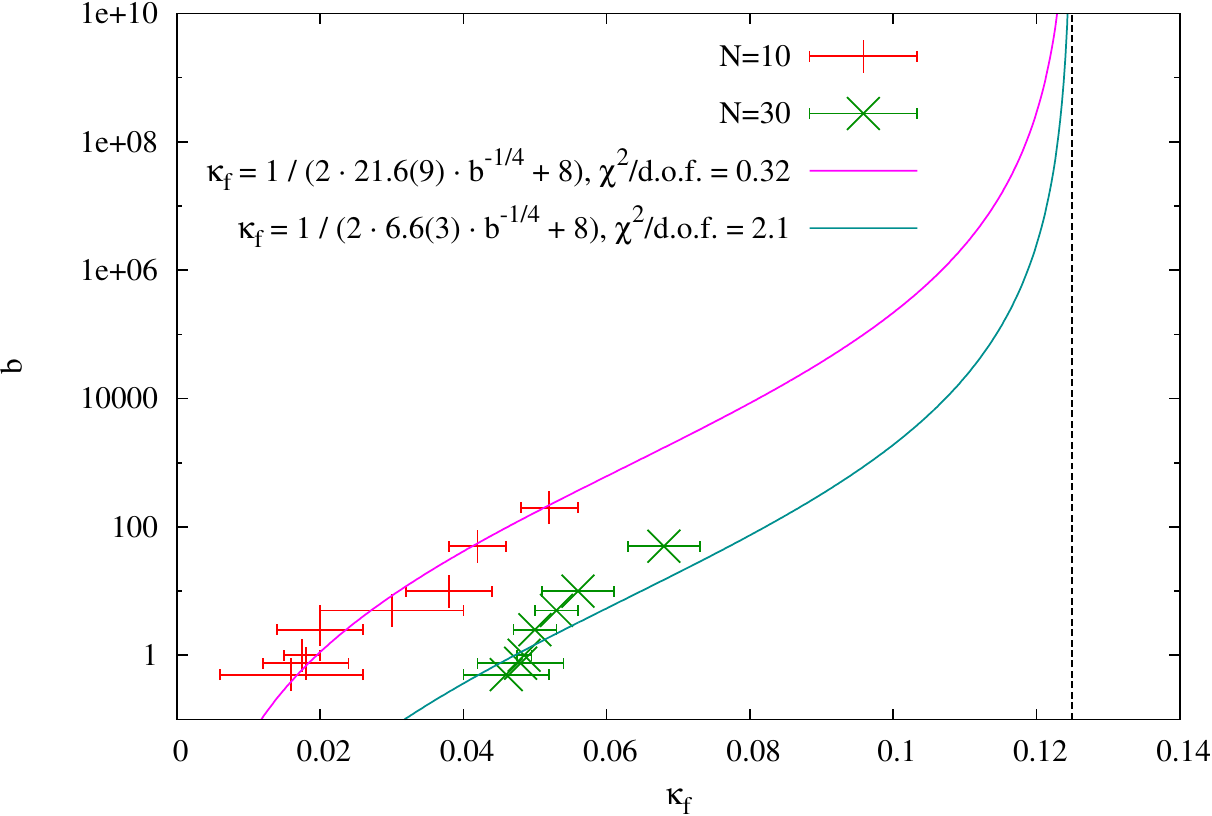}
\caption{The dependence of $\kappa_f$ on $b$, for $N=10$ and 30, $N_f=2$. The dashed vertical line marks
$\kappa_c(b=\infty)=1/8$. The fit functions are discussed in the text.}
\label{fig:kappa_f_b}
\end{figure}

Therefore, we limit ourselves to the analysis of the data for $N=10$ and 30, presented in Fig.~\ref{fig:kappa_f_b}. Along with
the data we present the fits to the form predicted by Ref.~\cite{ahu10} with $\kappa_c=0.125$ for all $b$, for
simplicity. The fit at $N=10$ is good while the fit at $N=30$ has some tension to the obtained data (the data seems to
be better reproduced if we insert the estimate for the true value of $\kappa_c$ but that also significantly increases
the uncertainties in $am_f$ as the precision of our estimate of $\kappa_c$ is rather limited). Overall, our conclusion
is that the obtained results are consistent with the predicted dependence on $b$.

\chapter{Physical measurements in the AEK model}
\chaptermark{AEK model: Physical measurements}
\label{ch:obs}

\section{Plaquette}
\label{sec:plaq}

The basic quantity to analyze in any lattice gauge theory is the average value of the plaquette
\begin{equation}
U_\text{plaq} = \frac1{6N}\sum_{\mu<\nu}\Tr U_\mu U_\nu U_\mu^\dagger U_\nu^\dagger.
\end{equation}
While the plaquette is a UV-dominated quantity, and thus does not relate directly to any continuum observable, it is
very helpful to study the gross features of the theory and to compare different lattice calculations.

\begin{figure}[tbp!]
\centering
\subbottom[$b=0.5$] {\label{fig:plaq_b05} \includegraphics[width=12cm]{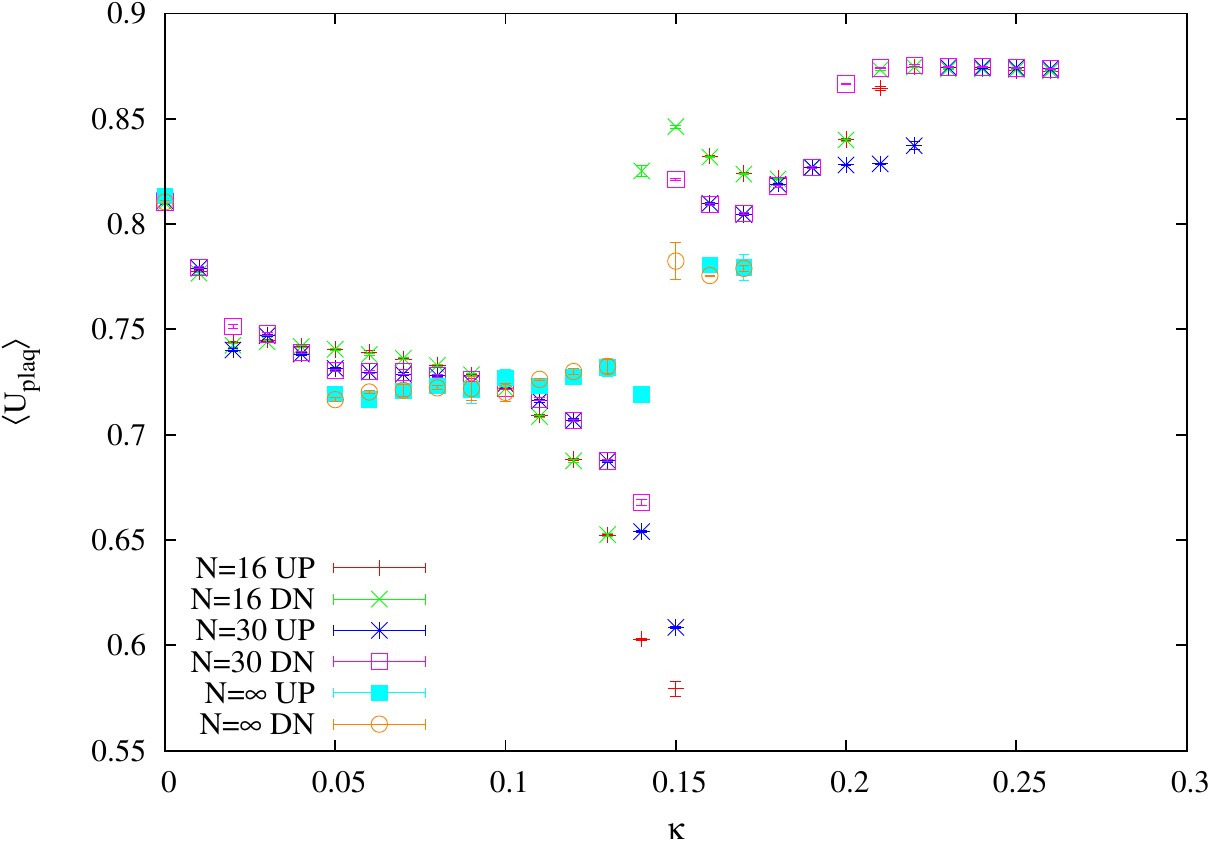}}\\[12pt]
\subbottom[$b=1.0$] {\label{fig:plaq_b1} \includegraphics[width=12cm]{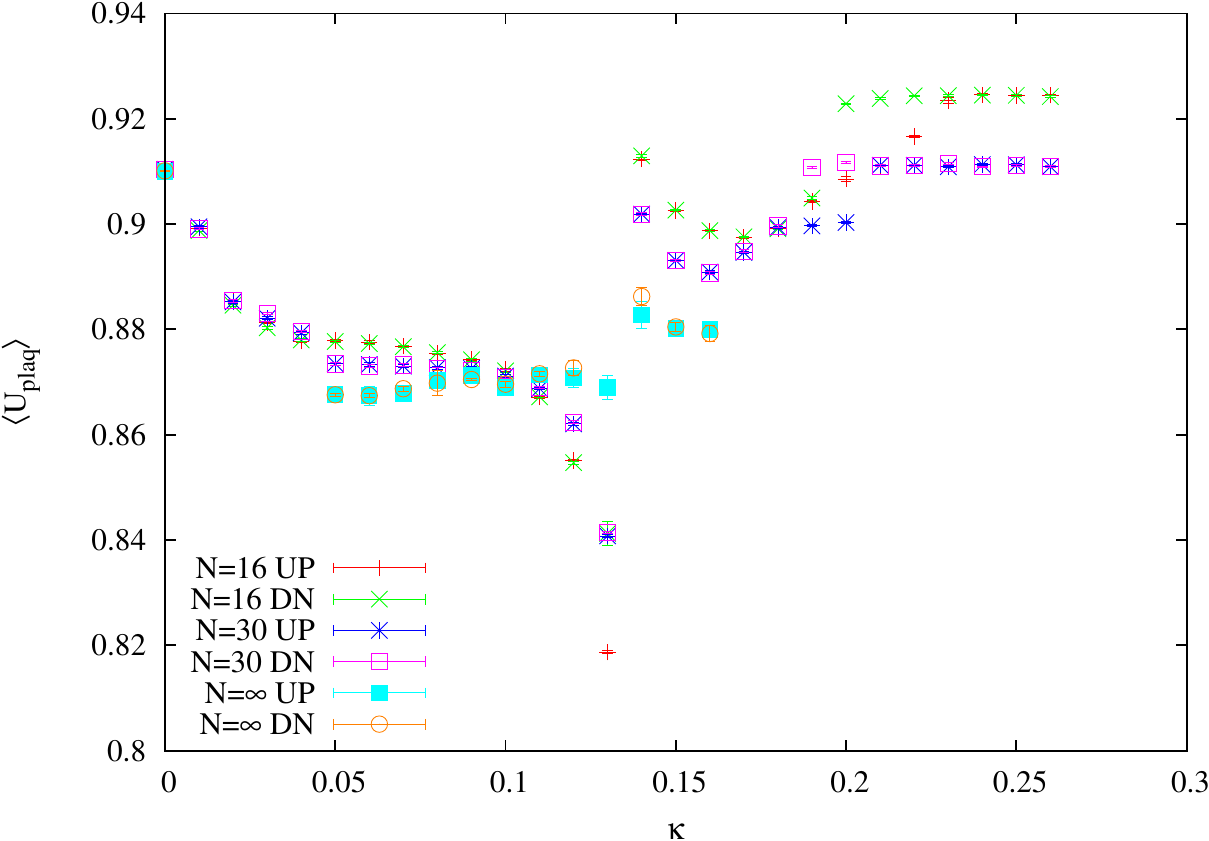}}\\[6pt]
\caption{Horizontal scans (both UP and DOWN) of the average plaquette at $N_f=2$ for $N=16$, 30 and the large-$N$
extrapolations (where possible), and two values of the 't Hooft coupling.}
\label{fig:plaq_scans}
\end{figure}

Fig.~\ref{fig:plaq_scans} presents the results of the plaquette in the horizontal scans at $b=0.5$ and $b=1.0$ (with the
parameters described in Sec.~\ref{sec:scans}). We also plot, where possible, the approximate large-$N$ results obtained
by fitting the results at the four values of $N$ (or more, where available) to the function $c_0+c_1/N+c_2/N^2$. The
more precise fits done in selected points of the $\kappa-b$ plane are described further in this section.

In both plots there is a visible change of slope around $\kappa\approx0.05$, when we enter the funnel. Also in the
large-$\kappa$ region we see a change of behaviour at the edge of the funnel. Another interesting feature of both plots
is the behaviour in the vicinity of $\kappa_c\approx0.13$ -- there is a large jump in the plaquette value. The jump
becomes smaller with growing $N$, however it seems to persist at large $N$. While a more precise (and very expensive
computationally) scan in the near-$\kappa_c$ region could resolve this issue, this data suggests that the phase
transition in $\kappa_c$ is likely to be of first order.

At $N_f=1$ we see the same picture, which was already reported in Ref.~\cite{bs09}. In the $N_f=1$ case this is the
expected behaviour because the large-volume theory is expected to be confining \cite{ds07,san09}. For $N_f=2$ the
supposed first order transition in $\kappa_c$ is however somewhat surprising as the large-volume theory is expected to
be conformal (in the massless case).

The argument goes as follows: based on analysis in chiral perturbation theory one can argue that for a confining theory
one expects (close to the continuum limit) a first order transition or two second order transitions separated by the
Aoki phase \cite{ss98,dfp08}. On the other hand if the theory is conformal one expects a single second order phase
transition at $\kappa_c$ -- that is the case for $N=2$ where the single second order phase transition is observed for
$b\gtrsim0.25$ \cite{cgs08,hrt09}. A different behaviour of the large-$N$ theory is possible although unexpected (see
Sec.~\ref{sec:def}).

\begin{figure}[tbp!]
\centering
\subbottom[$b=0.35, \kappa=0.09$] {\label{fig:plaq_extr_b035}
\includegraphics[width=12cm]{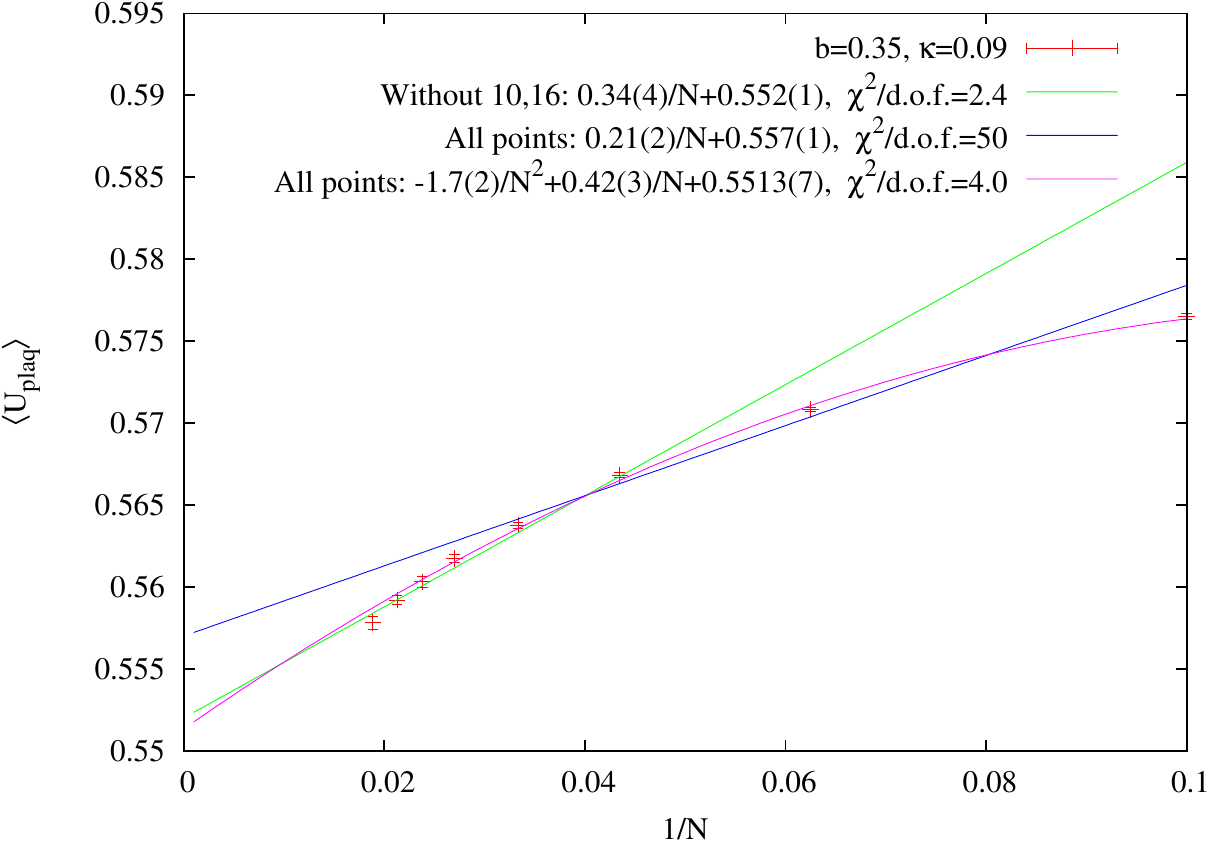}}\\[12pt]
\subbottom[$b=1.0, \kappa=0.12$] {\label{fig:plaq_extr_b1}
\includegraphics[width=12cm]{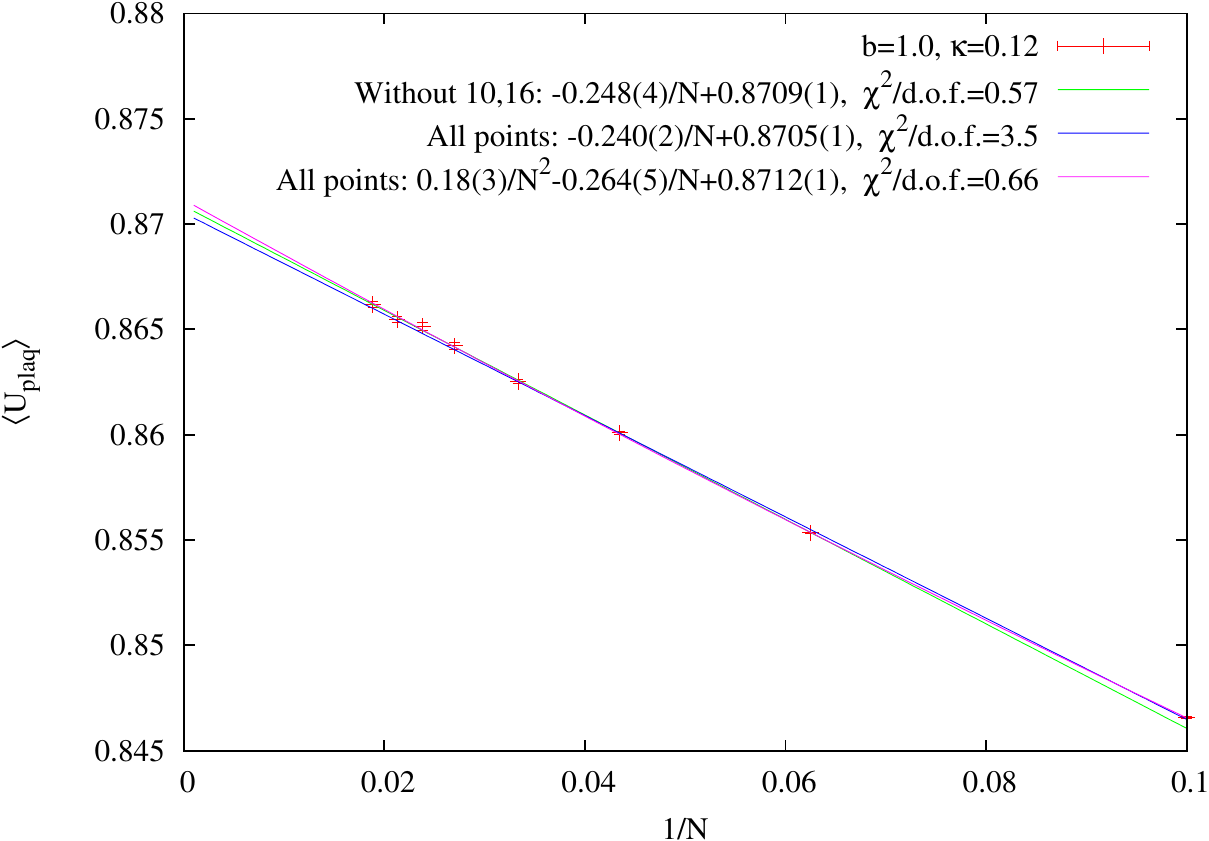}}\\[6pt]
\caption{Two examples of large-$N$ extrapolations of the plaquette for $N_f=2$. Various fits are shown in addition to
the data.}
\label{fig:plaq_extr}
\end{figure}

Let us now discuss the large-$N$ extrapolations of the average plaquette. Fig.~\ref{fig:plaq_extr} shows two example
plots for different parameters. As already pictured in Fig.~\ref{fig:plaq_scans}, the finite-$N$ corrections are quite
substantial and can have either sign, depending on the position in the $\kappa-b$ plane. We find that the leading
finite-$N$ correction is $\mathcal{O}(1/N)$. This is different from the large-volume results where one expects
$\mathcal{O}(1/N^2)$. In the reduced model one can however observe this kind of behaviour -- the one-loop analysis of
the one-site pure-gauge model around the center-symmetric vacuum gives \cite{oka82}:
\begin{equation}
\langle U_\text{plaq}\rangle = 1-\frac1{8b}(1-1/N) + \mathcal{O}(1/b^2).
\label{eq:plaq_pert}
\end{equation}
In fact, comparing Eq.~\ref{eq:plaq_pert} to the data in Fig.~\ref{fig:plaq_extr_b1} we see that at $b=1$ the large-$N$
one-loop result, 0.875, is a reasonable approximation to the limit obtained from the data, even with moderately light
fermions at $\kappa=0.12$.

\begin{table}[tbp!]
\setlength{\tabcolsep}{3.5mm}
\begin{tabular}{ccccccc}
\hline\hline
$b$ & $\kappa$ & $\chi^2/\text{d.o.f.}$ & $c_1$ & $c_0$ & pure-gauge value 
\\ \hline
0.35 & 0.06 & 1.8 & 0.75(4) & 0.549(1) & 0.550 
\\
0.35 & 0.09 & 2.4 & 0.34(4) & 0.552(1) & 0.550 
\\
0.35 & 0.12 & 1.5 & -0.92(3) & 0.565(1) & 0.550 
\\ \hline
1.0  & 0.06 & 0.2 & 0.120(3) & 0.8694(1) & 0.8692 
\\
1.0  & 0.09 & 1.1 & 0.076(3) & 0.8697(1) & 0.8692 
\\
1.0  & 0.12 & 0.6 & -0.248(4) & 0.8709(1) & 0.8692
\\
1.0  & 0.15 & 2.3 & 0.39(1) & 0.8795(4) & 0.8692
\\ \hline
\end{tabular}
\caption{Results from large-$N$ extrapolation of plaquette expectation values at $N_f=2$. We choose the fit function
$c_0+c_1/N$ to results at $N=23$, 30, 37, 42, 47 and 53 -- we find that in all cases analyzed this choice leads to
reasonable fit quality (the largest obtained $\chi^2/\text{d.o.f}$ is 2.4 which corresponds to p-value $p\approx0.04$).
The inclusion of $N=10$ and 16 leads to worse fit quality and in some cases even the inclusion of $1/N^2$ term does not
lead to acceptable fits (see e.g.\ Fig.~\ref{fig:plaq_extr_b035}). The table quotes the values of $c_0$, $c_1$ and
$\chi^2/\text{d.o.f.}$, with errors being statistical. Systematic errors (from different choices of fit function) are a
few times larger than the statistical errors. Our best estimates of the pure-gauge large-volume expectation value are
also quoted. The $b=1$ value is a perturbative result obtained from Ref.~\cite{bs09}, while that at $b=0.35$ is obtained
from the $N=8$ pure-gauge simulation at $b=0.3504$, from Ref.~\cite{att08}.}
\label{tab:plaq_extr}
\end{table}

In general, inside the center-symmetric funnel but far away from $\kappa_c$ (for the heavy fermions) we expect that the
large-$N$ extrapolation of the plaquette will be close to the large-volume pure-gauge result with a growing discrepancy
from the pure-gauge behaviour as $\kappa\to\kappa_c$. To verify this hypothesis, in Table \ref{tab:plaq_extr} we collect
the results of the extrapolations together with the estimates of the pure-gauge large-volume results.

We see that the extracted large-$N$ results, $c_0$, confirm our semi-quanti\-tative prediction. For $\kappa=0.06, 0.09$
the results are consistent with the pure-gauge results while for $\kappa=0.12$ (closer to $\kappa_c$) they begin to
differ. We also consider one point on the large-$\kappa$ side of $\kappa_c$. In this case the plaquette differs
substantially from the pure-gauge result. This shows that the difference of behaviour between the l.h.s.\ and r.h.s.\ of
$\kappa_c$ persists in the large-$N$ limit (see also the large-$\kappa$ regions in Fig.~\ref{fig:plaq_scans}).

\begin{figure}[tb!]
\centering
\includegraphics[width=12cm]{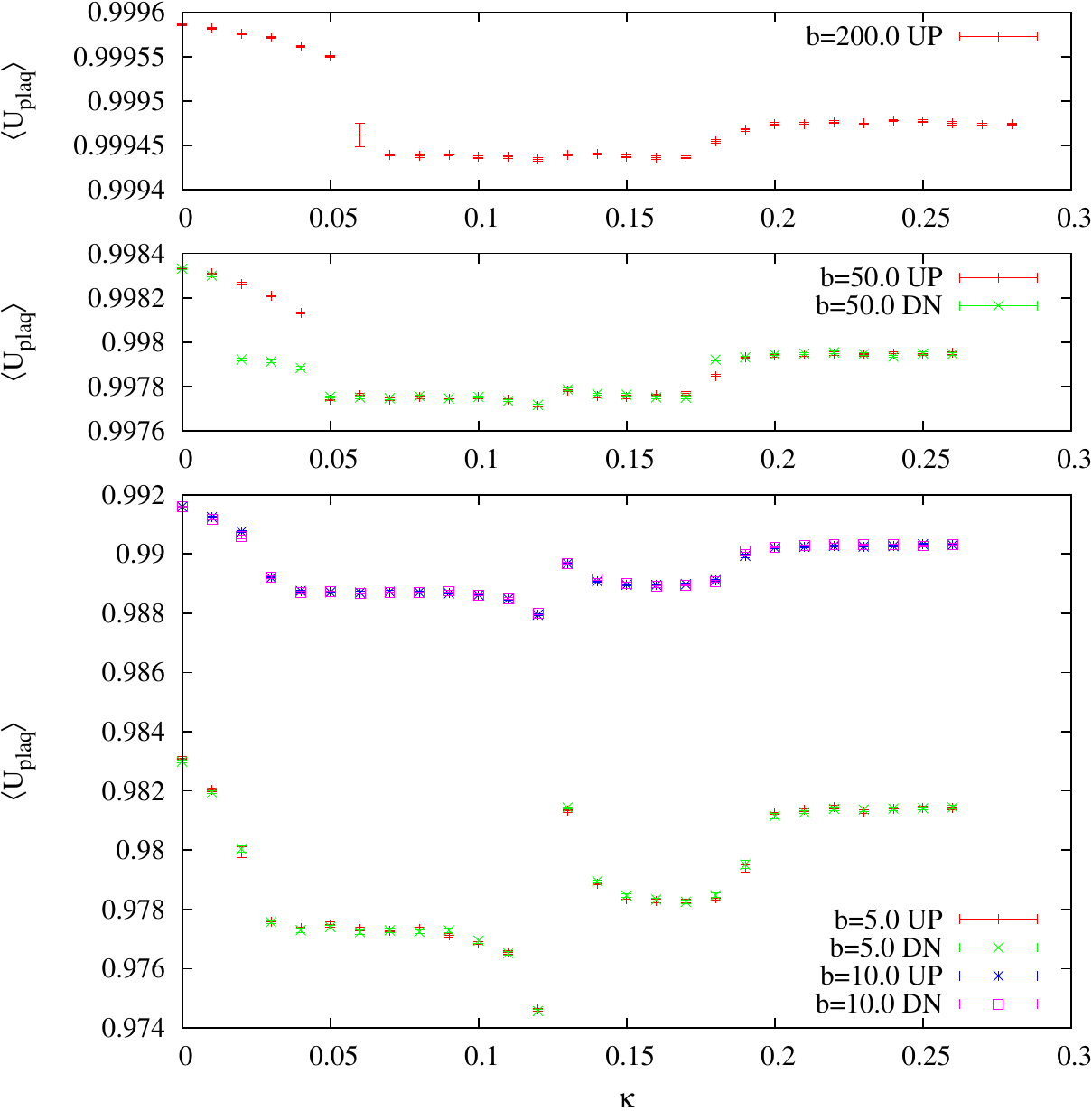}
\caption{The average plaquette in scans at extremely high $b$ for $N_f=2$, $N=10$. Note the very fine vertical scale at
large $b$. At such high values of $b$ we have $\kappa_c\approx0.125$, while $\kappa_f$ ranges approximately between 0.03
and 0.05, depending on the value of $b$ (cf.\ Fig.~\ref{fig:kappa_f_b}).}
\label{fig:plaq_high_b}
\end{figure}

We conclude this section with the results for very large $b$. The values of the plaquette in scans at $b=5, 10, 50, 200$
(which were already discussed in Sec.~\ref{sec:funnel_b}) are presented in Fig.~\ref{fig:plaq_high_b}. The general shape
of each curve resembles those at $b=1$ (compare Fig.~\ref{fig:plaq_b1}). There are two distinctive features:
\begin{enumerate}
  \item{The jump at $\kappa_c$ falls rapidly with increasing $b$ -- such behaviour is qualitatively consistent with the
  expectations of chiral perturbation theory in the confining scenario \cite{ss98}.}
  \item{The jump in the plaquette at $\kappa_f$ becomes more distinct, signalling that the transition in $\kappa_f$ is
  first order at very large $b$.}
\end{enumerate}

\section{Wilson loops and static-quark potential}
\label{sec:wilson}

While the plaquette is a very useful quantity to compare lattice simulations to one another, ultimately one would like
to be able to use the volume reduction to calculate physical (continuum) quantities. One such quantity that is
accessible in the single-site model is the heavy-quark potential. We calculate it using rectangular Wilson loops wrapped
around the $1^4$-torus, in accordance with the volume reduction prescription:
\begin{equation}
W(L_1,L_2) = \frac1{12}\sum_{\mu\ne\nu} \left\langle \tfrac1N \text{Re}\Tr U_\mu^{L_1}
U_\nu^{L_2} U_\mu^{\dagger L_1}U_\nu^{\dagger L_2}\right\rangle\,,
\end{equation}
where we have averaged over all orientations.

For $N\to\infty$, inside the funnel, the single-site result should be equal to the large-volume value, for all $L_1,
L_2$. We can extract the potential from the large $L_2$ behaviour:
\begin{equation}
W(L_1,L_2)\xrightarrow{L_2\to\infty}c(L_1)e^{-V(L_1)L_2}.
\end{equation}
In the confining regime we expect linear behaviour of the potential at large $L_1$, with non-vanishing string tension
$\sigma$:
\begin{equation}
\frac{dV(L_1)}{dL_1}\xrightarrow{L_1\to\infty}\sigma.
\end{equation} 

As we are always dealing with finite $N$ in the simulations, we need to keep $L\ll N$ to keep the finite-$N$ effects
under control -- the key question is how large $L_i$ one can achieve at given $N$. This is an exploratory study and we
only used unsmeared links, as opposed to the state-of-art large-volume simulations which use various kinds of smearing
as well as other noise reduction techniques. In this way we can identify the finite-$N$ effects in a more transparent
way.

\begin{figure}[tbp!]
\centering
\includegraphics[width=12cm]{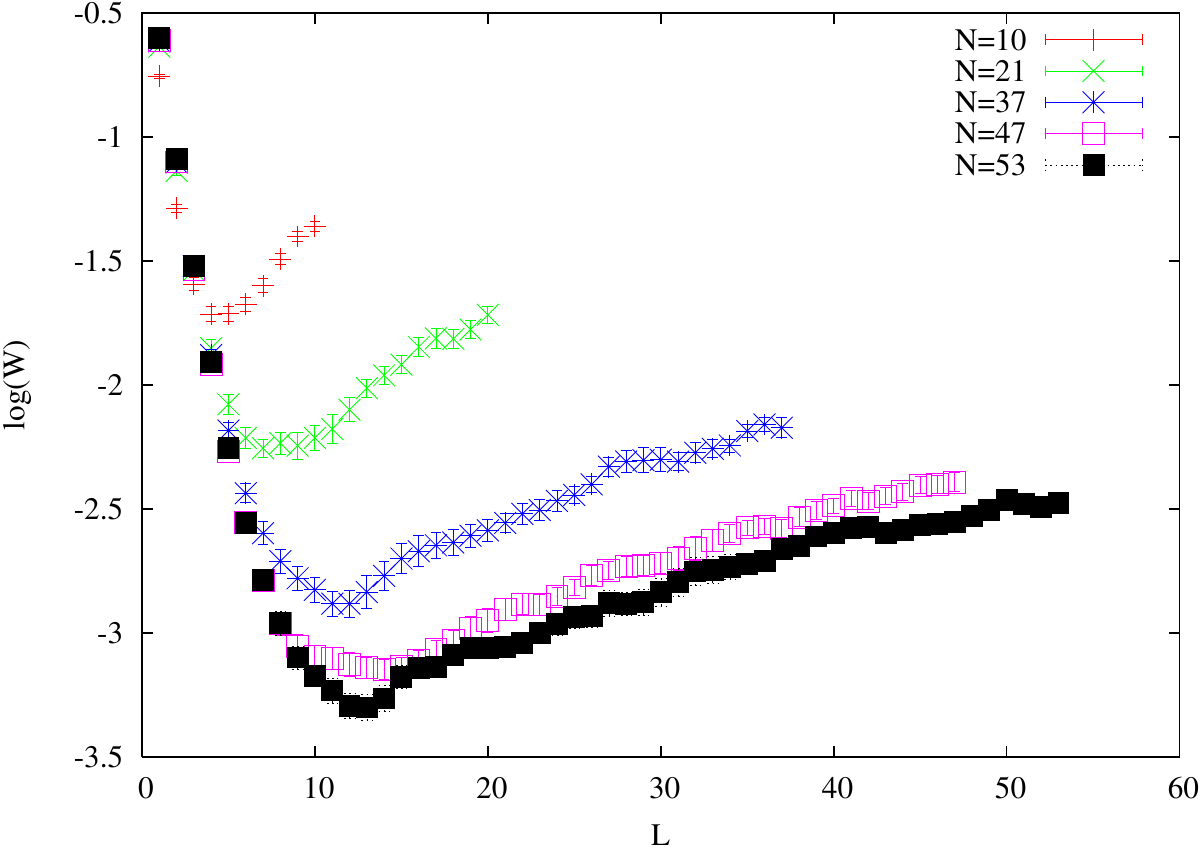}
\caption{Log-linear plot of $1\times L$ Wilson loop versus $L$ for $L\le N$. Results are from $N_f=2$, $b=0.35$,
$\kappa=0.12$ and for $N=10$, $21$, $37$, $47$ and $53$, using $20$ configurations except for $N=10$ where 150
configurations were used.}
\label{fig:wloop_1xL}
\end{figure}

In Fig.~\ref{fig:wloop_1xL} we present the log-linear plot of $1\times L$ Wilson loops for several values of $N$. For
each $N$ we find an approximately exponential decrease followed by a slow, approximately linear, rise. Since the rise is
an unphysical behaviour (in the large-volume language) and it begins at larger $L$ as $N$ increases we interpret it as a
finite-$N$ effect. The exponential drop-off seems to converge to a common envelope (linear on the logarithmic plot).
For example, at $L=6$ the $N=37$ point has already peeled off from the envelope while the $N=47$ and 53 are in good
agreement. We can thus reliably extract the large-$N$ behaviour up to some $L_{max}(N)$ -- the crucial question is how
exactly $L_{max}$ depends on $N$. To estimate this for every $N$ one can look at the minimum value of the loop for given
$L$. Then we find that it falls approximately as $1/N$ thus $L_{max}$ grows only logarithmically with $N$.

This is not unexpected as we are trying to extract an exponentially falling contribution from a quantity that has
finite-$N$ corrections, however this poses a significant numerical challenge. It is also worth stressing that it is not
the statistical errors that are limiting but the size of $N$ that needs to be taken to extract the potential for large
enough $L$.

\begin{figure}[tbp!]
\centering
\includegraphics[width=12cm]{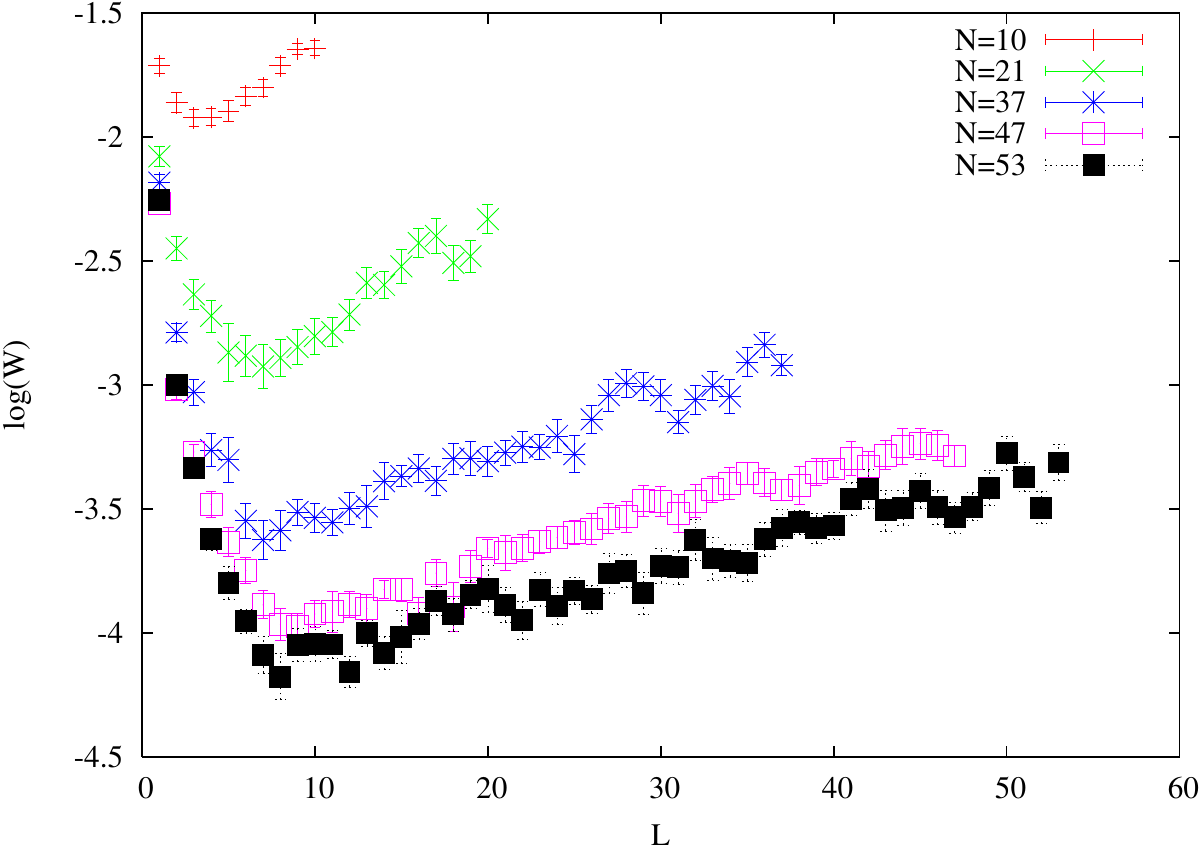}
\caption{As for Fig.~\ref{fig:wloop_1xL}, but for $5\times L$ loops.}
\label{fig:wloop_5xL}
\end{figure}

Having said that, we see from the Fig.~\ref{fig:wloop_1xL} that we can extract the value of the potential $V(1)$ with
reasonably small errors. To find $\sigma$ we also need to extract the potential at larger separations. In
Fig.~\ref{fig:wloop_5xL} we show the results for $5\times L$ loops. The overall pattern is similar to the one in
Fig.~\ref{fig:wloop_1xL} but the convergence to the common exponentially-decaying envelope is much poorer and only
results for $L\leq 2$ appear converged. This is by no means unexpected as the signal for $W(5,L)$ is significantly
smaller than for $W(1,L)$ while the finite-$N$ background is little changed. This type of behaviour only allows us to
calculate the potential up to separations of 2-3 lattice units which is too small to reliably estimate the string
tension. We were also unable to extract the Creutz ratios in a reliable way. Again, note that the problem is not in the
statistics (the 20 configurations are sufficient to measure the exponentially-falling part of the curves) but rather in
the $1/N$ corrections.

We have also carried out similar calculations in the other points inside the funnel. Our results are qualitatively the
same although we find that the slope of the rising part at large values of $L$ generally decreases with growing $b$.
Also, in the large-$\kappa$ part of the funnel (where the plaquette is closer to unity) the slope decreases even further
and becomes almost $L$-independent.

The unphysical rise at large $L$ can be understood, at least qualitatively, in the strong coupling limit of the pure
gauge theory (i.e.\ at $\kappa=0, b=0$). The links are then distributed according to the Haar measure. In the large
volume this results in a zero signal due to the vanishing integrals of type:
\begin{equation}
\int U dU = 0.
\end{equation}
On the single site, however, the zero-th order of the strong coupling expansion does not vanish because for every closed
contour every $U_\mu$ in the integral comes in pair with a corresponding $U_\mu^\dagger$. In fact, one can show (using
Refs.~\cite{die01,br08}) that for $b=0, \kappa=0$:
\begin{equation}
W(L_1,L_2) = \frac1{N^2-1}(L_1+L_2-1-L_1L_2/N^2)\quad \text{for} \quad 0<L_1,L_2\leq N.
\end{equation}
This result shows that the linear rise of the signal with $L$ is not unnatural in volume-reduced models at finite $N$.
On the quantitative level, however, the above model does not fit the data obtained in the simulations in the moderate
and weak coupling region. For example, one would expect that $W_{min}\sim1/N^2$ while we observe an approximate $1/N$
dependence.

\begin{figure}[tbp!]
\begin{adjustwidth}{-1.2cm}{-1.2cm}
\centering
\subbottom[All Wilson loops] {\label{fig:wloops_24_all} \includegraphics[width=7.5cm]{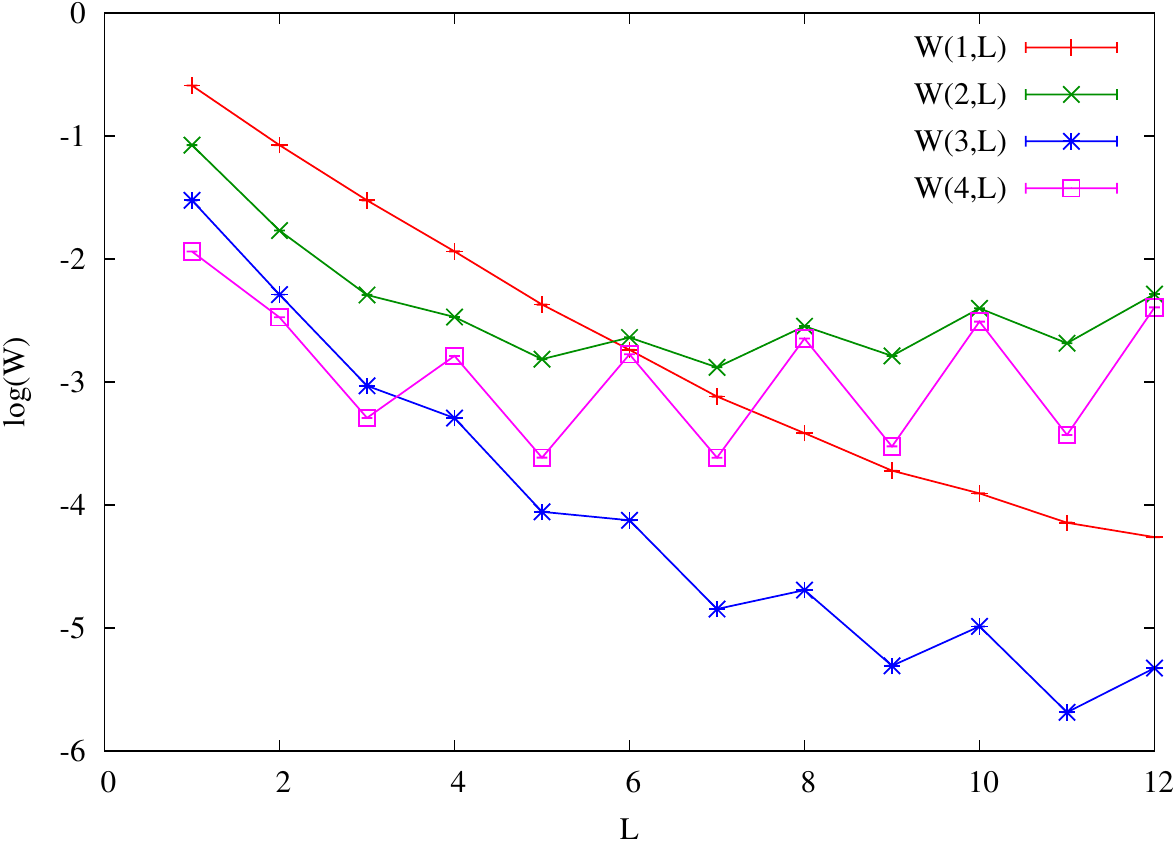}}
\hspace{0.2cm}
\subbottom[Only odd-sized Wilson loops] {\label{fig:wloops_24_odd} \includegraphics[width=7.5cm]{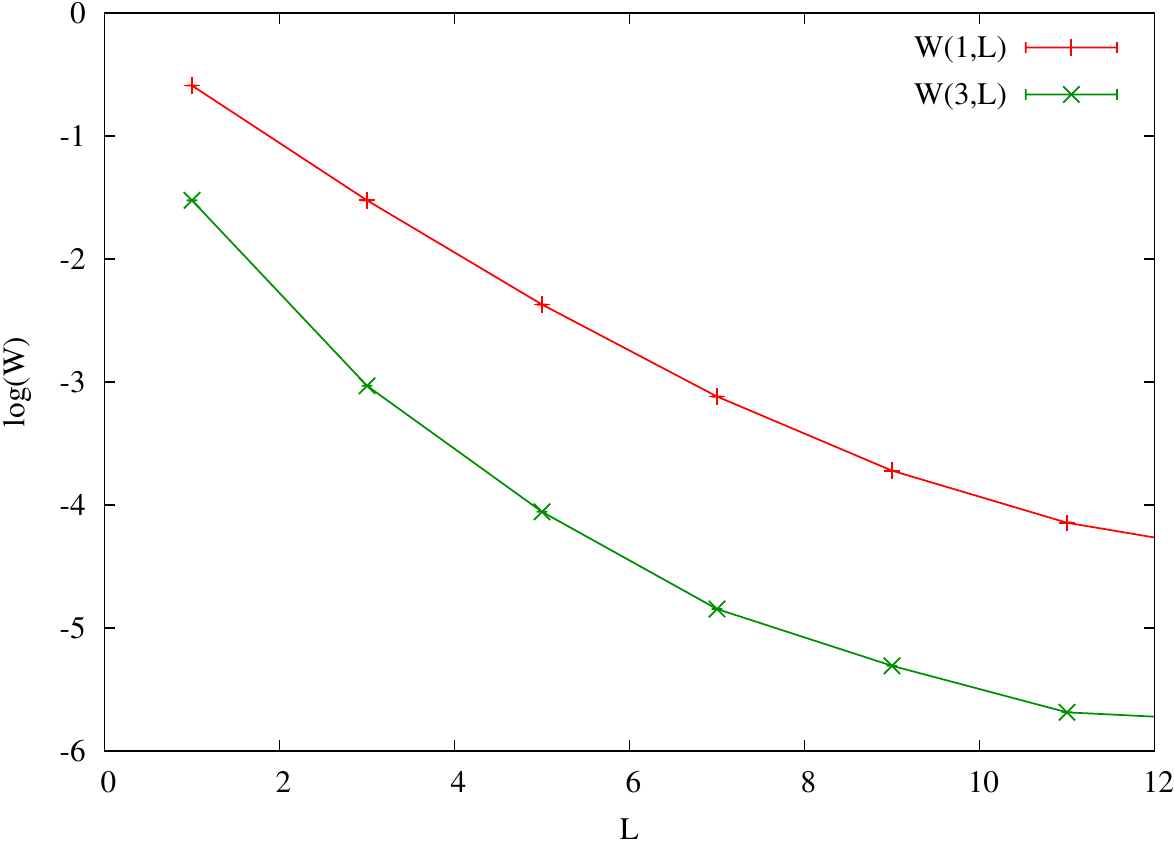}}
\caption{Wilson loops on $2^4$ lattice. $N_f=2$, $N=10$, $b=0.35$, $\kappa=0.1$. The lines between the datapoints are
added to guide the eye.}
\label{fig:wloops_24}
\end{adjustwidth}
\end{figure}

One idea on how to make the corrections smaller is to use the $2^4$ lattice and only measure the Wilson loops with odd
$L_1$ and $L_2$. In this way the loop never contains pairs $U$, $U^\dagger$ of the same link and the zero-th order in
the strong coupling expansion vanishes. That this in fact reduces the finite-$N$ corrections is pictured in
Fig.~\ref{fig:wloops_24}. This further confirms that the finite-$N$ behaviour can be understood, at least on the
qualitative level. However, our values of $N$ to date ($N\geq15$) are too small to allow a reliable extraction of the
string tension from the $2^4$ data.

\section{Spectrum of the Dirac operator}

\subsection{Spectrum of the Wilson Dirac operator $D_W$}

As discussed in Sec.~\ref{sec:l_eff}, one can analyze the effective size of the volume-reduced system at finite-$N$. One
of the observables that are useful to analyze this quantity is the spectrum of the Dirac operator. The orbifold
construction predicts the effective length $L_\text{eff}$ to scale proportionally to $N^{1/4}$  (cf.\
Eq.~\ref{eq:l_eff_orb} and the preceding discussion).

However, as we show further in this section, some features of the Dirac operator spectrum suggest that such scaling may
be perhaps overly pessimistic thus we also compare our data to a less conservative possibility, discussed in
Refs.~\cite{amn99,uns04} in the context of theories with twisted boundary conditions -- in this case every element of
the the volume-reduced link matrices is used in the packaging of the link matrices of the large-volume theory. This
leads to $L_\text{eff}\propto N^{1/2}$. There is also an even more optimistic scenario, motivated in the Appendix of
Ref.~\cite{bks11}, which leads to $L_\text{eff}\propto N$.

\begin{figure}[tbp!]
\centering
\includegraphics[width=13.5cm]{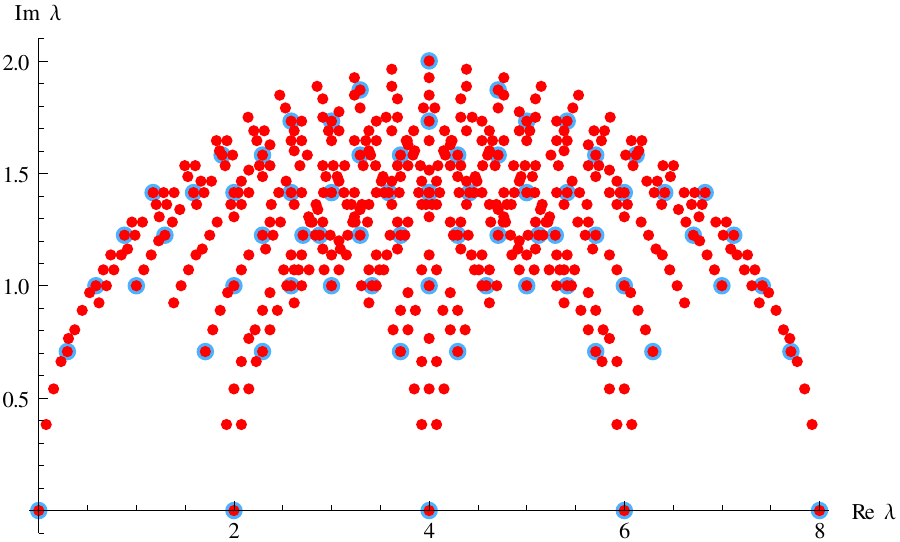}
\caption{Free spectrum of the Wilson Dirac operator at $8^4$ (large blue dots) and $16^4$ (small red dots), $m_0=0$. The
spectrum is symmetric with respect to the real axis, only the non-negative imaginary part is shown.}
\label{fig:freespect}
\end{figure}

We expect that in the center-symmetric funnel the spectrum of the Wilson Dirac operator $D_W$ (cf.\ Eq.~\ref{eq:D_W})
should resemble that of the large-volume four-dimensional lattice gauge theory on $L_\text{eff}^4$. In particular, in
the weak coupling, $b\gtrsim1$, where the spectrum of a free fermion is a valid approximation, the spectrum should have
the characteristic five ``fingers'' that reach down to the real axis \cite{cre07}. The number of fingers is a direct
indicator of the dimensionality of the system (in $d$ dimensions there are $d+1$ fingers). Thus observing a smaller
number of fingers is an indicator of the correlations between the lattice directions (supplementing the discussion in
Sec.~\ref{sec:scans}).
Also, the distance of the fingers to the real axis should scale like $1/L_\text{eff}$. As an example, the spectrum of
the free Wilson Dirac operator on lattices $8^4$ and $16^4$ is presented in Fig.~\ref{fig:freespect}.

We now show some representative results for the spectrum of \linebreak $D_W(m^\text{(valence)}_0=0)$ from our
simulations. Note that the Dirac operator in the determinant of Eq.~\ref{eq:Z_AEK} (which determines the masses of the
sea quarks) is
\begin{equation}
D_W(m_0) = 2\kappa\big(4 D_W(0) + \tfrac1{2\kappa}-4\big) = \frac{1}{4+am_0}\big(4D_W(0) + am_0\big),
\end{equation}
so that the eigenvalues of $4D_W(0)$ close to $\lambda=4-\tfrac1{2\kappa}$ are suppressed. The spectrum is bounded to
$0\leq \text{Re}\lambda\leq8$ so the determinant suppression is important only for $\kappa>1/8$.

Another thing to remember is that on a lattice with even number of sites in each direction the Dirac operator
spectrum is symmetric under reflection about the $\text{Re}\lambda=4$ axis \cite{gh98}. On the single site there is no
such symmetry but if the reduction holds we expect its emergence (approximate at finite-$N$).

\begin{figure}[tbp!] \centering
\begin{adjustwidth}{-1.2cm}{-1.2cm}
\subbottom[$\ \kappa=0.01$]
{\label{fig:dw_N30_b1_k0.01}
\includegraphics[width=7.5cm]{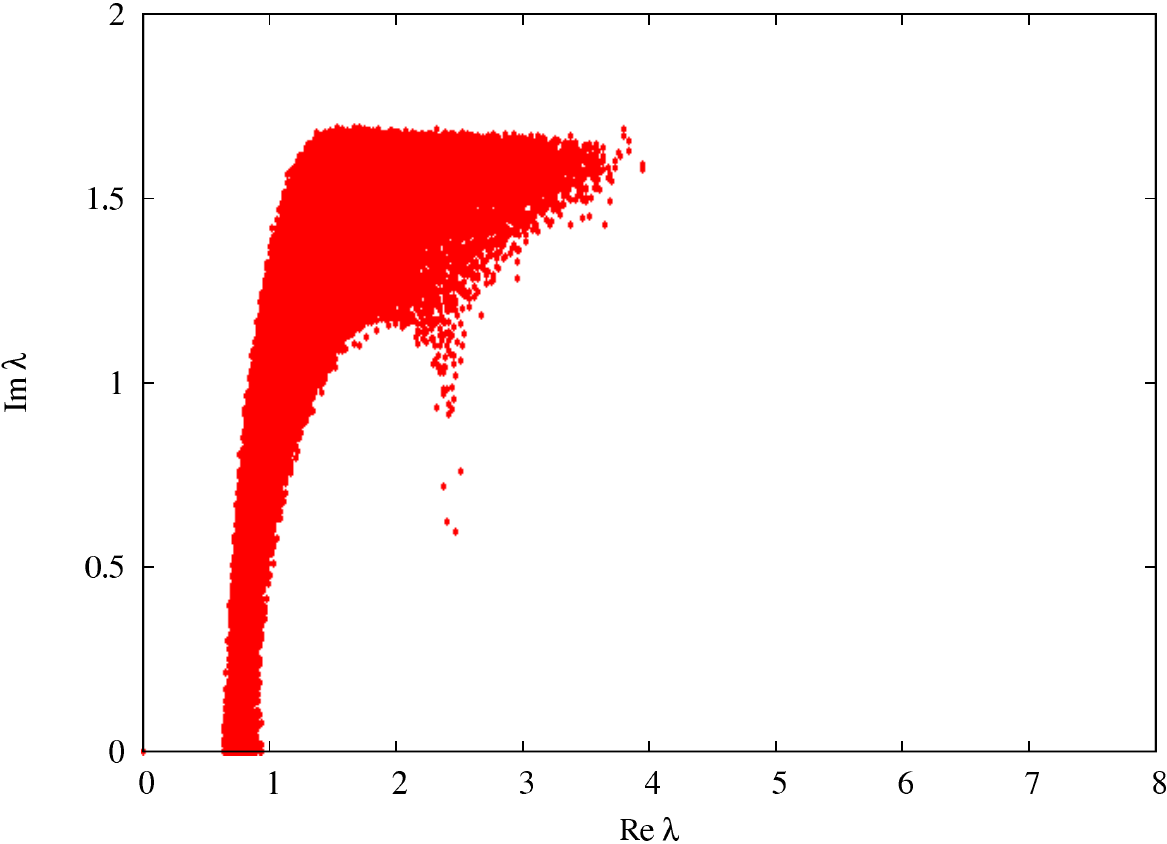}}
\hspace{0.3cm}
\subbottom[$\ \kappa=0.03$]
{\label{fig:dw_N30_b1_k0.03}
\includegraphics[width=7.5cm]{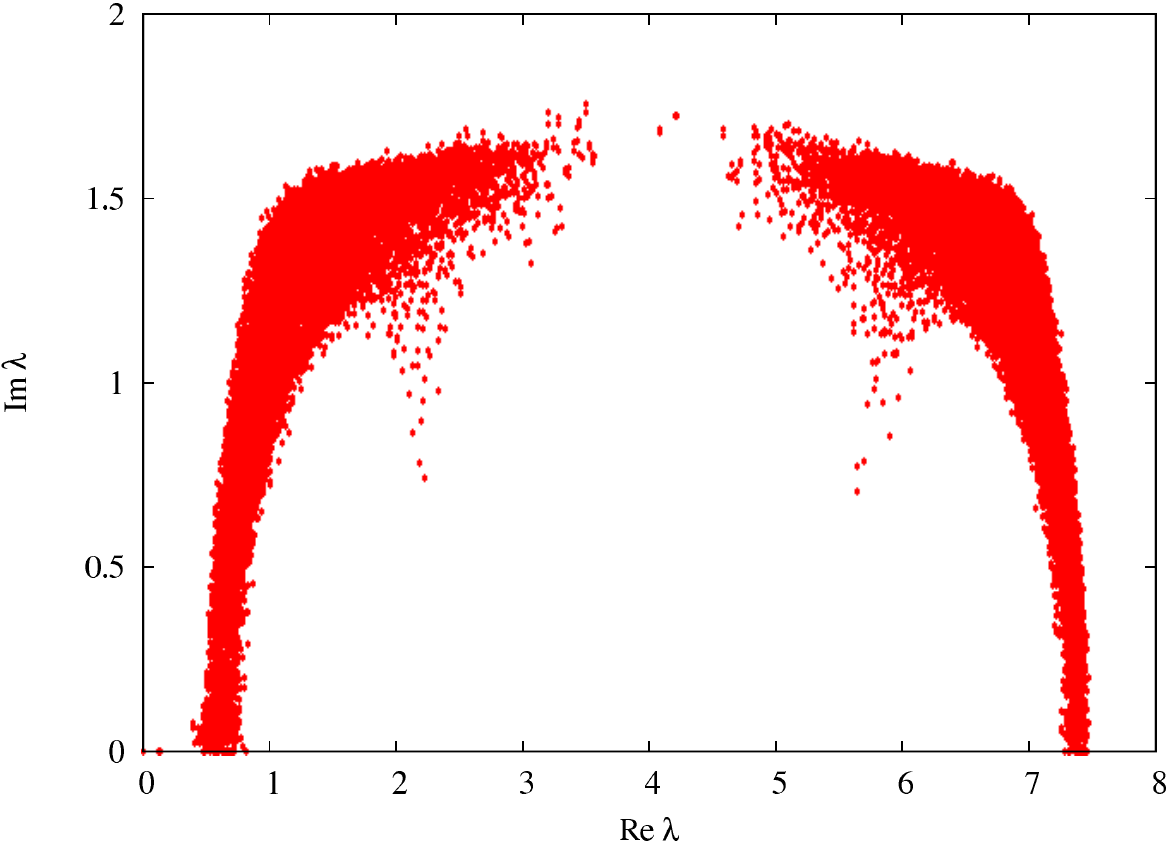}}\\[6pt]
\subbottom[$\ \kappa=0.12$]
{\label{fig:dw_N30_b1_k0.12}
\includegraphics[width=7.5cm]{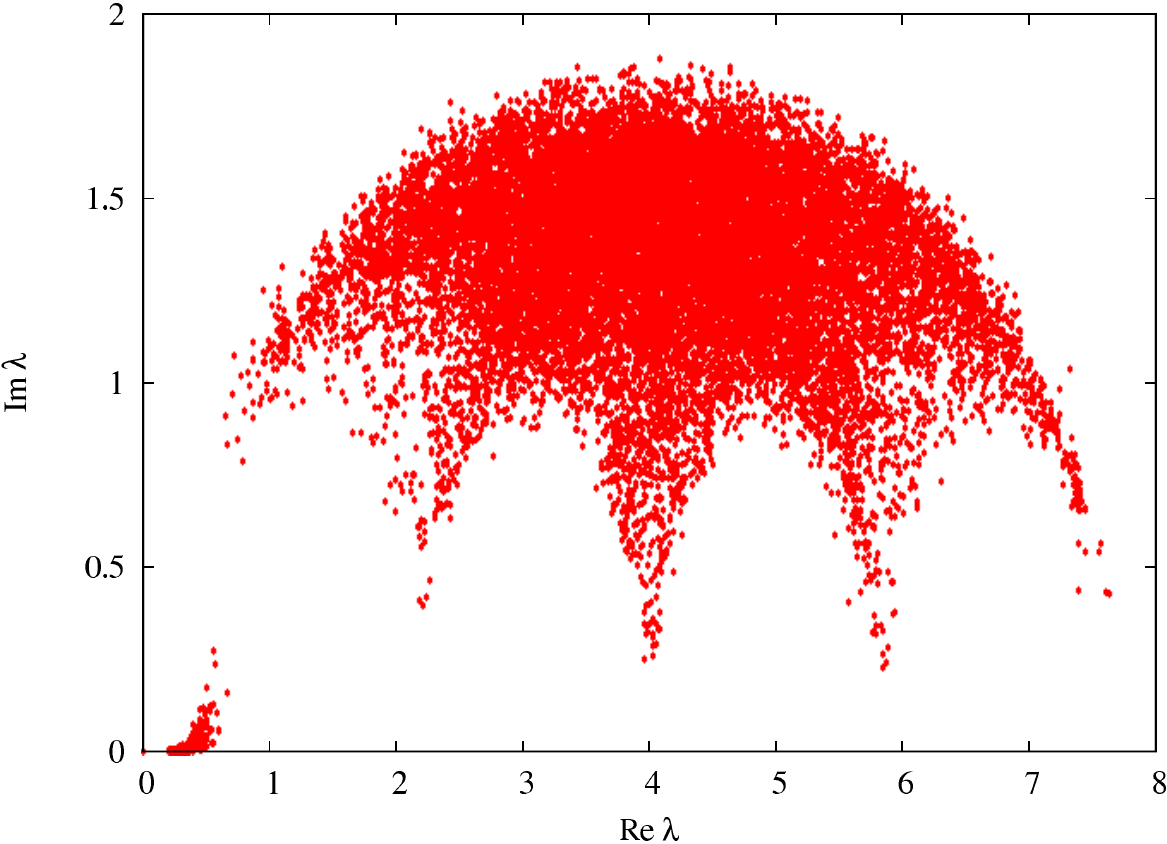}}
\hspace{0.3cm}
\subbottom[$\ \kappa=0.14$]
{\label{fig:dw_N30_b1_k0.14}
\includegraphics[width=7.5cm]{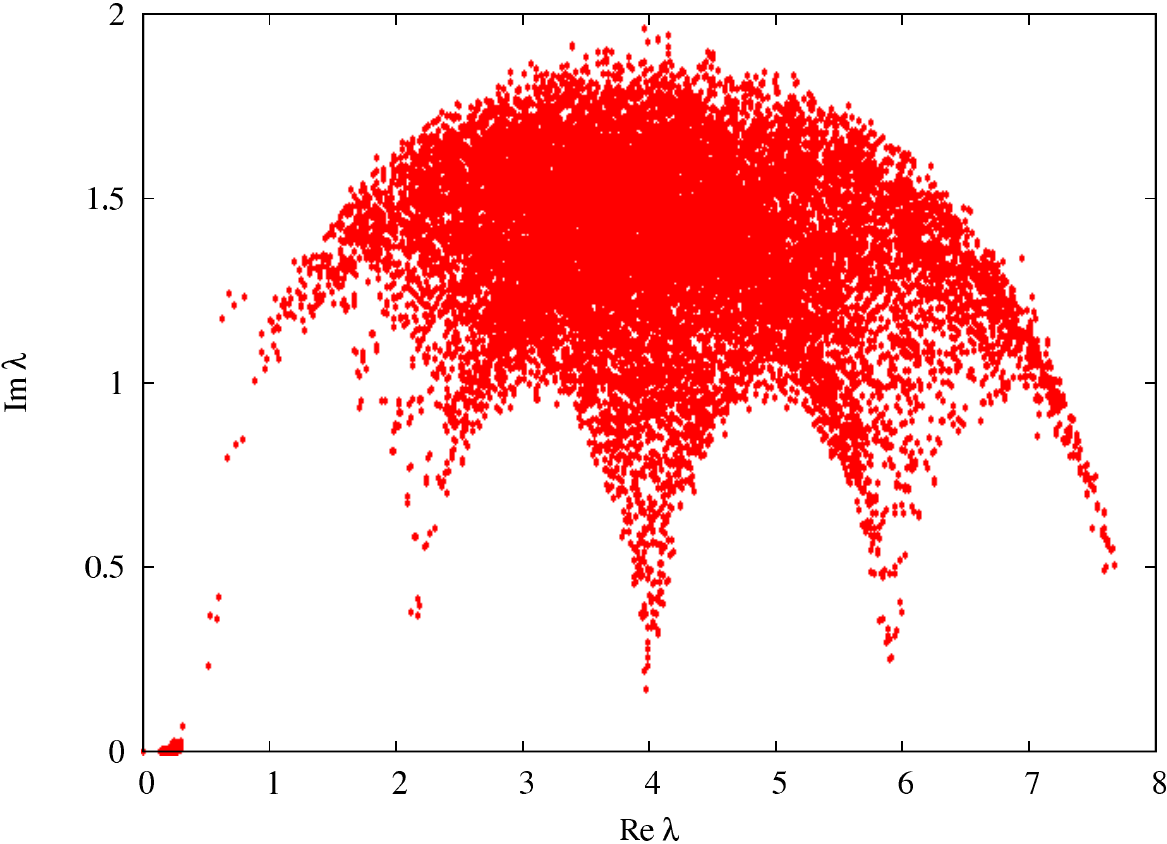}}\\[6pt]
\subbottom[$\ \kappa=0.17$]
{\label{fig:dw_N30_b1_k0.17}
\includegraphics[width=7.5cm]{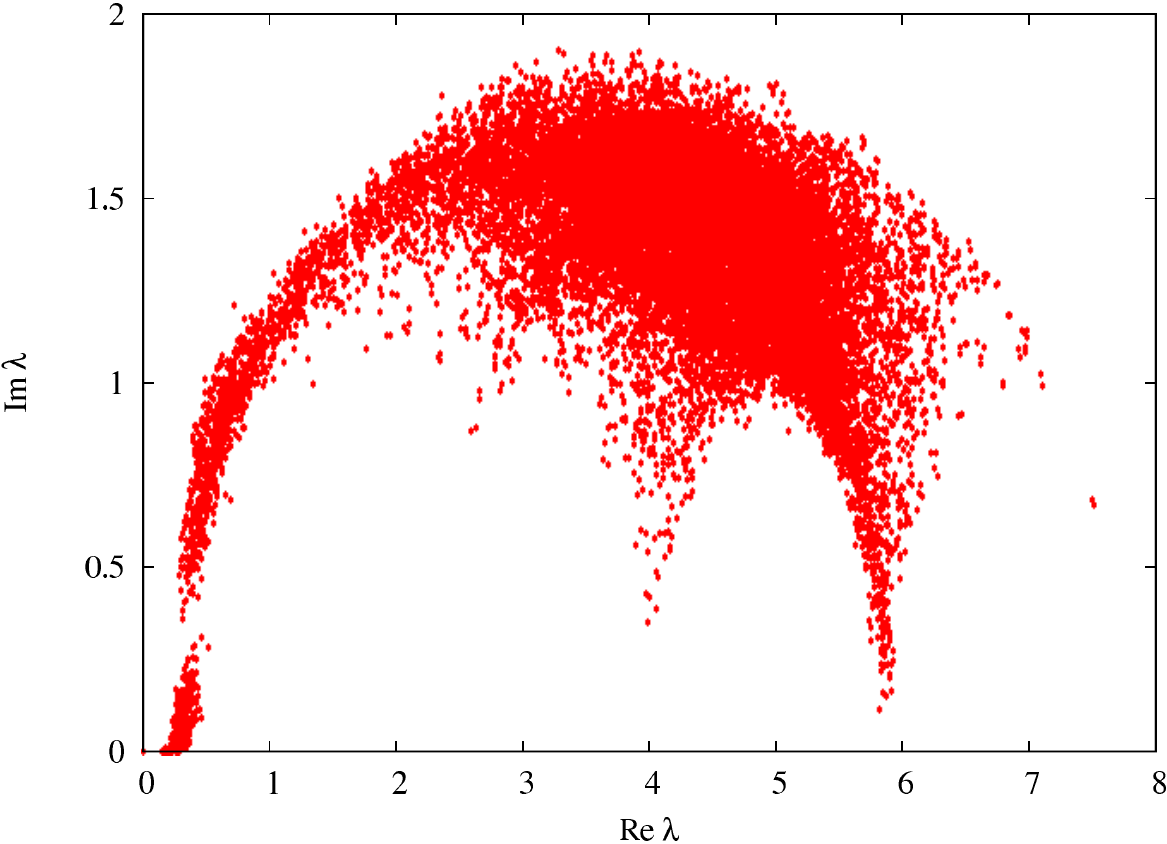}}
\hspace{0.3cm}
\subbottom[$\ \kappa=0.24$]
{\label{fig:dw_N30_b1_k0.24}
\includegraphics[width=7.5cm]{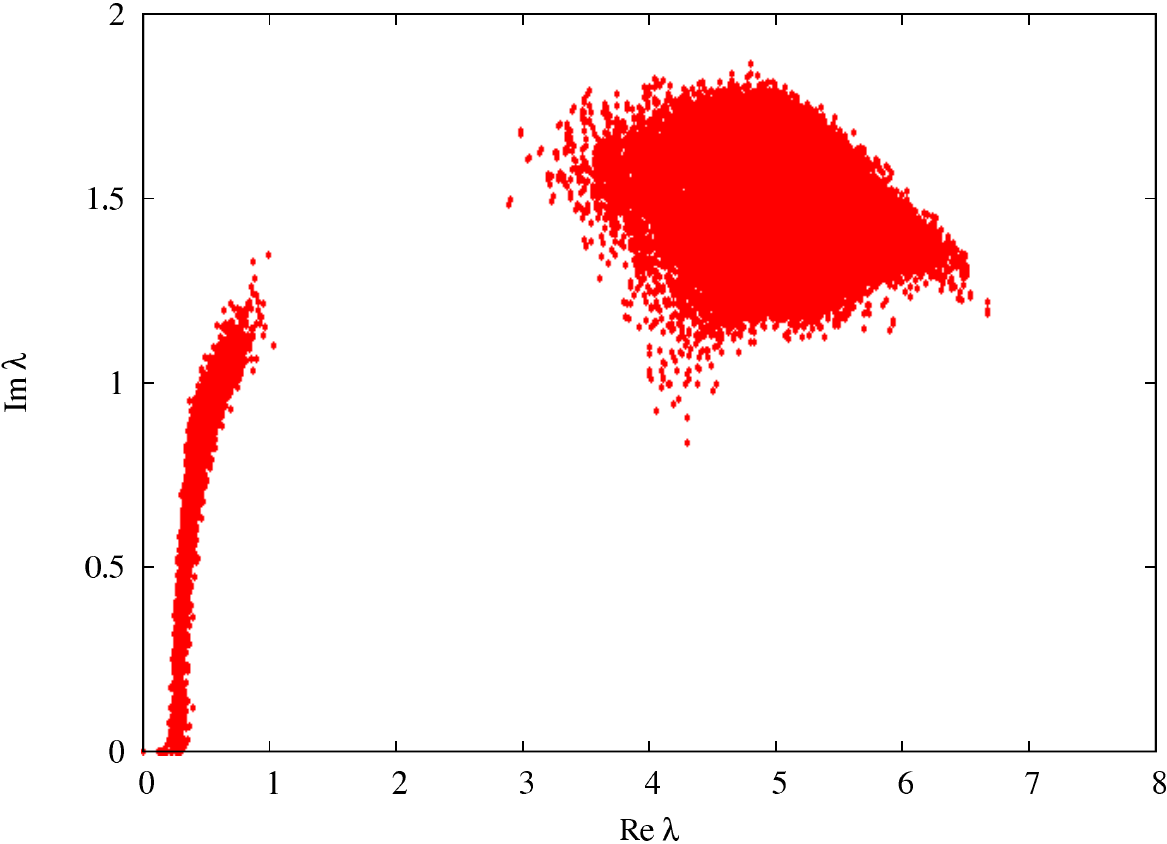}}
\caption{Spectrum of $4D_W(0)$ from simulations at $N_f=2$, $N=30$, $b=1.0$ and six representative values of $\kappa$
(i.e.\ masses of the sea quarks). The plots are made using 20 configurations each, only the eigenvalues with
non-negative imaginary part are shown.}
\label{fig:dirac_scan}
\end{adjustwidth}
\end{figure}

In Fig.~\ref{fig:dirac_scan} we show the spectrum for six representative values of $\kappa$ in the scan at $N_f=2$,
$N=30$, $b=1$. At $\kappa=0.01$, Fig.~\ref{fig:dw_N30_b1_k0.01}, which is in the $\zz_1$ phase (compare
Figs.~\ref{fig:phase_diag} and \ref{fig:corner_scan_nf2}) we see a well-formed first finger and a small indication of
the second one. This is consistent with the link eigenvalues forming a single clump, so that the ``momenta'', given by
eigenvalue differences are small (the Dirac operator is constructed out of the adjoint links and thus it gives the
information about the differences of eigenvalues of the fundamental links). At $\kappa=0.03$,
Fig.~\ref{fig:dw_N30_b1_k0.03}, we are in the $\zz_2$ phase. The two clumps of link eigenvalues allow the eigenvalue
differences to reach $\pi$ and to form the fifth finger, however the second and fourth finger are only slightly
indicated and the middle finger is absent. Also the ``rectangular'' shaped envelope shows that the spectrum is very
different from the large-volume one.

The next value, $\kappa=0.12$ pictured in Fig.~\ref{fig:dw_N30_b1_k0.12}, is well inside the center-symmetric funnel.
The distribution of the eigenvalues of $D_W$ is qualitatively very similar to that of a free fermion on a large lattice
with a rounded top and five fingers -- these features are present for all $\kappa<\kappa_c$ inside the funnel. A
particularly interesting feature is the presence of the comet-shaped clump of eigenvalues near the origin. We find that
there are exactly $4(N-1)$ eigenvalues per configuration in this clump. We thus interpret them as the would-be zero
modes i.e.\ the eigenvalues that would be zero in the $b\to\infty$ limit. In the weak-coupling analyses these modes are
neglected \cite{br091} because they do not impact the dynamics (i.e.\ they do not depend on $\vartheta^a_\mu$) and
because they only form a $\mathcal{O}(1/N)$ fraction of the total number of modes.

In our calculations, however, they form the lowest non-zero eigenvalues of the Dirac operator and thus can have a
significant impact on the finite-$N$ dynamics, despite their relative paucity. Recall that the smallest eigenvalues of
the large-volume Dirac operator determine the long-range behaviour of the theory, such as chiral symmetry breaking. For
very large values of $N$ we expect the smallest eigenvalues to come dominantly from the first finger, which should
approach close to the real axis -- we see that $N=30$ is rather far from this description and we conclude that the
would-be zero modes are a potential source of $\mathcal{O}(1/N)$ corrections whose contribution can be sizeable (given
how far the ``true'' low-energy modes in the first finger are from the real axis).

For $\kappa=0.14$, which is inside the funnel and above $\kappa_c$, the spectrum, shown in
Fig.~\ref{fig:dw_N30_b1_k0.14}, is very similar to the one at $\kappa=0.12$. The would-be zero modes look somewhat
different -- almost all of them are squeezed to the left of $\text{Re}\lambda=0.43$ which is the point suppressed by the
determinant. As a whole though, the picture looks similar on both sides of the funnel. Moving on further to
$\kappa=0.17$, Fig.~\ref{fig:dw_N30_b1_k0.17}, which is also expected to be inside the funnel, we see that the second
and fifth finger have almost disappeared while the first finger has become more distinct and reaches down closer to the
real axis. This may raise some concerns whether this point is in fact in the funnel where volume reduction holds (an
analysis of the width of the r.h.s.\ of the funnel, similar to the one in Sec.~\ref{sec:funnel_n}, would help to clarify
this issue).

The last plot, Fig.~\ref{fig:dw_N30_b1_k0.24}, is taken at $\kappa=0.24$, in the $\zz_3$ phase. This is reflected by the
spectrum being divided into three distinct regions (the one with the negative imaginary part is not shown in the plot)
resulting from link eigenvalue differences distributed around 0 and $\pm2\pi/3$.

\begin{figure}[tbp!]
\centering
\begin{adjustwidth}{-1.2cm}{-1.2cm}
\subbottom[$\ N=37$, 300 configs] {\label{fig:dw_N37}
\includegraphics[width=7.5cm]{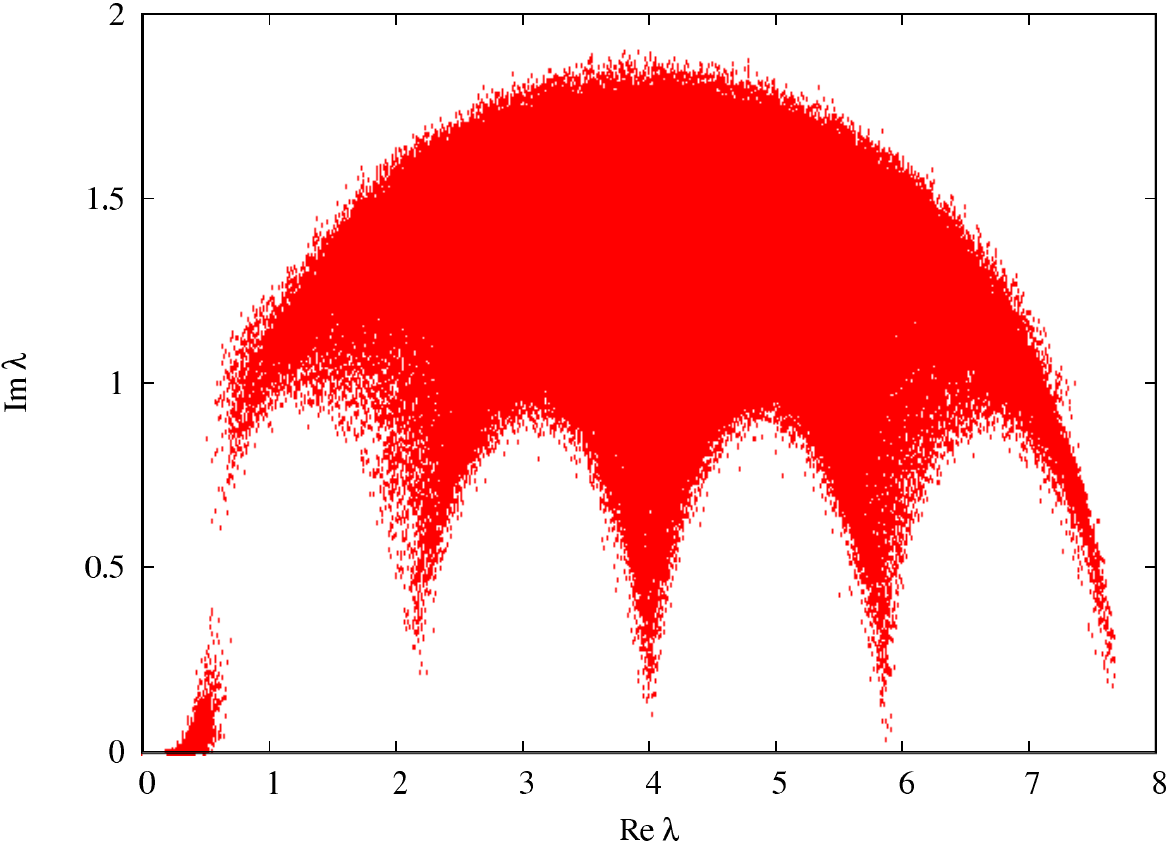}}
\hspace{0.2cm}
\subbottom[$\ N=53$, 150 configs] {\label{fig:dw_N53}
\includegraphics[width=7.5cm]{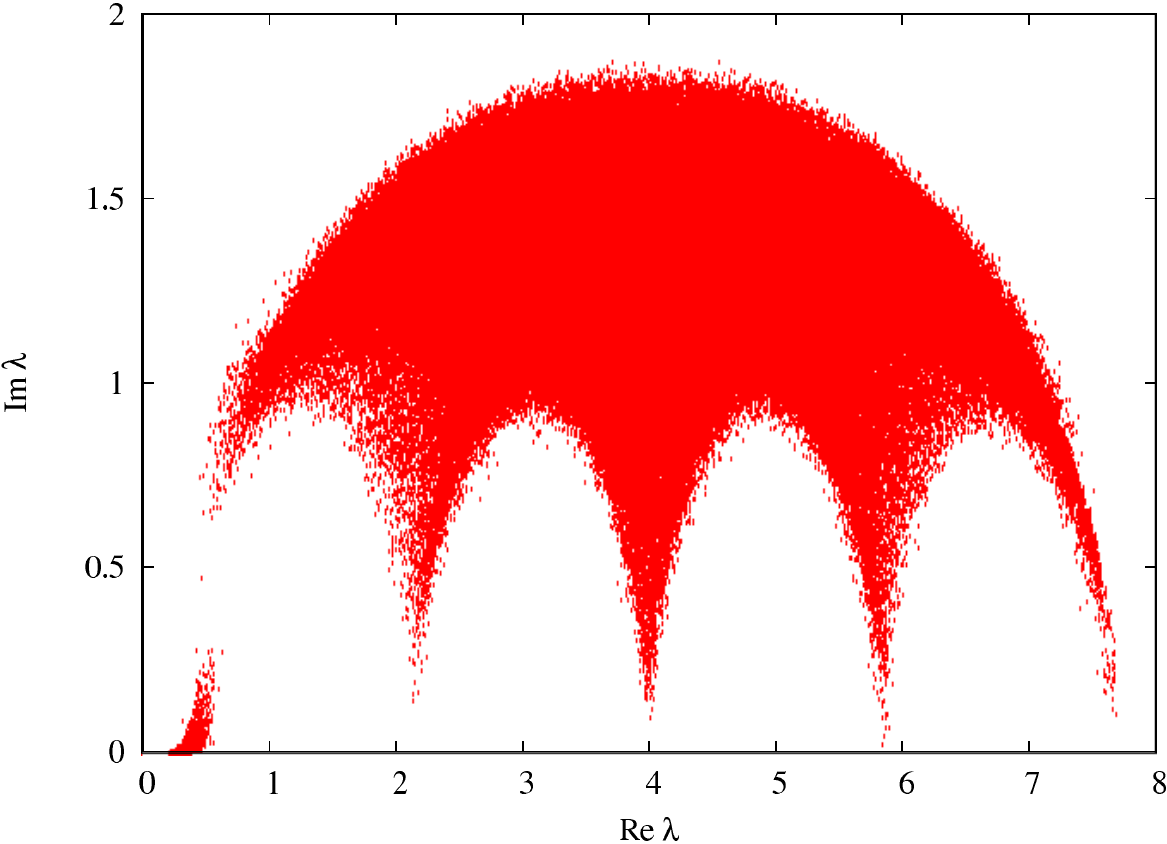}}
\caption{Spectrum of $4D_W(0)$ at $N_f=2$, $b=1.0$ and $\kappa=0.12$, for $N=37$ and $53$. Note that since
$(53/37)^2\approx2$ the number of points is approximately the same in both plots.}
\label{fig:dirac_n}
\end{adjustwidth}
\end{figure}

We have also studied the $N$-dependence of the spectrum at $b=1$ and $\kappa=0.12$. Fig.~\ref{fig:dirac_n} compares the
results for $N=37$ and $N=53$. The number of eigenvalues is approximately the same in both panels (and a lot bigger than
in Fig.~\ref{fig:dirac_scan}). There is not much difference in the spectra -- the clump of the would-be zero modes
decreases with growing $N$ and the first and fifth finger move a little downwards. The tips of the other fingers,
however, barely move.

The conclusions from these results can be summarized as follows:
\begin{enumerate}
  \item{There is a qualitative agreement of the spectrum of $D_W$ inside the funnel with that of a theory on a four-dimensional
  large-volume lattice that supports our claim that the reduction holds therein.}
  \item{The dependence of the distance of the fingertips to the real axis is inconsistent with $L_\text{eff}\propto N$. On
  our level of precision it allows both $N^{1/2}$ and $N^{1/4}$ although the presence of well-formed fingers disfavors the
  latter possibility. At $N\leq53$ we would not expect the presence of the fingers in that case, unless some effects that we
  have not taken into account make the spectrum look ``surprisingly'' good\footnote{The $N=48$ would correspond to
  $L_\text{eff}=2$ at $N_\text{eff}=3$ which shows no signs of fingers.}.}
  \item{The would-be zero modes are a potential source of $\mathcal{O}(1/N)$ corrections and can significantly influence
  the finite-$N$ dynamics, provided that these $1/N$ corrections are not exactly cancelled by some mechanism among the
  $4(N^2-N)$ ``bulk'' eigenvalues.}
\end{enumerate}

We have also calculated the spectrum of the Dirac operator in the fundamental representation. This gives the information
about the link eigenvalues themselves, rather than their differences. The obtained results confirm the results presented
above and are not presented for the sake of brevity.

\subsection{Spectrum of the hermitian Dirac operator}

The scans similar to the ones in the previous section were also done for $b$ lower than 1. In that case however, the
picture is harder to compare with the known-results of the large-volume free fermion case. In particular, at $b=0.35$
the spectrum fills the whole allowed region and there is no sign of fingers. A better approach in this case is to
analyze the spectrum of the hermitian Dirac operator $Q$ (cf.\ Eq.~\ref{eq:Q}) or, equivalently, $Q^2 = D_W(m_0)
D_W(m_0)^\dagger$.

Analyzing the eigenvalues of $Q^2$ can teach us many interesting things about the theory. In the continuum limit the
spectrum has a gap
\begin{equation}
\lambda^\text{min}_{Q^2} = (am_\text{phys})^2,
\end{equation}
where $m_\text{phys}$ is the physical quark mass. Away from the continuum the gap is smoothed but the spectrum still
begins approximately at the square of the physical mass \cite{glu08}. Also, if the theory shows the spontaneous breaking
of the chiral symmetry, for small enough quark masses the density of the spectrum above the gap is approximately
constant and equal to the condensate.

\begin{figure}[tbp!]
\centering
\includegraphics[width=12cm]{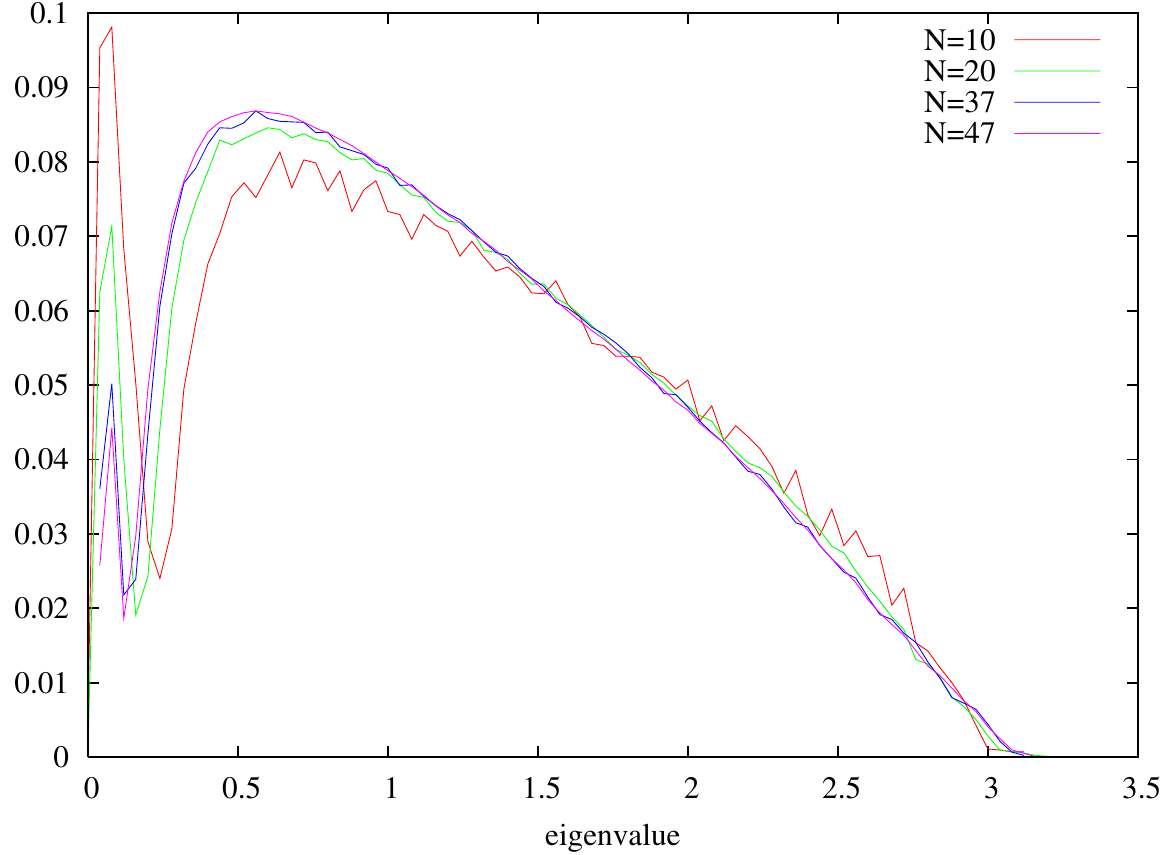}
\caption{Density of eigenvalues of $Q^2$ for $N_f=2$ at $b=0.35$ and $\kappa=0.12$, for $N=10$, $20$, $37$ and $47$,
using 150, 150, 20 and 60 configurations, respectively. The vertical scale is arbitrary, but the relative scales for
different $N$ are chosen so that the area under each spectrum is the same. Errors are not shown, but can be estimated
from the kinks in the spectra.}
\label{fig:Q2spect}
\end{figure}

In Fig.~\ref{fig:Q2spect} we present the results for the density of eigenvalues of $Q^2$ for $N_f=2$, $b=0.35$ and
$\kappa=0.12$ for several values of $N\leq47$ (the results for $N=53$ are very similar to those of $N=47$ and were
dropped due to lower statistics). The plots are normalized to have the same integral so that we can study the
$N\to\infty$ limit. The peak at small eigenvalues around 0.01 has exactly the area to contain the $4(N-1)$ would-be zero
modes. Its area drops as $1/N$ and we expect it to disappear completely in the large-$N$ limit.

The $N$-dependence of the bulk part of the spectrum seems to be divided into two parts -- for the eigenvalues above
$\lambda_{Q^2}\approx1.5$ the density is approximately $N$-independent while the density of the smaller eigenvalues has
substantial $N$-dependence for smaller values of $N$. The values for $N=37$ and $N=47$ (and unplotted $N=53$) seem to be
pretty consistent and there is not much area of the would-be zero modes to be ``redistributed'' to the bulk part. Thus
we conclude that the form of the bulk density at $N=47$ should be a good approximation to the large-$N$ density.

These results give us another estimate of how large $N$ should one use in simulations to obtain the results that are
close enough to the large-$N$ limit. On one hand the bulk part of the spectrum is little changed beyond $N\approx40$. On
the other hands the $\mathcal{O}(1/N)$ contribution of the would-be zero modes is dominant in the physically-important
low mode region (and will most likely remain so also for much larger values of $N$). It is not easy to determine a
priori how important are the corrections due to the would-be zero modes -- one needs to calculate physical observables,
such as meson masses, and study their $N$-dependence.

We can now also give a very crude estimate of the physical mass by neglecting the would-be zero modes and linearly
extrapolating the shape of the bulk density towards zero density. This analysis suggests that the gap is
$\lambda^\text{min}_{Q^2}\approx0.1$ which corresponds to the quark mass $am_\text{phys}\approx0.3$ modulo an unknown
multiplicative renormalization factor (which we however expect to be $\mathcal{O}(1)$ at $b=0.35$ \cite{bs09}). This
rough estimate shows that the quarks are relatively heavy at the given parameters, not much below $m_\text{phys}=1/a$.

\chapter{Summary and outlook}
\label{ch:sum}

In this work we have presented the analysis of large-$N$ volume-reduced gauge theories using the lattice regularization.
The volume reduction allows one to non-perturbatively analyze theories that are of great importance. With heavy fermions
(both for $N_f=1$ and $N_f=2$) one can compare the results to the large-volume pure gauge theory in the large-$N$ limit.
On the other hand, with light fermions one can analyze a theory ($N_f=1$ AEK) that is within $1/N$ of the physical QCD
with two light quarks or a theory that is close to the conformal window and is very important for the walking
technicolor models ($N_f=2$ AEK)\footnote{Also the supersymmetric $N_f=1/2$ case, that was not analyzed in this work, is
a theory of great interest for the theorists.}.

We have reviewed the concept of volume reduction showing how it arises from the comparison of large-$N$ Dyson-Schwinger
equations on the infinitely large and single-site lattice \cite{ek82}. We have discussed the lack of spontaneous
breaking of the center symmetry in the volume-reduced model as the key condition for the volume reduction to hold -- and
how it fails in the pure-gauge model originally discussed by Eguchi and Kawai in Ref.~\cite{ek82}. We have reviewed the
ideas to cure the model and we have focused on the concept of adding fermions in the adjoint representation of the gauge
group. We have then presented a pedagogical review of the large-$N$ orbifold equivalences that constitute a modern
theoretical background for the volume reduction with adjoint fermions.

Next, we have introduced the volume-reduced system to be analyzed in this paper -- the single-site $SU(N)$ lattice gauge
theory with $N_f$ adjoint Wilson Dirac fermions -- called the Adjoint Eguchi-Kawai model. We have discussed the
theoretical importance of this model for comparisons with standard large-volume calculations and predictions. We have
presented the perturbative predictions of the behaviour of the model as well as the semi-quantitative arguments going
beyond the perturbation theory presented in Ref.~\cite{ahu10}. The perturbative analysis predicts that the center
symmetry in the large-$N$ limit is intact only for the massless fermions. On the other hand, Ref.~\cite{ahu10} argues
that in the massive case the perturbative attraction of eigenvalues of the Polyakov loop may be overwhelmed by the
non-perturbative fluctuations and result in a center-symmetric ground state, perhaps even with heavy fermions of $m\sim
1/a$. In such case we expect that the observables will be influenced mostly by the gluon dynamics and that one can
compare AEK with heavy fermions directly to the pure-gauge large-volume large-$N$ results. That would be a prescription
to obtain a working Eguchi-Kawai reduction.

After introducing the subject we went on to the numerical analysis of the AEK model with the Monte Carlo method. We have
presented the numerical setup used in this work -- the Hybrid Monte Carlo algorithm for the $N_f=2$ case as well as the
Rational Hybrid Monte Carlo for the $N_f=1$ case. We have discussed the $N$-scaling and the performance of the
algorithm.

After that we have analyzed the phase diagram of the AEK model. To establish the realization of the center symmetry we
performed, using $N\leq60$ and a robust set of observables, scans in the $\kappa-b$ plane going to $b$ as high as 200.
We have shown that the model has a rich phase structure that is in agreement with the predictions of Ref.~\cite{ahu10}.
The center symmetry is fully broken for very heavy fermions -- as one decreases the fermion mass the theory undergoes a
series of phase transitions with partial breakings of the center symmetry. Finally, at small enough mass it enters the
``funnel'' of the unbroken center symmetry. The funnel exists on both sides of the $\kappa_c$ (the critical value of
$\kappa$ where the physical quark mass disappears). However, the pattern of the partial breakings is different on the
two sides of $\kappa_c$. On the l.h.s.\ (this is the region typically used in the large-volume analyses) we see partial
breakings up to $\zz_3$ phase while on the r.h.s.\ -- up to $\zz_5$. We believe that at larger values of $N$ it is
possible to obtain groups with even higher partial breakings. Also, in accordance with Ref.~\cite{ahu10} it is possible
to see higher partial breakings when one goes to larger $b$ -- e.g.\ for $N=30$ we see at most $\zz_2$ at $b=1$ while
for $b\geq10$ we observe also a $\zz_3$ phase. We have also found that whenever the center symmetry is broken one
observes strong correlations between different lattice directions.

We have found that $N_f=1$ and $N_f=2$ cases have a similar behaviour and thus we have mostly focused on presenting
results for the $N_f=2$ case (described in Ref.~\cite{bks11}). The main difference is the width of the funnel -- the
funnel is expected to be narrower in the $N_f=1$ case (i.e.\ going to lower fermion mass is necessary to obtain the
center-symmetric phase). We have found that this is in fact observed in the simulations.

We have also observed that for both numbers of flavours the funnel becomes narrower as $N$ grows. It is thus a matter of
crucial importance to establish whether it remains finite in the large-$N$ limit. This is a significant numerical
challenge and we have limited ourselves to prove it for a single value of coupling, $b=1$, on the l.h.s.\ of $\kappa_c$.
We have found that both for $N_f=2$ and $N_f=1$ the large-$N$ extrapolation of quark mass $m_f$ (the largest mass
sufficient to restore the center symmetry) is rather large ($m_f\sim1/a$), allowing to treat the dynamics at the verge
of the funnel as essentially pure-gauge. Also, in both cases the fits to the closed-funnel hypothesis ($m_f=0$) have
very large values of $\chi^2/\text{d.o.f.}$ (23 for $N_f=2$ and 6.3 for $N_f=1$). This allows us to conclude that (at
$b=1$) the funnel does not vanish in the large-$N$ limit and that massive fermions stabilize the center symmetry. We
have also found that $m_f$ has a $b$-dependence that is consistent with the predictions of Ref.~\cite{ahu10}.

Finally, we have used the volume reduction in order to analyze physical observables. First we have analyzed the average
plaquette. We have performed the large-$N$ extrapolations in selected points of the $\kappa-b$ plane and shown that for
heavy quarks the large-$N$ value is consistent with large-volume pure-gauge calculations while for light quarks it
begins to differ. We have also presented data suggesting that for both $N_f=1$ and $N_f=2$ there is a first order
transition at $\kappa_c$ for $b\lesssim1$. While this result is expected for $N_f=1$, it is rather surprising for
$N_f=2$ because it supports the confining scenario in contrast to large-volume calculations at $N=2$ which show that the
theory is conformal. A discrepancy of behaviour between $N=2$ and $N\to\infty$ is unexpected in the literature and the
situation would definitely benefit from further study focusing particularly on the light quark region.

We have also analyzed the heavy-quark potential using large Wilson loops. We found that the accuracy that one can obtain
is limited by the finite-$N$ effects -- using $N\leq53$ we were only able to obtain $V(L)$ for $L\lesssim3$ and so we
were unable to reliably extract the string tension. We have provided a qualitative description of the $1/N$ corrections
using the strong coupling expansion and proposed a way to reduce them significantly using odd-sized loops on a lattice
of size $2^4$.

The last analyzed observable was the spectrum of the Wilson Dirac operator. By using the spectrum of $D_W$ we have
investigated the measure of ``effective lattice size'' $L_\text{eff}$ at finite values of $N$. We have shown that the
spectrum favours the $L_\text{eff}\propto N^{1/2}$ although we cannot exclude the $L_\text{eff}\propto N^{1/4}$
hypothesis. We have also argued how the would-be zero modes, which are a $\mathcal{O}(1/N)$ effect typically excluded in
the perturbative analyses, can have a significant impact on the dynamics of the theory for finite $N$ -- especially when
discussing the properties of the theory described by the low eigenvalues of the Dirac operator. In fact, they could be
one possible explanation of the discrepancy with $N=2$ simulations in the conformal/confining scenario that we observe
at $N_f=2$. We have also analyzed the square of the hermitian Wilson Dirac operator $Q^2$, argued how one can
distinguish the would-be zero modes from the bulk part of the spectrum and given an estimate of the physical quark mass
using the bulk part of the $Q^2$ spectrum.
\vspace{8pt}
\fancybreak*{{* * *}\\[4pt]}

There are several ways in which one can extend the results presented in this work. A first and foremost is the need to
go to larger systems -- that can be realized by using larger values of $N$ and/or using lattices with more than one
site. The latter possibility has the advantage of being easier to parallelize -- there is a lot of room for improvement
in this respect as almost all simulations performed in this work were executed on a single CPU core.

We also find finite-$N$ corrections that are of order $\mathcal{O}(1/N)$ and that are rather large (see e.g.\
Fig.~\ref{fig:plaq_scans}). $\mathcal{O}(1/N)$ correction are to be expected from the perturbation theory in the
volume-reduced case (as opposed to $\mathcal{O}(1/N^2)$ that are expected in the large-volume models). They can also
arise from the would-be zero modes that we observe in the spectrum of the Dirac operator.

All these problems can be healed by using twisted boundary conditions -- there the system has $\mathcal{O}(1/N^2)$
corrections. Also as shown in Refs.~\cite{ahu10,go122} the finite-$N$ corrections to the plaquette are very small
compared to the untwisted case. Another great advantage of the twisted boundary conditions is the effective size
$L_\text{eff}(N)\propto N^{1/2}$ \cite{amn99,uns04}.

To see how crucial it is to obtain this scaling rather than the more pessimistic one let us compare the CPU-time scaling
of the volume-reduced model with the large-volume large-$N$ simulations. For the large-volume simulations we have
\begin{equation}
t_\text{CPU}(L^4)\propto L^5 N^3,
\end{equation}
where $L^5$ is the standard volume scaling of the HMC algorithm \cite{cre88,gks88} and $N^3$ is the cost of the matrix
multiplication. On the single site and for light quarks we have:
\begin{equation}
t_\text{CPU}(1^4)\propto N^{4.5} = \left\{
  \begin{array}{l l}
   L_\text{eff}^5 N^2 & \text{if }\ L_\text{eff}\propto N^{1/2}\\
   L_\text{eff}^5 N^{3.25}  & \text{if }\ L_\text{eff}\propto N^{1/4}
  \end{array} \right.
\end{equation}
Thus the volume reduction is computationally much more competitive in the case of the former scaling. Our calculations
suggest that the optimistic scenario is also possible in the untwisted model, however the matter is far from being as
well-established as it is in the twisted case\footnote{To even further confirm this we have made some preliminary
calculations of the Wilson Dirac spectrum with twisted boundary conditions and we find that at large $b$ the spectrum
resembles very closely that of the free fermion on the corresponding $L_\text{eff}^4=(\sqrt{N})^4$ lattice. We also see
no sign of the would-be ``zero modes'', which are present in the untwisted case.}.

It is also crucial to calculate more physical quantities, in particular the meson masses. This can be done on the single
site using the Quenched Momentum Prescription as described in Ref.~\cite{nn05}. However, as this method is quite novel
and it raises some controversies (see e.g.\ Ref.~\cite{tep09}), it would be very beneficial to compare its results to
the standard calculations that can be done by introducing one extended lattice direction. Calculation of the pion mass
would be an important crosscheck on our findings of the confining/conformal scenario at $N_f=2$. A setup with one
elongated direction also allows the calculation of glueball masses.

\backmatter

\end{document}